\documentclass[11pt]{article}
\usepackage[letterpaper,margin=1in,headheight=14pt]{geometry}
\usepackage{microtype}
\usepackage{amsmath,amssymb,amsthm,mathtools,bm}
\usepackage{booktabs,array}
\usepackage{graphicx}
\usepackage{tikz}
\usepackage{enumitem}
\usepackage{caption}
\usepackage{placeins}
\usepackage{xcolor}
\usepackage{xurl}
\usepackage[numbers,sort&compress]{natbib}
\usepackage{hyperref}
\usepackage[nameinlink,noabbrev]{cleveref}
\hypersetup{colorlinks=true,linkcolor=blue!45!black,citecolor=blue!45!black,
  urlcolor=blue!45!black,pdftitle={A Poisson Transition and Bias Law for KSG Mutual-Information Estimation},
  pdfauthor={Abhiram Gorle, Tsachy Weissman}}
\setlist{leftmargin=1.6em,itemsep=0.25em,topsep=0.4em}
\newtheorem{theorem}{Theorem}[section]
\newtheorem{proposition}[theorem]{Proposition}
\newtheorem{corollary}[theorem]{Corollary}
\newtheorem{lemma}[theorem]{Lemma}
\theoremstyle{definition}

\theoremstyle{remark}
\newtheorem{remark}[theorem]{Remark}
\newcommand{\R}{\mathbb R}
\newcommand{\E}{\mathbb E}
\newcommand{\Pp}{\mathbb P}
\newcommand{\Var}{\operatorname{Var}}

\newcommand{\Ihat}{\widehat I}
\newcommand{\phip}{\varphi_p}

\newcommand{\GP}{\operatorname{Gamma}}
\newcommand{\Pois}{\operatorname{Pois}}
\newcommand{\ind}{\mathbf 1}
\newcommand{\dd}{\,\mathrm d}

\title{A Poisson Transition and Bias Law for KSG\\Mutual-Information Estimation}
\author{Abhiram Gorle, Tsachy Weissman\\
\normalsize Stanford University, Stanford, CA, USA\\
\normalsize\texttt{\{abhiramg,tsachy\}@stanford.edu}}
\date{}

\begin{document}
\maketitle
\begin{abstract}
We quantify the error of the strict-radius
Kraskov--St\"ogbauer--Grassberger (KSG-1) mutual-information
estimator on a Gaussian channel as the sample size \(n\) grows
and the channel becomes nearly deterministic.
For fixed dimension \(p\) and neighbor index \(k\), we define
the resolution parameter \(u_n=ne^{-I_n}\), where \(I_n\) is
the true mutual information in nats.
When \(u_n\to u\in(0,\infty)\), we prove that the estimation
error converges in mean square to a computable deterministic
bias \(B_{p,k}(u)\). Sampling fluctuations therefore vanish,
while an additive bias can persist.
We derive this bias from an explicit marked-Poisson law for
the neighborhood counts, characterize its small-\(u\)
asymptotics, and show that it tends to zero as \(u\to\infty\).
The same law describes how often neighborhoods saturate,
meaning that they make the largest possible contribution
to the estimate. The limiting saturation probability decreases
continuously and strictly from one to zero, allowing a
Gaussian-model bias correction on fixed critical intervals.
Finite-sample experiments show substantial bias even when
the true information lies below the estimator's logarithmic
ceiling and saturated neighborhoods are rare.
Saturation-based correction reduces estimation error in these experiments,
though recovering the resolution parameter from the observed saturation
fraction can remain difficult when saturation is rare.
\end{abstract}

\section{Introduction}
\label{sec:introduction}

We study mutual information (MI) estimation for a Gaussian channel whose output becomes nearly determined by its input. As the channel noise vanishes, the mutual information grows and the joint distribution concentrates near the diagonal. A nearest-neighbor estimator must resolve
the shrinking width of this distribution from finitely many observations. We ask how the estimation error behaves when the sample size grows at the same time as the noise vanishes.

The Kraskov--St\"ogbauer--Grassberger estimator (KSG)
\citep{kraskov2004} measures dependence by comparing joint and marginal nearest-neighbor counts at a shared radius. For a fixed distribution, existing analyses establish consistency and bounds on bias and variance \citep{gao2018,zhao2020}. Here the channel noise shrinks with sample size, changing the density, so those fixed-distribution bias bounds need not apply uniformly. Fundamental limits on KSG estimation are well studied: the estimator has a ceiling of order \(\log n\), and estimating large MI accurately can require a sample size exponential in that information \citep{gao2015}. The ceiling bounds what KSG can report, but does not quantify its error when the true information lies below it. Our question is therefore not whether an
exponential sample requirement exists, but how KSG behaves at that scale.

Geometric methods address the local nonuniformity underlying this behavior \citep{lord2018}. Using nearest-neighbor limits, \citet{gao2018bandwidth} derive a bias correction for local-likelihood entropy estimators. They analyze a fixed distribution and obtain a limiting bias
independent of its density, so the correction can be precomputed. Here the limiting bias of strict-radius KSG depends on \(u\) because the Gaussian channel changes with sample size; our count law also gives the saturation probability. Practical bias diagnostics and error bars~\citep{holmes2019} assess estimation reliability, while distribution-free results~\citep{mcallester2020} limit the strength of high-confidence information lower bounds. Here we characterize KSG's limiting error within a specified Gaussian family. The channel's mutual information is known exactly, \(I_n=(p/2)\log(1/\delta_n)\), where \(p\) is the dimension and \(\delta_n\) controls the noise level. This lets us measure KSG's error against a known reference as the channel becomes nearly deterministic.
The resolution parameter we use throughout is
\[
u_n=n e^{-I_n}=n\delta_n^{p/2}.
\]
Here \(I_n\) is measured in nats. This compares the sample size \(n\) with \(e^{I_n}\), as the known exponential sample requirement suggests. It also has a geometric interpretation. The noise width is of order
\(\sqrt{\delta_n}\), while the source-neighbor spacing is of order \(n^{-1/p}\), up to a local density factor. Their ratio is of order \(u_n^{1/p}\). When \(u_n\) tends to a positive constant, neither scale becomes negligible relative to the other; this is the critical regime. For fixed dimension \(p\) and neighbor index \(k\), when
\(u_n\to u\in(0,\infty)\), the neighborhood counts converge to an explicit mixed-Poisson law. We use this law to prove that
\[
\Ihat_{n,k}-I_n\longrightarrow B_{p,k}(u)
\text{ }\text{in }L^2,
\]
where \(B_{p,k}(u)\) is the computable limiting bias. Along this shrinking-noise sequence, the estimation error concentrates around \(B_{p,k}(u)\), which need not be zero. \\
KSG chooses its radius from neighbors that are close in both \(X\) and \(Y\). At the same radius, the \(X\)-count requires closeness only in
\(X\), and the \(Y\)-count only in \(Y\). Both can therefore include
additional observations outside the joint neighborhood. In the critical
regime, rescaling distances by the noise width gives a marked Poisson
process: each source point carries a Gaussian noise value that determines
its output location. For fixed source and noise values of the query,
this process describes the selected joint radius and the extra marginal
counts. Averaging over the query values then gives the limiting count
law for a randomly chosen observation. \\
A neighborhood is \textit{saturated} when it makes the largest possible contribution to the estimate. The count law gives the limiting saturation fraction \(q_{p,k}(u)\). This curve is continuous and strictly decreasing, approaching one as \(u\to0\) and zero as \(u\to\infty\). Within the Gaussian model, this gives a bias correction: on a fixed compact critical interval, we invert the curve to estimate \(u\) from the observed saturation fraction and subtract the corresponding bias. \\
We also characterize the counts on either side of the critical regime. If \(u_n\to0\), a typical neighborhood saturates and the estimate grows like \(\log n\). If \(\delta_n\to0\) and \(u_n\to\infty\), the marginal counts grow on the \(\sqrt{u_n}\) scale. Separately, in the limiting Poisson law, the small-\(u\) bias approaches the gap between the ceiling and the true information, and the bias tends to zero as \(u\to\infty\).
Finite-sample experiments compare KSG's mean error and saturation
probability with independently computed Poisson predictions. They show substantial error even below the estimator's ceiling, when saturation is rare. The correction reduces this error, although estimating \(u\) from finite-sample saturation can limit its accuracy.

Our results build on local Poisson approximation and geometric bounds for
nearest-neighbor statistics \citep{bickelbreiman,penroseyukich,lastpenrose}.
We control marginal counts over regions that are unbounded in the other
coordinate, and establish convergence of the full estimator despite
dependence among neighborhoods. 
Section \ref{sec:problem} fixes the channel and estimator. Sections
\ref{sec:poisson} and \ref{sec:bias} derive the critical count law
and estimator limit. Section \ref{sec:results} treats the thin and
regular regimes and the bias endpoints, and Section
\ref{sec:saturation-calibration} derives saturation-based bias correction.
Numerical results appear in Section \ref{sec:evidence}.
Proofs of the supporting tail, geometric, and
integrability estimates appear in Appendices
\ref{app:local}--\ref{app:saturation}.

\section{Channel, estimator, and resolution scale}
\label{sec:problem}

\subsection{The Gaussian channel}

\begin{figure}[htbp]
\centering
\input{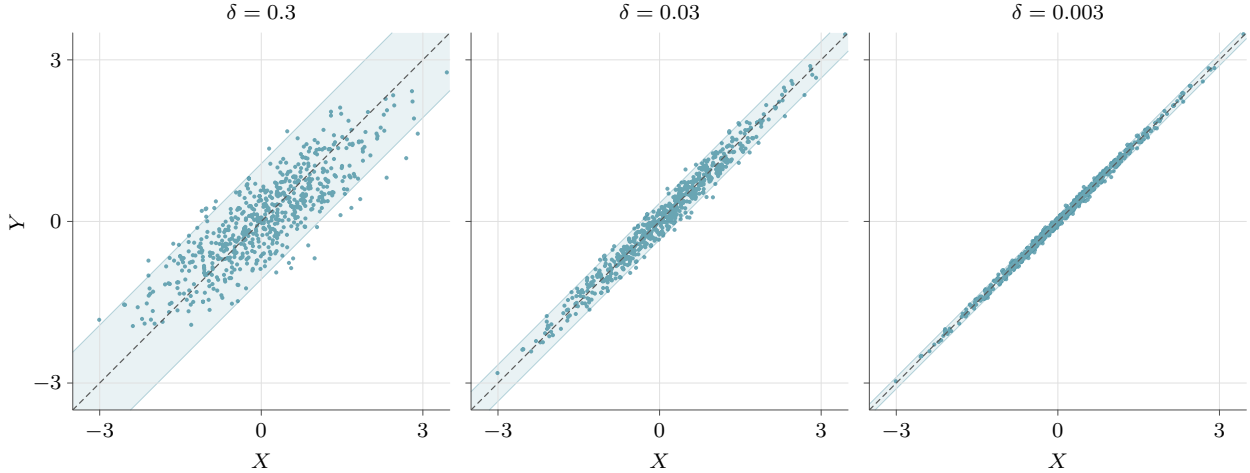}
\caption{The scalar Gaussian channel approaches \(Y=X\) as
\(\delta\) decreases. Each panel shows 600 pairs generated by
\cref{eq:channel} with \(p=1\), using the same source and noise draws across panels.}
\label{fig:gaussian-diagonal}
\end{figure}

We fix integers $p,k\ge1$, denoting the dimension and neighbor index, respectively. Both remain fixed in every limit. For each sample size $n>k$, draw $X_1,\ldots,X_n,Z_1,\ldots,Z_n$ independently from the standard Gaussian distribution $N(0,I_p)$. For $0<\delta_n<1$, we define
\begin{equation}
Y_{i,n}=(1-\delta_n)X_i+\sqrt{\delta_n(1-\delta_n)}Z_i
\label{eq:channel}
\end{equation}
The difference $X_i-Y_{i,n}$ is independent of $Y_{i,n}$ and has distribution $N(0,\delta_n I_p)$. Thus $\delta_n$ is the mean squared difference per coordinate.
\(\operatorname{Cov}(Y_n)=(1-\delta_n)I_p\), while
\(\operatorname{Cov}(Y_n| X)=\delta_n(1-\delta_n)I_p\).
Taking the difference of the corresponding Gaussian differential
entropies therefore gives the mutual information
\[
I_n:=I(X;Y_n)=\frac12\log
\frac{\det\{(1-\delta_n)I_p\}}
     {\det\{\delta_n(1-\delta_n)I_p\}}
=\frac p2\log\frac1{\delta_n}.
\]
As \(\delta_n\to0\), the output approaches the input and \(I_n\)
diverges. We use natural logarithms in the theory and report numerical
information errors in bits. \Cref{fig:gaussian-diagonal} illustrates how
the joint distribution concentrates around \(Y=X\) as the noise decreases.
The coordinate difference \(Y-X\) has variance \(\delta\), so the standard deviation perpendicular to \(Y=X\) is \(\sqrt{\delta/2}\).

\subsection{Strict-radius KSG-1}

For query \(i\) and competitor \(j\ne i\), let
\[
A_{ij}=\|X_j-X_i\|_\infty,
\text{ } B_{ij}=\|Y_{j,n}-Y_{i,n}\|_\infty,
\text{ } J_{ij}=\max\{A_{ij},B_{ij}\}.
\]
Under the max norm, the joint ball of radius \(r\) is a cube in
\(\R^{2p}\), or an axis-aligned square around the query when \(p=1\).
Let \(\varepsilon_{i,n}\) be the \(k\)th smallest
joint distance. The two marginal counts exclude the query and use a
strict inequality:
\begin{equation}
N_{x,i}=\#\{j\ne i:A_{ij}<\varepsilon_{i,n}\},
\text{ }
N_{y,i}=\#\{j\ne i:B_{ij}<\varepsilon_{i,n}\}.
\label{eq:counts}
\end{equation}
The estimator is
\begin{equation}
\Ihat_{n,k}=\psi(k)+\psi(n)
-\frac1n\sum_{i=1}^n
\{\psi(N_{x,i}+1)+\psi(N_{y,i}+1)\},
\label{eq:ksg}
\end{equation}
where \(\psi\) is the digamma function and
\(\psi(m+1)-\psi(m)=1/m\). We use this strict-radius KSG-1
convention \citep{kraskov2004} throughout, including its norm,
query exclusion, and \(+1\) shifts.

For the Gaussian sample, each joint or marginal distance list is tie-free
almost surely, and each competitor's two marginal distances are unequal.
The unique \(k\)th joint neighbor is
on the boundary in exactly one marginal. It is therefore excluded from one
marginal count and included in the other. The first \(k-1\) joint
neighbors contribute to both. After possibly exchanging the marginals, every
neighborhood has counts at least \((k-1,k)\), which gives
\begin{equation}
\Ihat_{n,k}\le C_{n,k}:=\psi(n)-\psi(k+1).
\label{eq:ceiling}
\end{equation}
A neighborhood makes the largest possible contribution to the estimate
exactly when its counts are \((k-1,k)\) or \((k,k-1)\).
We call such a neighborhood \emph{saturated} and define
\[
S_{i,n}=\ind\{\{N_{x,i},N_{y,i}\}=\{k-1,k\}\},
\text{ } \overline S_{n,k}=\frac1n\sum_{i=1}^nS_{i,n}.
\]
The estimator equals \(C_{n,k}\) exactly when every neighborhood is saturated.
The order-\(\log n\) ceiling and exponential sample obstruction are prior
results \citep{gao2015}. Their two-variable algebraic ceiling agrees with
\cref{eq:ceiling}, although their derivation uses a rectangle-based KSG-2
convention. Our count-pair argument and every result below concern
\cref{eq:counts,eq:ksg}.

\subsection{Noise width and source-neighbor spacing}

The noise width is of order \(\sqrt{\delta_n}\). Near a query
\(X_1=a\), a small cube of radius \(r\) contains approximately
\(n\phip(a)(2r)^p\) source observations on average. Keeping this
number fixed gives a source-neighbor spacing of order
\(\{n\phip(a)\}^{-1/p}\). Apart from the query-dependent density
factor, noise width divided by source-neighbor spacing is
\(u_n^{1/p}\). We use its \(p\)th power, \(u_n\), as the resolution
parameter:
\begin{equation}
u_n^{1/p}=\frac{\sqrt{\delta_n}}{n^{-1/p}}
\implies
u_n=n\delta_n^{p/2}=ne^{-I_n}.
\label{eq:u}
\end{equation}

The ratio \(u_n^{1/p}\) distinguishes three regimes:
\begin{itemize}
\item \emph{Thin regime:} \(u_n\to0\).
The noise width becomes small compared with the source-neighbor spacing,
so the source and output neighborhoods nearly coincide.

\item \emph{Critical regime:} \(u_n\to u\in(0,\infty)\).
The two scales remain comparable, so noise affects which observations
enter the joint and marginal neighborhoods.

\item \emph{Regular regime:} \(\delta_n\to0\) and \(u_n\to\infty\).
The source-neighbor spacing becomes small compared with the noise width.
Increasingly many source observations therefore lie within that width,
even though the noise itself vanishes.
\end{itemize}


\section{Limiting Neighborhood Counts}
\label{sec:poisson}
Beyond the \(k\) nearest joint neighbors, an observation can still contribute to a marginal count if it is close in \(X\) but not in \(Y\), or close in \(Y\) but not in \(X\). We characterize these extra counts in the critical regime \(u_n\to u\in(0,\infty)\).


\subsection{Rescaling the Neighborhood}

Condition on the query's source and noise values, \(X_1=a\) and
\(Z_1=b\). For every competitor, use noise-scale coordinates
\begin{equation}
\xi_{j,n}=\frac{X_j-a}{\sqrt{\delta_n}},
\text{ }
\eta_{j,n}=\frac{Y_{j,n}-Y_{1,n}}{\sqrt{\delta_n}}.
\label{eq:local-coordinates}
\end{equation}
Subtracting the query channel equation gives the exact relation
\begin{equation}
\eta_{j,n}=(1-\delta_n)\xi_{j,n}
+\sqrt{1-\delta_n}(Z_j-b).
\label{eq:local-channel}
\end{equation}

Rescaling both coordinates by the same factor leaves the nearest-neighbor ordering unchanged, while keeping the noise displacement on a fixed scale. In the critical regime, the point process formed by these rescaled observations converges to a Poisson point process on \(\R^p\times\R^p\) with intensity
\begin{equation}
\lambda_{a,b}(s,t)=u\phip(a)\phip(b+t-s).
\label{eq:raw-intensity}
\end{equation}
Here \(\phip\) is the standard \(p\)-dimensional Gaussian density. In the limiting process, the source points have constant intensity \(u\varphi_p(a)\). A source displacement \(s\) produces an output displacement \(t=s+W-b\), where \(W\sim N(0,I_p)\) is that point's noise value. This explains the second factor \(\phip(b+t-s)\) in the intensity.

We now change units so that a cube of radius \(r\) contains \((2r)^p\) points on average in either marginal. This makes the marginal counts easier to characterize. Set \(h=\{u\phip(a)\}^{1/p}\) and scale both coordinates by \(h\): \(S'=hs\), \(T'=ht\). The source points \(S'\) form a
homogeneous Poisson process of unit intensity. Attach to each point a mark \(W\sim N(0,I_p)\), independently across points and independently of the source process, and put
\begin{equation}
T'=S'+h(W-b).
\label{eq:marked}
\end{equation}
This also gives unit intensity in the output
marginal: the normalized joint intensity is
\(h^{-p}\phip\{b+(t'-s')/h\}\), whose integral in \(s'\) is
one. Thus a radius-\(r\) cube has mass \((2r)^p\) in either
marginal.

A point lies in the joint cube only when both \(\|S'\|_\infty\le r\)
and \(\|T'\|_\infty\le r\). For a fixed noise value \(W\), each source
coordinate must lie in two intervals of length \(2r\), whose centers are
\(h|W_q-b_q|\) apart. Their intersection has length
\((2r-h|W_q-b_q|)_+\), where \(x_+=\max(x,0)\). Multiplying these
lengths gives the volume of source locations satisfying both constraints.
Averaging over the independent Gaussian coordinates gives the mean number
of points in the joint cube
\begin{equation}
\Lambda_{h,b}(r)=\prod_{q=1}^p
\E_W[(2r-h|W_q-b_q|)_+].
\label{eq:Lambda}
\end{equation}
This function is continuous and strictly increasing from zero to infinity.
Indeed, the joint intensity is positive on every nonempty joint annulus,
and \(\Lambda_{h,b}(r)/(2r)^p\to1\) by dominated convergence.
As the joint cube expands, its expected number of points is \(\Lambda_{h,b}(r)\). Using this expected count, rather than the radius, to index the process gives a unit-rate Poisson counting process. The enclosed intensity when the \(k\)th point enters therefore has a \(\operatorname{Gamma}(k,1)\) distribution, so the selected radius \(R\) satisfies
\begin{equation}
\Lambda_{h,b}(R)=G_k,
\label{eq:gamma-radius}
\end{equation}
where \(G_k\sim\GP(k,1)\), with shape \(k\) and rate one.
Subtracting the joint mass from either marginal mass gives
\begin{equation}
\mu=\mu_u=(2R)^p-G_k.
\label{eq:mu}
\end{equation}
This is the intensity in each marginal-only region, and hence the conditional mean of each extra marginal count given the query values and \(G_k\). It is strictly positive and finite almost surely for every \(u>0\).


\begin{proposition}[Critical count law]
\label{prop:countlaw}
Draw mutually independent \(A,B\sim N(0,I_p)\), representing the query's
source and noise values, and \(G_k\sim\GP(k,1)\). Set \(h=\{u\phip(A)\}^{1/p}\), solve
\cref{eq:gamma-radius} with \(b=B\), and define \(\mu_u\) by
\cref{eq:mu}. Conditional on \((A,B,G_k)\), draw independent
\(P_x,P_y\sim\Pois(\mu_u)\) and an independent fair Bernoulli
variable \(J\). Put
\begin{equation}
(M_x,M_y)=
\begin{cases}
(k-1+P_x,\ k+P_y),&J=0,\\
(k+P_x,\ k-1+P_y),&J=1.
\end{cases}
\label{eq:count-functional}
\end{equation}
If \(u_n\to u\in(0,\infty)\), then
\[
(X_1,Z_1,N_{x,1},N_{y,1})\Rightarrow(A,B,M_x,M_y).
\]
In particular,
\begin{equation}
\Pp\{\{M_x,M_y\}=\{k-1,k\}\mid A,B,G_k\}=e^{-2\mu_u}.
\label{eq:sat-prob}
\end{equation}
Here \(P_x,P_y\) count the extra marginal-only neighbors, while \(J\) records which marginal includes the boundary neighbor.
\end{proposition}

\begin{proof}
Fix the query values \((X_1,Z_1)=(a,b)\). The exact density in
\cref{lem:app-local-density}, multiplied by \(n-1\), converges
locally uniformly to \cref{eq:raw-intensity}. Conditional independence
therefore gives the Poisson process limit in
\cref{lem:app-conditional-ppp}. To recover the marginal counts, we must also account for points close in \(X\) but far away in \(Y\), or vice versa. The radius and Gaussian tail bounds let us restrict to a large bounded region without changing the selected radius or either count, with arbitrarily high probability (\cref{lem:app-radius-tight,lem:app-remote}). Within this region, small perturbations preserve the selected neighbor, which marginal includes it, and every other point's contribution to the counts (\cref{lem:app-generic}). \Cref{lem:app-noncompact} therefore gives convergence of the marginal counts to those of the process with intensity \cref{eq:raw-intensity}.


Scale that process by \(h=\{u\phip(a)\}^{1/p}\) to obtain the marked representation \cref{eq:marked}. The Gamma arrival determines the joint radius through \cref{eq:gamma-radius}. Conditional on this radius and the points inside or on the boundary of the joint cube, the points outside form a Poisson process with its intensity restricted to the exterior (\cref{lem:app-post-arrival}). The two marginal-only regions are disjoint and each has intensity \(\mu_u\). Their count law depends on the interior and boundary points only through \(R\). Thus, given the query values and \(G_k\), these counts are independent \(\Pois(\mu_u)\) variables and are independent of
the boundary orientation.

The first \(k-1\) joint neighbors enter both marginal balls, while the boundary neighbor enters exactly one. The symmetry
\((s',t')\mapsto(-t',-s')\) preserves the conditional intensity and joint cube while exchanging the boundary marginals, so the two orientations are equally likely. This gives the independent fair Bernoulli variable \(J\) and the count law \cref{eq:count-functional}, as established in
\cref{prop:app-countlaw}.


For a bounded continuous test function of the query values and counts, the conditional expectation now converges at each fixed \((a,b)\). Integrating against the fixed Gaussian query law by bounded convergence gives the joint convergence with \((A,B)\). Finally, saturation in \cref{eq:count-functional} is exactly \(P_x=P_y=0\). Conditional independence gives its probability \(e^{-2\mu_u}\), proving \cref{eq:sat-prob}.
\end{proof}

\section{Estimator Bias}
\label{sec:bias}

The limiting counts are random, but their expected KSG score determines the estimator's limiting bias:
\begin{equation}
B_{p,k}(u)=\log u+\psi(k)
-\E[\psi(M_x+1)+\psi(M_y+1)].
\label{eq:B-definition}
\end{equation}
For computation, define the Poisson-averaged score, for integers \(m\ge1\),
\begin{equation}
g_m(\mu)=\E[\psi\{m+\Pois(\mu)\}].
\label{eq:gm}
\end{equation}
Either orientation in \cref{eq:count-functional} has conditional
expected score sum \(g_k(\mu_u)+g_{k+1}(\mu_u)\). Hence the
equivalent computational representation is
\[
B_{p,k}(u)=\log u+\psi(k)
-\E[g_k(\mu_u)+g_{k+1}(\mu_u)],
\]
where this outer expectation is over \((A,B,G_k)\). To compute the bias, draw \(A,B,G_k\), solve one scalar equation for
\(R\), and evaluate the one-dimensional overlap integrals. There is no
need to simulate the entire Poisson process.

\begin{theorem}[Critical estimator bias]
\label{thm:bias}
Under \cref{eq:channel,eq:counts,eq:ksg}, with fixed \(p,k\), if
\(u_n\to u\in(0,\infty)\), then
\begin{equation}
\Ihat_{n,k}-I_n\longrightarrow B_{p,k}(u)
\text{ }\text{in }L^2.
\label{eq:bias-limit}
\end{equation}
\end{theorem}

\begin{proof}
We first show that the mean error converges to \(B_{p,k}(u)\), then
show that the estimator's variance tends to zero.

\noindent\emph{Limiting mean.} For observation \(i\), write
\[
\ell_{i,n}=\psi(k)-\psi(N_{x,i}+1)-\psi(N_{y,i}+1),
\]
and let
\(\ell^{(u)}=\psi(k)-\psi(M_x+1)-\psi(M_y+1)\).
The channel has \(I_n=\log n-\log u_n\), so the exact estimator
formula gives
\begin{equation}
\Ihat_{n,k}-I_n
=\psi(n)-\log n+\log u_n
 +\frac1n\sum_{i=1}^n\ell_{i,n}.
\label{eq:app-exact-centered-ksg}
\end{equation}

\Cref{prop:countlaw} gives convergence in distribution of the counts in a typical neighborhood. The finite-sample moment bound \cref{lem:app-fractional-moments} supplies, for every \(0<\alpha<1\),
\[
\sup_{n>k}\E\{(N_{x,1}+1)^\alpha+(N_{y,1}+1)^\alpha\}<\infty.
\]
For every finite \(q\), \( |\psi(m+1)|^q\le C_{q,\alpha}\{1+(m+1)^\alpha\}\) for \(m\ge0\). Choosing \(q>2\) makes the neighborhood scores and their squares uniformly integrable. These bounds allow us to pass from convergence in distribution
to convergence of the expected scores
(\cref{lem:app-score-moment}): \(\E\ell_{1,n}\to\E\ell^{(u)}\). By exchangeability,
\[
\E\left[\frac1n\sum_i\ell_{i,n}\right]
=\E\ell_{1,n}\longrightarrow\E\ell^{(u)}.
\]
Since \(\psi(n)-\log n\to0\) and \(\log u_n\to\log u\),
\cref{eq:app-exact-centered-ksg,eq:B-definition} imply
\(\E(\Ihat_{n,k}-I_n)\to B_{p,k}(u)\).

\noindent\emph{Variance.} The neighborhood scores are computed from the same sample and are not independent. The replacement bound \cref{thm:app-ksg-variance} gives

\[
\Var(\Ihat_{n,k})
\le C_{p,k}\frac{(1+\log n)^2}{n}\longrightarrow0.
\]
Replacing one observation changes only a bounded number of joint \(k\)th radii, but can change many more marginal counts. For a unit marginal-count change, the digamma score changes by
\(1/m\), where \(m\) is the larger count. Summing these changes by marginal rank using \cref{lem:app-harmonic-reverse} bounds the change in the sum of
neighborhood scores by \(O_{p,k}(\log n)\). Applying the Efron--Stein inequality gives the variance bound above,
with a constant \(C_{p,k}\) that does not depend on \(\delta_n\).



So,
\[
\E|\Ihat_{n,k}-I_n-B_{p,k}(u)|^2
=\Var(\Ihat_{n,k})
 +\{\E(\Ihat_{n,k}-I_n)-B_{p,k}(u)\}^2
\longrightarrow0.
\]
This proves \cref{eq:bias-limit}.
\end{proof}

\begin{remark}[Additive error and sampling fluctuations]
The theorem concerns a fixed critical limit \(u\in(0,\infty)\), with
\(p,k\) fixed and scores measured in nats. In particular,
\[
\begin{aligned}
\E[\Ihat_{n,k}-I_n]&\longrightarrow B_{p,k}(u),\\
\E[(\Ihat_{n,k}-I_n)^2]&\longrightarrow B_{p,k}(u)^2.
\end{aligned}
\]
Thus \(B_{p,k}(u)\) is the limiting bias, and its square is the limiting mean-squared error. Although \(I_n=\log n-\log u_n\) diverges, its additive estimation error can persist after sampling fluctuations vanish. A relative-error plot can
therefore conceal the bias. The variance order above is known
\citep{gao2018,zhao2020}; the exact-convention proof makes its
validity at every positive noise level explicit.
\end{remark}

The same law gives the limiting saturation probability
\(q_{p,k}(u)=\E e^{-2\mu_u}\). In a fixed critical regime,
\(\overline S_{n,k}\to q_{p,k}(u)\) in \(L^2\), as proved with
the curve's continuity, strict decrease, and endpoints in
\cref{cor:saturation}. Saturation records whether both extra marginal-only counts are zero and the bias depends on the full count distribution. A low saturation fraction need not imply a small error.


\section{Bias Endpoints and Count Regimes}
\label{sec:results}

We first study \(B_{p,k}(u)\) as \(u\to0\) and \(u\to\infty\).
Additionally, we establish count limits for finite Gaussian samples along sequences with \(u_n\to0\) or \(u_n\to\infty\).

\subsection{Endpoints of the limiting bias}

\begin{theorem}[Endpoints of the critical bias]
\label{thm:endpoints}
For every fixed \(p,k\ge1\),
\begin{align}
B_{p,k}(u)&=\log u-\psi(k+1)+O_{p,k}(u^{1/p}),
&u\downarrow0,
\label{eq:thin-endpoint}\\
B_{p,k}(u)&\longrightarrow0,
&u\uparrow\infty.
\label{eq:regular-endpoint}
\end{align}
\end{theorem}

\begin{proof}
Use the common \((A,B,G)\) construction in \cref{prop:countlaw}, with \(G\sim\GP(k,1)\), and write \(\mu=\mu_u\). By \cref{eq:gm,eq:B-definition},
\[
B_{p,k}(u)=\log u+\psi(k)
-\E\{g_k(\mu)+g_{k+1}(\mu)\}.
\]
For the thin endpoint, condition on \(\mu\) and let
\(P\sim\Pois(\mu)\). The digamma recurrence gives, for every integer
\(m\ge1\),
\begin{equation}
0\le\psi(m+P)-\psi(m)
=\sum_{r=0}^{P-1}\frac1{m+r}\le\frac Pm.
\label{eq:app-digamma-small-mu}
\end{equation}
Thus \(0\le g_m(\mu)-\psi(m)\le\mu/m\), and
\[
-\left(\frac1k+\frac1{k+1}\right)\E\mu
\le B_{p,k}(u)-\{\log u-\psi(k+1)\}\le0.
\]
The bound \(\E\mu\le C_{p,k}u^{1/p}\) from
\cref{lem:app-small-u-mu} proves \cref{eq:thin-endpoint}, including \(p=1\). For the regular endpoint, \cref{lem:app-large-u-scaling} gives
\[
\frac{\mu}{\sqrt u}\longrightarrow
\left\{\frac{G\phip(A)}{\phip(B)}\right\}^{1/2}
\text{ almost surely}.
\]
The logarithmic convergence in \cref{lem:app-log-mu} then gives
\[
2\E\log\mu-\log u
\longrightarrow
\E\log G+\E\log\phip(A)-\E\log\phip(B)=\psi(k).
\]
Here the Gaussian terms cancel because \(A,B\) are identically
distributed, and the Gamma integral gives \(\E\log G=\psi(k)\).
To control the difference between the Poisson scores and their logarithms,
\cref{lem:app-poisson-digamma} gives, for \(N\sim\Pois(\mu)\),
\[
g_m(\mu)-\log\mu
=E_1(\mu)+\sum_{j=1}^{m-1}\E\frac1{N+j},
\text{ } E_1(\mu)=\int_\mu^\infty\frac{e^{-t}}t\,\dd t.
\]
This nonnegative correction tends to zero as \(\mu\to\infty\) and is
bounded by \(C_m+(-\log\mu)_+\). For \(u\ge1\), set
\(M=G+\mu\) and \(\vartheta=\mu/M\). Then
\[
(-\log\mu)_+
\le\left(-\log\frac M{\sqrt u}\right)_+
+(-\log\vartheta)_+.
\]
The bounds in the proof of \cref{lem:app-log-mu} give a nonnegative random variable \(H=H(A,B,G)\), independent of \(u\), such that
\(\E H<\infty\) and \((-\log\mu_u)_+\le H\) for all \(u\ge1\).
Since \(\mu\to\infty\) almost surely, dominated convergence proves
\begin{equation}
\E\{g_k(\mu)+g_{k+1}(\mu)-2\log\mu\}\longrightarrow0.
\label{eq:app-poisson-correction-vanishes}
\end{equation}
Consequently,
\[
B_{p,k}(u)=\log u+\psi(k)-2\E\log\mu+o(1)
\longrightarrow0,
\]
which proves \cref{eq:regular-endpoint}.
\end{proof}
The leading thin-endpoint term is the limiting difference between the ceiling \(C_{n,k}\) and the truth \(I_n\). For every finite \(u>0\), \(\mu_u>0\) and strict increase of \(\psi\) also give
\[
B_{p,k}(u)<\log u-\psi(k+1).
\]
The expansion therefore implies negative bias for all sufficiently small \(u\), but this ceiling comparison does not establish a global sign or monotonicity for \(B_{p,k}\).

\subsection{Thin and Regular Regimes}

\begin{theorem}[Neighborhood count regimes]
\label{thm:phase}
For the channel and counts in \cref{eq:channel,eq:counts}, with fixed \(p,k\):
\begin{enumerate}[label=\textup{(\alph*)}]
\item If \(u_n\to0\), then
\[
\Pp(S_{1,n}=1)\to1,
\text{ } \overline S_{n,k}\to1\text{ in }L^1,
\text{ } \frac{\Ihat_{n,k}}{\log n}\to1\text{ in }L^1.
\]
\item If \(u_n\to u\in(0,\infty)\), the query values and count
pair converge jointly as in \cref{prop:countlaw}.
\item If \(\delta_n\to0\) and \(u_n\to\infty\), then
\begin{equation}
\left(\frac{N_{x,1}}{\sqrt{u_n}},
\frac{N_{y,1}}{\sqrt{u_n}}\right)
\Rightarrow(L_{p,k},L_{p,k}),
\text{ }
L_{p,k}=\left\{G_k\frac{\phip(U)}{\phip(V)}\right\}^{1/2},
\label{eq:regular-count}
\end{equation}
where \(U,V\sim N(0,I_p)\) and \(G_k\sim\GP(k,1)\) are
mutually independent. Consequently, \(\overline S_{n,k}\to0\) in
\(L^1\).
\end{enumerate}
\end{theorem}

\begin{proof}
Part~\textup{(a)} is proved in \cref{thm:tr-thin-collapse}. On the
source-neighbor scale, the noise vanishes and the extra marginal-only
counts disappear. The fraction of nonsaturated neighborhoods tends to
zero, and each has score deficit \(O_k(\log n)\); together these give
the normalized estimator limit. Part~\textup{(b)} is \cref{prop:countlaw}.

For part~\textup{(c)}, \cref{thm:tr-regular-counts} applies under the stated assumptions \(\delta_n\to0\) and \(u_n\to\infty\). It proves joint convergence of the query values, scaled joint radius, and scaled marginal counts. Keeping only the two marginal counts gives \cref{eq:regular-count}, with \(L_{p,k}\) as stated. Since \(L_{p,k}>0\) almost surely, the count limit implies divergence of each marginal count in probability: for any fixed \(M\) and \(\epsilon>0\), eventually \(M/\sqrt{u_n}<\epsilon\), so convergence in distribution gives
\[
\limsup_n\Pp(N_{x,1}\le M)\le\Pp(L_{p,k}\le\epsilon).
\]
Let \(\epsilon\downarrow0\), and use the same argument for
\(N_{y,1}\). Saturation forces \(N_{x,1}\le k\), so
exchangeability yields
\[
\E|\overline S_{n,k}|=\Pp(S_{1,n}=1)
\le\Pp(N_{x,1}\le k)\longrightarrow0,
\]
completing part~\textup{(c)}.
\end{proof}

Part~\textup{(a)} concerns a typical neighborhood and the estimator divided by \(\log n\). It does not assert that every neighborhood saturates or that the unnormalized gap from the ceiling vanishes. The shrinking-noise assumption in part~\textup{(c)} is essential: fixed positive noise can give different marginal limit constants, as \cref{rem:tr-delta-necessary} shows.

\begin{proposition}[Moments of the regular count limit]
\label{prop:tr-regular-moments}
The variable \(L_{p,k}\) in \cref{eq:regular-count} is positive and finite almost surely. For \(0<q<2\),
\begin{equation}
\E L_{p,k}^q
=\frac{\Gamma(k+q/2)}{\Gamma(k)}
(1-q^2/4)^{-p/2}.
\label{eq:tr-L-moments}
\end{equation}
For every \(q\ge2\), this moment is infinite. In particular,
\(\E L_{p,k}<\infty\) but \(\E L_{p,k}^2=\infty\).
\end{proposition}

\begin{proof}
The normalizing constants cancel in the density ratio:
\[
\frac{\phip(U)}{\phip(V)}
=\exp\left\{\frac{\|V\|_2^2-\|U\|_2^2}{2}\right\}.
\]
Consequently,
\begin{equation}
L_{p,k}
=G_k^{1/2}
\exp\left\{\frac{\|V\|_2^2-\|U\|_2^2}{4}\right\},
\label{eq:tr-L-exponential}
\end{equation}
which is positive and finite almost surely. For \(q>0\), Tonelli's theorem and independence give
\begin{equation}
\E L_{p,k}^q
=\E G_k^{q/2}\,
\E\exp\{q\|V\|_2^2/4\}\,
\E\exp\{-q\|U\|_2^2/4\}.
\label{eq:tr-L-factorization}
\end{equation}
The Gamma moment is
\[
\E G_k^{q/2}=\frac{\Gamma(k+q/2)}{\Gamma(k)}.
\]
For a standard \(p\)-variate Gaussian \(W\),
\[
\E e^{t\|W\|_2^2}=(1-2t)^{-p/2}
\text{ if }t<\frac12,
\]
and the expectation is infinite when \(t\ge1/2\). Thus, for \(0<q<2\),
\[
\E e^{q\|V\|_2^2/4}=(1-q/2)^{-p/2},
\text{ }
\E e^{-q\|U\|_2^2/4}=(1+q/2)^{-p/2}.
\]
Multiplication in \cref{eq:tr-L-factorization} proves
\cref{eq:tr-L-moments}. If \(q\ge2\), the \(V\)-factor is infinite, while the Gamma and \(U\)-factors are strictly positive and finite. Tonelli's theorem therefore gives \(\E L_{p,k}^q=\infty\). Taking \(q=1\) and \(q=2\) proves the final assertion.
\end{proof}

\begin{corollary}[Polynomial shrinking-noise paths]
\label{cor:paths}
If \(\delta_n=n^{-2a/p}\), \(a>0\), then \(I_n=a\log n\) and
\(u_n=n^{1-a}\). For \(a>1\),
\[
\frac{\Ihat_{n,k}}{I_n}\longrightarrow\frac1a
\text{ }\text{in }L^1.
\]
For \(a=1\), \(\Ihat_{n,k}-I_n\to B_{p,k}(1)\) in \(L^2\).
For \(0<a<1\), the regular count law \cref{eq:regular-count} applies.
\end{corollary}

\begin{proof}
If \(a>1\), then \(u_n=n^{1-a}\to0\). By \cref{thm:phase},
\(\Ihat_{n,k}/\log n\to1\) in \(L^1\), so
\[
\frac{\Ihat_{n,k}}{I_n}
=\frac1a\frac{\Ihat_{n,k}}{\log n}
\longrightarrow\frac1a\text{ in }L^1.
\]
If \(a=1\), then \(u_n=1\), and \cref{thm:bias} gives the stated \(L^2\) bias limit. If \(0<a<1\), then \(u_n=n^{1-a}\to\infty\) and \(\delta_n=n^{-2a/p}\to0\), exactly the two assumptions of \cref{thm:phase} part~\textup{(c)}.
\end{proof}

\begin{remark}[Count limits and score limits]
The infinite second moment of \(L_{p,k}\) concerns the regular count coefficient; it does not contradict the \(L^2\) convergence of the logarithm-like estimator score in a fixed critical window. Neither the regular count law nor the endpoint \(B_{p,k}(u)\to0\) proves estimator consistency along every sequence with \(u_n\to\infty\). The endpoint takes \(n\to\infty\) at fixed \(u\) before increasing \(u\).
To deduce consistency along such sequences, one needs bounds on the estimator’s error that remain valid as \(u\) grows with \(n\).
\end{remark}

\section{Saturation-Based Bias Correction}
\label{sec:saturation-calibration}

Within the Gaussian model, the limiting saturation fraction determines \(u\). We estimate \(u\) from the observed saturation fraction, then subtract the predicted bias at that estimate. The correction is consistent on a fixed critical interval.

\subsection{The saturation curve}

\begin{corollary}[Critical saturation curve]
\label{cor:saturation}
Let \(q_{p,k}(u)=\E e^{-2\mu_u}\). This function is continuous and strictly decreasing on \((0,\infty)\), takes values in \((0,1)\), and satisfies
\begin{align}
1-q_{p,k}(u)&=O_{p,k}(u^{1/p}),&&u\downarrow0,
\label{eq:q-small}\\
q_{p,k}(u)&\longrightarrow0,&&u\uparrow\infty.
\label{eq:q-large}
\end{align}
If \(u_n\to u\in(0,\infty)\), then
\begin{equation}
\overline S_{n,k}\longrightarrow q_{p,k}(u)
\text{ }\text{in }L^2.
\label{eq:q-empirical}
\end{equation}
\end{corollary}

\begin{proof}
Use one common realization of mutually independent \(A,B\sim N(0,I_p)\) and \(G_k\sim\GP(k,1)\) for all \(u>0\). Put \(h_u=\{u\phip(A)\}^{1/p}\), let \(R_u\) solve \(\Lambda_{h_u,B}(R_u)=G_k\), and set \(\mu_u=(2R_u)^p-G_k\). Fix finite query values \((a,b)\), a positive arrival \(G_k\), and \(r>0\). For \(0<h_1<h_2\), each difference
\[
(2r-h_1|W_j-b_j|)_+-(2r-h_2|W_j-b_j|)_+
\]
is nonnegative and is strictly positive on
\(0<|W_j-b_j|<2r/h_2\), an event of positive Gaussian probability.
Every overlap factor is positive and strictly decreasing in \(h\),
so \(\Lambda_{h_2,b}(r)<\Lambda_{h_1,b}(r)\).
Because \(\Lambda_{h,b}(r)\) decreases with \(h\) and increases
with \(r\), the radius satisfying \(\Lambda_{h,b}(R)=G_k\) must
increase with \(h\). Thus \(R_u\), and hence \(\mu_u\), is strictly
increasing in \(u\).

The overlap factors are jointly continuous in \((h,r)\), by dominated
convergence on compact sets. To check root continuity at \(h_0\), let
\(R_0\) be its root and choose \(0<\epsilon<R_0\). Strict
radial increase puts \(G_k\) strictly between
\(\Lambda_{h_0,b}(R_0-\epsilon)\) and
\(\Lambda_{h_0,b}(R_0+\epsilon)\). Joint continuity preserves
these two inequalities for \(h\) near \(h_0\), trapping the root
in \((R_0-\epsilon,R_0+\epsilon)\). Thus \(R_u\) and \(\mu_u\)
depend continuously on \(u\).

Because \(0<\mu_u<\infty\) almost surely,
\(e^{-2\mu_u}\in(0,1)\). Bounded convergence gives continuity of
\(q_{p,k}\). If \(u_1<u_2\), then
\(e^{-2\mu_{u_1}}-e^{-2\mu_{u_2}}>0\) almost surely;
its expectation is strictly positive. Hence \(q_{p,k}\) is strictly
decreasing and lies in \((0,1)\). For \(0<u\le1\), the small-strip-mass bound
\cref{lem:app-small-u-mu} gives
\[
0\le1-q_{p,k}(u)
=\E(1-e^{-2\mu_u})
\le2\E\mu_u\le C_{p,k}u^{1/p},
\]
proving \cref{eq:q-small} and the limit one at zero. The large-\(u\)
overlap scaling \cref{lem:app-large-u-scaling} gives
\[
\frac{\mu_u}{\sqrt u}\longrightarrow
\left\{G_k\frac{\phip(A)}{\phip(B)}\right\}^{1/2}>0
\text{ }\text{almost surely}.
\]
Thus \(e^{-2\mu_u}\to0\), and bounded convergence proves
\cref{eq:q-large}.

Since both the finite-sample and limiting counts are integer-valued,
\cref{prop:countlaw} gives convergence of the saturation probability.
By exchangeability,
\[
\E\overline S_{n,k}
=\Pp(S_{1,n}=1)\longrightarrow q_{p,k}(u).
\]
\Cref{lem:saturation-variance} gives
\[
\Var(\overline S_{n,k})\le\frac{D_{p,k}^2}{2n},
\text{ }
D_{p,k}=1+2\kappa_{2p}k+4\kappa_p(k+1).
\]
Here \(\kappa_d\) depends only on the dimension: one observation
can be among the first \(m\) nearest neighbors of at most
\(\kappa_d m\) other observations
(\cref{lem:app-reverse-knn}).

After one observation is replaced, at most \(2\kappa_{2p}k\)
common query radii change.
At an unchanged radius, an indicator can change only if one configuration
is saturated. Its marginal counts are at most \(k\), and the other
configuration's counts are at most \(k+1\). The old or new point
responsible for a change therefore has marginal rank at most \(k+1\).
The reverse-neighbor bound in each marginal space controls these queries
by \(4\kappa_p(k+1)\); adding the replaced query gives \(D_{p,k}\).
Efron--Stein yields the variance inequality without assuming independent neighborhood scores.
Finally,
\[
\E|\overline S_{n,k}-q_{p,k}(u)|^2
=\Var(\overline S_{n,k})
 +\{\E\overline S_{n,k}-q_{p,k}(u)\}^2\longrightarrow0,
\]
which proves \cref{eq:q-empirical}.
\end{proof}

\subsection{Bias Correction on a Fixed Interval}

The continuous, strictly decreasing saturation curve \(q=q_{p,k}\)
maps \((0,\infty)\) onto \((0,1)\), so it can be inverted to estimate
\(u\). For a fixed interval \(K=[a,b]\), we first clip the
observed saturation fraction to \([q(b),q(a)]\), moving values outside
this range to its nearest endpoint. This keeps the estimated parameter
within \(K\). The same sample is used to estimate \(u\) and to compute KSG.

\begin{corollary}[Bias correction on a fixed interval]
\label{cor:calibration}
Fix \(p,k\), and choose \(K=[a,b]\) with \(0<a<b<\infty\).
Suppose the channel and estimator satisfy \cref{eq:channel,eq:counts,eq:ksg}
and \(u_n\to u\in K\). Write \(q=q_{p,k}\), \(B=B_{p,k}\),
and let \(\Pi_{[c,d]}(s)=\min\{d,\max\{c,s\}\}\) denote clipping.
The function \(B\) is continuous on \((0,\infty)\). Define
\begin{equation}
\widehat u_n
=q^{-1}\!\left(\Pi_{[q(b),q(a)]}(\overline S_{n,k})\right),
\text{ }
\widehat I^{\mathrm{cal}}_{n,k}
=\Ihat_{n,k}-B(\widehat u_n).
\label{eq:calibration}
\end{equation}
Then
\begin{equation}
\widehat u_n\longrightarrow u,
\text{ }
\widehat I^{\mathrm{cal}}_{n,k}-I_n\longrightarrow0
\text{ }\text{in }L^2.
\label{eq:calibration-limit}
\end{equation}
\end{corollary}

\begin{proof}
\Cref{lem:bias-continuity} proves that \(B_{p,k}\) is finite and
continuous. It is therefore bounded and uniformly continuous on the
fixed interval \(K\). We use these properties below to show that
estimating \(u\) consistently also gives a consistent bias correction.

Let \(J=[q(b),q(a)]\). By \cref{cor:saturation}, the restriction
of \(q\) to \(K\) is a continuous, strictly decreasing bijection
onto \(J\). A continuous strictly monotone function on a closed
interval has a continuous inverse. Therefore
\[
F(s)=q^{-1}(\Pi_J(s)),\text{ }s\in[0,1],
\]
is continuous, takes values in \(K\), and satisfies \(F(q(u))=u\),
including when \(u=a\) or \(u=b\).

Given \(\epsilon>0\), choose \(\eta>0\) such that
\( |F(s)-u|<\epsilon\) whenever \( |s-q(u)|<\eta\).
The empirical mean-square limit \cref{eq:q-empirical} and Markov's
inequality give
\begin{align*}
\E|\widehat u_n-u|^2
&\le\epsilon^2+(b-a)^2
\Pp\{|\overline S_{n,k}-q(u)|\ge\eta\}\\
&\le\epsilon^2+\frac{(b-a)^2}{\eta^2}
\E|\overline S_{n,k}-q(u)|^2.
\end{align*}
Hence \(\limsup_n\E|\widehat u_n-u|^2\le\epsilon^2\);
letting \(\epsilon\downarrow0\) proves \(\widehat u_n\to u\)
in \(L^2\).

For the bias subtraction, let \(M=\sup_{v\in K}|B(v)|<\infty\).
Uniform continuity gives, for any \(\epsilon>0\), an \(\eta>0\)
such that \( |B(v)-B(u)|<\epsilon\) when \( |v-u|<\eta\).
Then
\[
\E|B(\widehat u_n)-B(u)|^2
\le\epsilon^2+4M^2\Pp\{|\widehat u_n-u|\ge\eta\}.
\]
Thus \(\limsup_n\E|B(\widehat u_n)-B(u)|^2\le\epsilon^2\).
Letting \(\epsilon\downarrow0\) proves
\(B(\widehat u_n)\to B(u)\) in \(L^2\). Finally,
\begin{align*}
\E|\widehat I^{\mathrm{cal}}_{n,k}-I_n|^2
&\le2\E|\Ihat_{n,k}-I_n-B(u)|^2
   +2\E|B(\widehat u_n)-B(u)|^2\\
&\longrightarrow0,
\end{align*}
by \cref{thm:bias}. No independence between the estimated bias and
the original KSG score is used. The argument needs only \(u_n\to u\in K\),
not that every \(u_n\) belongs to \(K\).
\end{proof}

\begin{remark}[Scope of bias correction]
The interval \(K\) is fixed before seeing the sample. Clipping
keeps the estimated parameter within \(K\), but finite-sample
saturation may still give an inaccurate estimate of \(u\),
particularly when saturation is rare. The corollary uses exact
curves. For numerical lookup tables, \cref{prop:numerical-calibration}
preserves consistency when both curve approximations converge uniformly
on \(K\) and each approximating saturation curve is continuous and
strictly decreasing. The fixed tables used in the experiments are
checked numerically in Appendix~\ref{app:repro}.
\end{remark}


\section{Numerical Results}
\label{sec:evidence}

We test whether the Poisson law predicts KSG's finite-sample bias and saturation probability, and whether estimation errors concentrate around the predicted bias as the sample size grows. We then test saturation-based bias correction and examine the large-\(u\) behavior of the limiting bias. The Gaussian channel's exact mutual information provides the reference for the finite-sample comparisons. Numerical information errors are reported in bits; saturation differences are in percentage points.

\subsection{Bias and Saturation}
\label{sec:simulation-study}

We use dimensions \(p=1,2,4\), neighbor indices \(k=3,5,10\),
and six sample sizes between \(250\) and \(8000\), with
30 independent datasets per setting. Within each repetition, we
reuse source and noise draws across channel settings and values of \(k\). At fixed \(u<n\), we set \(\delta_n=(u/n)^{2/p}\), so increasing \(n\) reduces the channel noise and raises the true information. These comparisons test whether KSG's error approaches the predicted bias as the channel becomes increasingly deterministic. Appendix~\ref{app:repro} gives the simulation methods and numerical checks.

\begin{figure}[tb]
\centering
\includegraphics[width=\linewidth]{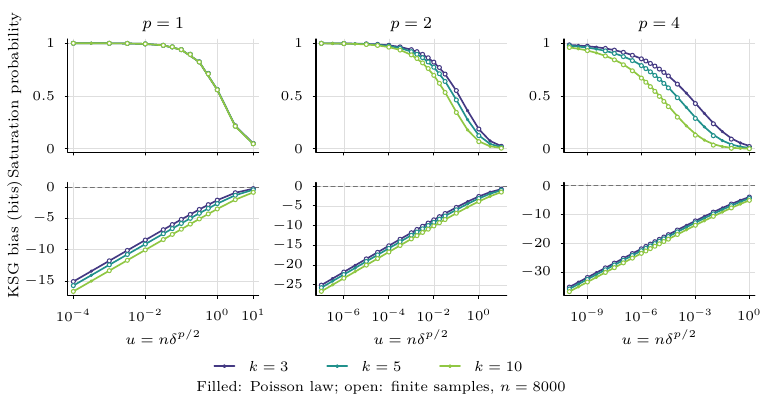}
\caption{Critical saturation probability (top) and additive bias (bottom). Columns separate dimensions \(p=1,2,4\); colors distinguish \(k=3,5,10\). Open finite-sample points are means at \(n=8000\), with 95\% Student-\(t\) intervals across 30 independent repetitions per setting. Each dimension uses the \(u\)-range shown;
the dimension-four extension is shown separately in
Figure \ref{fig:endpoint}.}
\label{fig:transition}
\end{figure}

\noindent Figure~\ref{fig:transition} compares Poisson predictions (curves) with finite-sample means at \(n=8000\) (open circles) as \(u\) varies. The upper panels show saturation probability; the lower panels show mean KSG error in bits. Columns distinguish dimensions and colors distinguish \(k\). The predictions are computed independently of the finite-sample datasets, using 30,000 Monte Carlo draws per reference setting.

Across 114 matched settings at \(n=8000\), the root-mean-square
differences between measured and predicted means are \(0.0024\) bit for bias and \(0.32\) percentage points for saturation probability. The theory therefore accurately predicts KSG's mean error over the tested grid. Some saturation discrepancies exceed their Monte Carlo uncertainty; their relative size is examined below.

\subsection{Bias Below the Ceiling}

KSG can remain substantially biased even when the true information lies below its ceiling. At \(n=8000\), \(k=5\), and \(u=10\), the true information lies about \(0.86\) bit below the ceiling in all three dimensions. Yet KSG underestimates it by about \(2.4\) bits on average
in dimension four, while almost no neighborhoods saturate
(Table~\ref{tab:bias-saturation}). This dimension-four setting lies outside the original plotted Poisson grid.

\begin{table}[tb]
\centering
\caption{Finite-sample bias and neighborhood saturation at \(n=8000\), \(k=5\),
\(u=10\). Intervals are 95\% Student-\(t\) intervals for the means across 30 independent repetitions; saturation is expressed as a percentage.}
\label{tab:bias-saturation}
\begin{tabular}{@{}ccc@{}}
\toprule
Dimension & Mean bias in bits (95\% CI) & Saturation (95\% CI)\\
\midrule
1 & $-0.4064 \pm 0.0052$ & $(4.737 \pm 0.102)\%$\\
2 & $-1.0474 \pm 0.0065$ & $(1.222 \pm 0.057)\%$\\
4 & $-2.3976 \pm 0.0066$ & $(0.0271 \pm 0.0078)\%$\\
\bottomrule
\end{tabular}
\end{table}

\noindent The three dimensions have the same ceiling gap but different mean errors. Saturation alone therefore misses substantial estimation error. The distinction is already present in the theory: \(q_{p,k}\) measures the probability of no extra marginal-only neighbors, whereas \(B_{p,k}\) depends on the full distribution of neighborhood scores.

\subsection{Convergence at Fixed \texorpdfstring{\(u\)}{u}}

\begin{figure}[tbp]
\centering
\includegraphics[width=\linewidth]{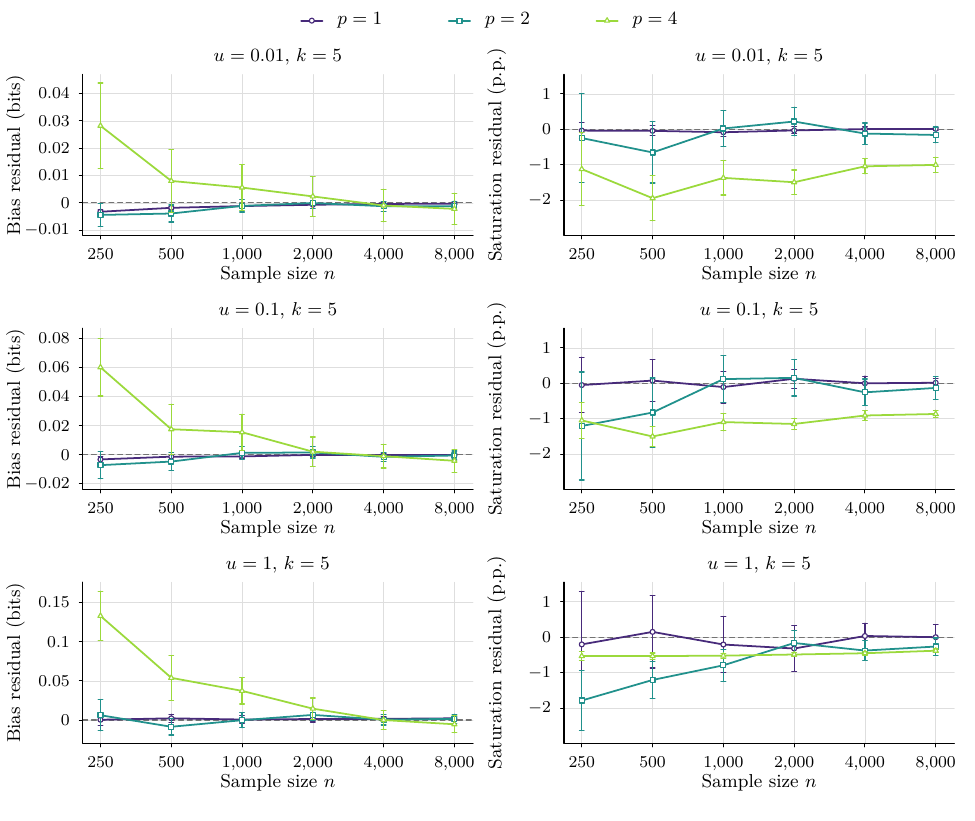}
\caption{Convergence at fixed \(u\), with \(k=5\).
Rows correspond to \(u=0.01,0.1,1\); colors distinguish \(p=1,2,4\).
Left: measured mean KSG bias minus its Poisson prediction, in bits.
Right: measured saturation fraction minus its Poisson prediction,
in percentage points. Zero indicates agreement with the Poisson predictions.
Pointwise 95\% intervals account for sampling uncertainty in both
means. The same Poisson reference is reused across sample sizes,
so its error is shared. Appendix~\ref{sec:fixed-transition-residuals}
gives the interval calculation.}
\label{fig:convergence}
\end{figure}

We next examine the fixed-\(u\) limit using the same simulations at \(k=5\), \(u=0.01,0.1,1\), and all six sample sizes. As above, the channel noise decreases with \(n\). This experiment tests whether the mean error approaches its predicted limit and variation across independent datasets decreases. The left panels of Figure~\ref{fig:convergence} show observed mean bias minus its Poisson prediction, \(B_{p,5}(u)/\log 2\). The right panels show the corresponding difference in saturation probability. Values approaching zero indicate agreement with the predicted limits. The intervals combine uncertainty across the 30 independent datasets with uncertainty in the independently computed Poisson reference. Each curve reuses the same Poisson reference at every sample size, so its Monte Carlo error is shared across sample sizes.

At \(u=1\), the measured mean bias approaches the Poisson prediction
as \(n\) increases, although not monotonically. In dimension four at
\(n=8000\), KSG still underestimates the information by about
\(4.38\) bits. The theory accurately predicts this persistent error.

At the same setting, the observed saturation fraction is about half
its predicted value. Thus accurate mean-bias predictions can coexist
with substantial relative error for rare saturation events.
Appendix~\ref{sec:fixed-transition-residuals} gives the uncertainty
calculations.

\subsubsection{Mean-Square Convergence}
\label{sec:mean-square-check}

The theorem also predicts that estimation errors across independent datasets
concentrate around the limiting bias. To check this, we measure their
mean-squared distance from that prediction.
Write \(T_r=(\widehat I_{n,5}^{(r)}-I_n)/\log 2\) and
\(b=B_{p,5}(u)/\log 2\), both in bits. For the independent numerical
reference \(\widetilde b\), let \(\bar d=\bar T-\widetilde b\)
and let \(s_T^2\) be the unbiased sample variance across the
\(R=30\) datasets. The empirical mean-square discrepancy
decomposes exactly as
\begin{equation}
\widetilde M_n
=\frac1R\sum_{r=1}^R(T_r-\widetilde b)^2
=\bar d^{\,2}+\frac{R-1}{R}s_T^2.
\label{eq:empirical-mean-square}
\end{equation}
The two terms separate the difference in means from variation across
datasets. Thus \(\sqrt{\widetilde M_n}\) measures distance from the
predicted bias, while \(s_T\) measures dataset-to-dataset spread.
The numerical reference also has sampling uncertainty, which
contributes to \(\widetilde M_n\) but does not affect \(s_T\).
Appendix~\ref{sec:fixed-transition-residuals} gives the corresponding
variance correction.

\begin{table}[tb]
\centering
\caption{Mean-square convergence diagnostic from the same 30 independent
datasets per setting, at \(k=5\). All entries are in bits. RMSE is
\(\sqrt{\widetilde M_n}\), before subtracting reference variance;
dataset SD is \(s_T\). The last column reports the standard error of
the independent 30,000-draw Poisson reference, shared across \(n\).
The table displays the first and last sample sizes; the analysis uses all
six.}
\label{tab:mean-square}
\begin{tabular}{@{}ccccccc@{}}
\toprule
 & & \multicolumn{2}{c}{RMSE about $\widetilde b$} & \multicolumn{2}{c}{Dataset SD} & Reference\\
$p$ & $u$ & $n=250$ & $n=8000$ & $n=250$ & $n=8000$ & SE\\
\midrule
1 & 0.01 & 0.00362 & 0.00040 & 0.00160 & 0.00027 & 0.00011\\
1 & 0.1 & 0.00662 & 0.00075 & 0.00587 & 0.00075 & 0.00038\\
1 & 1 & 0.01784 & 0.00427 & 0.01814 & 0.00371 & 0.00125\\
2 & 0.01 & 0.01114 & 0.00216 & 0.01040 & 0.00177 & 0.00079\\
2 & 0.1 & 0.02461 & 0.00335 & 0.02394 & 0.00337 & 0.00144\\
2 & 1 & 0.05136 & 0.00710 & 0.05187 & 0.00705 & 0.00273\\
4 & 0.01 & 0.04819 & 0.00742 & 0.03975 & 0.00721 & 0.00262\\
4 & 0.1 & 0.07740 & 0.01001 & 0.04953 & 0.00924 & 0.00366\\
4 & 1 & 0.15472 & 0.01309 & 0.08033 & 0.01212 & 0.00516\\
\bottomrule
\end{tabular}

\end{table}
\noindent
Both the root-mean-square discrepancy and the dataset standard
deviation decrease from \(n=250\) to \(n=8000\) in all nine
settings in Table~\ref{tab:mean-square}. In dimension four at
\(u=1\), the estimates become much less variable even though their
mean error at \(n=8000\) remains about \(-4.38\) bits. This is
the behavior described by the mean-square theorem. At the smallest
discrepancies, uncertainty in the Poisson reference limits how
precisely we can assess agreement.

\subsection{Bias Correction}

We test whether estimating \(u\) from saturation and subtracting
the predicted bias improves finite-sample accuracy. This separate
study uses \(p=1,2,4\), \(k=5\), \(n=1000,4000,8000\),
with 16 independent datasets per setting. In each repetition,
channel settings share source and noise draws.

The Poisson tables are computed independently of these datasets.
Test values lie between the table entries. We interpolate linearly in
\(\log u\), moving saturation fractions outside the tabulated
range to its nearest endpoint before estimating \(u\). Every
dataset is retained, including those with no saturated neighborhoods. An independently computed second table checks sensitivity to Monte Carlo error. Appendix~\ref{app:repro} gives the grids and implementation details.

We compare raw KSG with saturation-based correction, correction
evaluated at the true \(u\), and the Gaussian estimator based on
the variance of \(X-Y\) in \cref{eq:gaussian-comparator}.
The true-\(u\) correction, labeled \emph{Oracle}, uses the same
bias table but removes the need to estimate \(u\). It therefore
helps identify the source of any remaining error. The Gaussian
estimator uses the known channel model directly. We measure
root-mean-square error (RMSE), the square root of the average
squared information-estimation error across datasets.

\begin{figure}[tb]
\centering
\includegraphics[width=\linewidth]{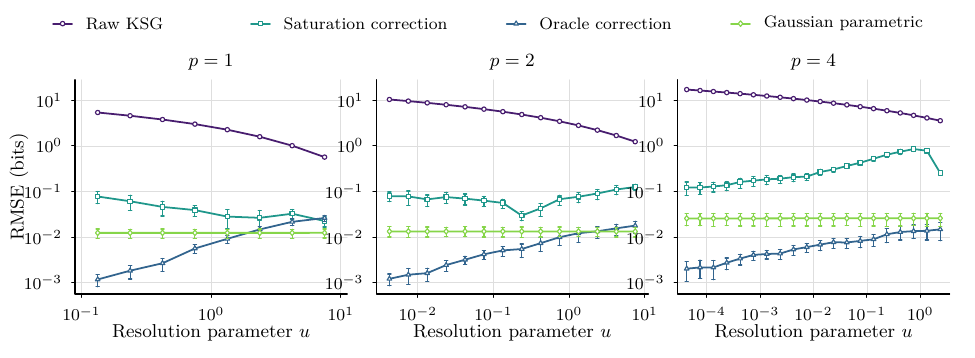}
\caption{Bias-correction test at \(n=8000,k=5\), retaining every trial. Each setting has 16 independent datasets; the horizontal settings reuse the same source/noise pools. Error bars are pointwise percentile bootstrap intervals for RMSE from 3,000 resamples of those 16 datasets, with shared resample indices across methods. Oracle correction uses the true \(u\), whereas saturation-based correction uses only the observed fraction and a prespecified Gaussian lookup. The parametric curve uses the exact mean correction in \cref{eq:gaussian-comparator}. The rightmost dimension-four drop is affected by clipping to the known window.}
\label{fig:calibration}
\end{figure}

Figure~\ref{fig:calibration} shows that correction substantially
reduces KSG's error. Table~\ref{tab:calibration-matched} compares
all three dimensions at \(n=8000\) and \(u\approx0.75\).
The remaining error is much larger in dimension four.
Additional results across sample sizes appear in
Appendix~\ref{sec:additional-correction-results}. To investigate the dimension-four error, we apply the same correction at the true \(u\). The RMSE falls from about \(0.85\) bit to \(0.014\) bit. Thus the predicted bias is accurate here when evaluated at the correct parameter.

\begin{table}[tb]
\centering
\begin{minipage}[t]{0.48\linewidth}
\vspace{0pt}
\centering
\caption{Information-estimation RMSE in bits at the matched setting
\(n=8000\), \(k=5\), \(u\approx0.7499\), over 16 independent
datasets. All three dimensions use the same value of \(u\).}
\label{tab:calibration-matched}
\small
\setlength{\tabcolsep}{4pt}
\begin{tabular}{@{}lrr@{}}
\toprule
Dimension \(p\) & Raw KSG & Corrected KSG\\
\midrule
1 & 3.000 & 0.0390\\
2 & 3.453 & 0.0677\\
4 & 4.665 & 0.8525\\
\bottomrule
\end{tabular}
\end{minipage}\hfill
\begin{minipage}[t]{0.48\linewidth}
\vspace{0pt}
\centering
\caption{Monte Carlo estimates of the limiting bias \(B_{4,5}(u)/\log2\), pooling
two independent seeds with 30,000 draws each. Intervals show pointwise
95\% Monte Carlo uncertainty.}
\label{tab:endpoint-bias}
\small
\setlength{\tabcolsep}{4pt}
\begin{tabular}{@{}lrr@{}}
\toprule
\(u\) & Bias in bits & 95\% interval\\
\midrule
\(10^4\) & \(-0.11341\) & \([-0.11957,-0.10725]\)\\
\(10^6\) & \(0.03728\) & \([0.03424,0.04031]\)\\
\(10^{12}\) & \(0.004254\) & \([0.004053,0.004455]\)\\
\bottomrule
\end{tabular}
\end{minipage}
\end{table}

The observed saturation fraction is only about half its prediction, leading the procedure to overestimate \(u\) and apply too little correction. No clipping occurs in this setting. Halving the lookup-grid spacing leaves the corrected mean error nearly unchanged, so coarse interpolation alone does not explain the discrepancy. The difficulty here is estimating \(u\) from finite-sample saturation.

Across the tested windows, the Gaussian estimator has lower pooled RMSE
than saturation-corrected KSG. The correction nevertheless provides a direct
test of the neighborhood law: its bias prediction is useful, while
recovering the required parameter from saturation can limit accuracy
at the sample sizes studied here. Appendix~\ref{sec:additional-correction-results}
reports the lookup and clipping checks.

\subsection{Large-\texorpdfstring{\(u\)}{u} Bias}
\label{sec:large-u}

\begin{figure}[tb]
\centering
\includegraphics[width=\linewidth]{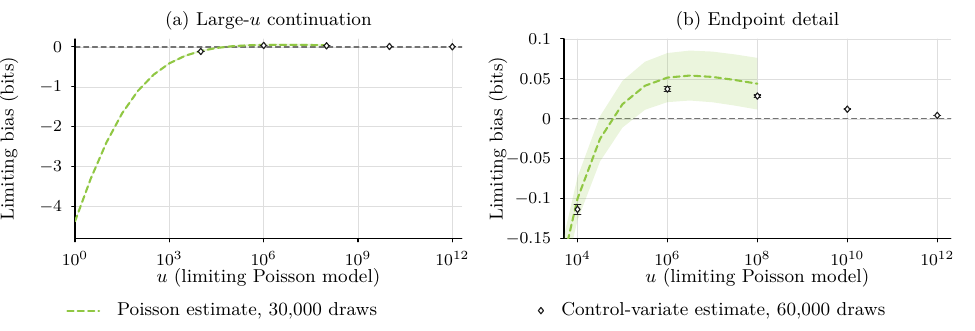}
\caption{Additional calculations of the limiting bias
\(B_{4,5}(u)/\log2\), in bits. The extended range exposes the approach
toward the zero-bias endpoint. The black control-variate points pool two
independent seeds, totaling 60,000 draws. Error bars show pointwise 95\%
Monte Carlo intervals.}
\label{fig:endpoint}
\end{figure}

The endpoint theorem gives \(B_{p,k}(u)\to0\), but does not
specify the sign of the bias or whether the approach is monotone.
We examine these features for \(p=4,k=5\), extending the Poisson
calculation to \(u=10^{12}\). We compute \(B_{4,5}(u)\) directly from the limiting Poisson law. To estimate the small bias more precisely, we subtract a random quantity with exactly zero expectation:
\begin{equation}
Z=\psi(k)-\log G_k+
\frac{\|A\|_2^2-\|B\|_2^2}{2},
\text{ } \E Z=0.
\label{eq:mc-control}
\end{equation}
Subtracting \(Z\) reduces Monte Carlo variation without changing the
expected bias. Table~\ref{tab:endpoint-bias} and
Figure~\ref{fig:endpoint} show a negative region, followed by positive
bias that decreases toward zero. Both independent sets of 30,000 draws
give the same signs. These are numerical findings about the limiting
bias function. Appendix~\ref{app:repro} gives the computational method
and numerical checks; the intervals describe pointwise Monte Carlo
uncertainty.

\FloatBarrier

\section{Discussion}
\label{sec:scope}

The critical Poisson law describes KSG's neighborhood counts and gives mean-square convergence of its estimation error to a computable limiting bias. This quantifies the estimator's behavior at the known exponential-in-information sample scale, including the effects of dimension and neighbor index. A low saturation fraction need not imply small bias, and sampling fluctuations can vanish while an additive error persists.

\noindent The saturation curve gives a consistent Gaussian-model correction on a fixed transition interval. In the dimension-four example, the remaining error comes mainly from estimating \(u\): applying the same correction at the true \(u\) is accurate even when estimating it from saturation is not. When the Gaussian family is known, the estimator based on the variance
of \(X-Y\) in \cref{prop:gaussian-comparator} is a more direct estimator.

\begin{remark}[One-sided interpretation]
If \(B_{p,k}(u)>0\) at any particular \(u\),
\cref{thm:bias} implies
\[
\Pp\{\Ihat_{n,k}>I_n\}\longrightarrow1
\text{ }\text{whenever }u_n\to u.
\]
Indeed, the probability of a nonpositive error is at most
\(\E|\Ihat_{n,k}-I_n-B_{p,k}(u)|^2/B_{p,k}(u)^2\), which tends
to zero. The calculations in Section~\ref{sec:large-u} provide numerical evidence for this positive-bias premise. A one-sided confidence guarantee for raw or bias-corrected KSG requires separate justification.
\end{remark}

\noindent The scope is fixed \(p\), fixed \(k\), the exact strict-radius convention, and the shrinking Gaussian channel \cref{eq:channel}. Other source and noise laws would require corresponding control of remote marks and score tails, and a replacement for the Gaussian overlap calculation. Consistency along all regular-side sequences would require centered-score control uniform in a growing \(u\). Such extensions would broaden the count and bias laws beyond the Gaussian, fixed-window setting studied here.


\section*{Acknowledgments}
We thank Aayush Rajesh and Matthew Ho for their helpful discussions and feedback.

\noindent\textit{AI statement:} OpenAI's ChatGPT and Codex tools assisted with drafting and checking the proofs underlying Sections~\ref{sec:poisson}--\ref{sec:saturation-calibration}, and with experimental code and plotting in Section~\ref{sec:evidence}. The authors are responsible for the final manuscript.

\bibliographystyle{plainnat}
\begingroup
\setlength{\bibsep}{0pt}
\bibliography{references}

@article{kraskov2004,
  author  = {Kraskov, Alexander and St{\"o}gbauer, Harald and Grassberger, Peter},
  title   = {Estimating Mutual Information},
  journal = {Physical Review E},
  volume  = {69},
  number  = {6},
  pages   = {066138},
  year    = {2004},
  doi     = {10.1103/PhysRevE.69.066138}
}

@inproceedings{gao2015,
  author    = {Gao, Shuyang and Ver Steeg, Greg and Galstyan, Aram},
  title     = {Efficient Estimation of Mutual Information for Strongly Dependent Variables},
  booktitle = {Proceedings of the Eighteenth International Conference on Artificial Intelligence and Statistics},
  series    = {Proceedings of Machine Learning Research},
  volume    = {38},
  pages     = {277--286},
  year      = {2015},
  url       = {https://proceedings.mlr.press/v38/gao15.html}
}

@article{gao2018,
  author  = {Gao, Weihao and Oh, Sewoong and Viswanath, Pramod},
  title   = {Demystifying Fixed $k$-Nearest Neighbor Information Estimators},
  journal = {IEEE Transactions on Information Theory},
  volume  = {64},
  number  = {8},
  pages   = {5629--5661},
  year    = {2018},
  doi     = {10.1109/TIT.2018.2807481}
}

@article{zhao2020,
  author  = {Zhao, Puning and Lai, Lifeng},
  title   = {Analysis of {KNN} Information Estimators for Smooth Distributions},
  journal = {IEEE Transactions on Information Theory},
  volume  = {66},
  number  = {6},
  pages   = {3798--3826},
  year    = {2020},
  doi     = {10.1109/TIT.2019.2945041}
}

@article{gao2018bandwidth,
  author  = {Gao, Weihao and Oh, Sewoong and Viswanath, Pramod},
  title   = {Breaking the Bandwidth Barrier: Geometrical Adaptive Entropy Estimation},
  journal = {IEEE Transactions on Information Theory},
  volume  = {64},
  number  = {5},
  pages   = {3313--3330},
  month   = may,
  year    = {2018},
  doi     = {10.1109/TIT.2018.2810313}
}

@article{lord2018,
  author  = {Lord, Warren M. and Sun, Jie and Bollt, Erik M.},
  title   = {Geometric $k$-Nearest Neighbor Estimation of Entropy and Mutual Information},
  journal = {Chaos},
  volume  = {28},
  number  = {3},
  pages   = {033114},
  year    = {2018},
  doi     = {10.1063/1.5011683}
}

@article{bickelbreiman,
  author  = {Bickel, Peter J. and Breiman, Leo},
  title   = {Sums of Functions of Nearest Neighbor Distances, Moment Bounds, Limit Theorems and a Goodness of Fit Test},
  journal = {The Annals of Probability},
  volume  = {11},
  number  = {1},
  pages   = {185--214},
  year    = {1983},
  doi     = {10.1214/aop/1176993668}
}

@article{penroseyukich,
  author  = {Penrose, Mathew D. and Yukich, Joseph E.},
  title   = {Limit Theory for Point Processes in Manifolds},
  journal = {The Annals of Applied Probability},
  volume  = {23},
  number  = {6},
  pages   = {2161--2211},
  year    = {2013},
  doi     = {10.1214/12-AAP897}
}

@book{lastpenrose,
  author    = {Last, G{\"u}nter and Penrose, Mathew},
  title     = {Lectures on the {Poisson} Process},
  publisher = {Cambridge University Press},
  year      = {2017},
  doi       = {10.1017/9781316104477}
}

@article{efronstein,
  author  = {Efron, Bradley and Stein, Charles},
  title   = {The Jackknife Estimate of Variance},
  journal = {The Annals of Statistics},
  volume  = {9},
  number  = {3},
  pages   = {586--596},
  year    = {1981},
  doi     = {10.1214/aos/1176345462}
}

@article{holmes2019,
  author  = {Holmes, Caroline M. and Nemenman, Ilya},
  title   = {Estimation of Mutual Information for Real-Valued Data with Error Bars and Controlled Bias},
  journal = {Physical Review E},
  volume  = {100},
  number  = {2},
  pages   = {022404},
  year    = {2019},
  doi     = {10.1103/PhysRevE.100.022404},
  url     = {https://arxiv.org/abs/1903.09280}
}

@inproceedings{mcallester2020,
  author    = {McAllester, David and Stratos, Karl},
  title     = {Formal Limitations on the Measurement of Mutual Information},
  booktitle = {Proceedings of the Twenty Third International Conference on Artificial Intelligence and Statistics},
  series    = {Proceedings of Machine Learning Research},
  volume    = {108},
  pages     = {875--884},
  year      = {2020},
  url       = {https://proceedings.mlr.press/v108/mcallester20a.html}
}

@book{kallenberg,
  author    = {Kallenberg, Olav},
  title     = {Random Measures, Theory and Applications},
  publisher = {Springer},
  year      = {2017},
  doi       = {10.1007/978-3-319-41598-7}
}

@book{resnick,
  author    = {Resnick, Sidney I.},
  title     = {Extreme Values, Regular Variation and Point Processes},
  publisher = {Springer},
  year      = {2008},
  doi       = {10.1007/978-0-387-75953-1}
}
\endgroup

\clearpage
\appendix
\numberwithin{equation}{section}
\section{Local Poisson Convergence and Marginal Counts}
\label{app:local}

Marginal neighborhoods can include points far away in the other
coordinate, so convergence on bounded regions alone does not establish
count convergence. We prove the process and count limits in
\cref{prop:countlaw}, including control of these distant points, then
establish the fractional-count moment bound needed for \cref{thm:bias}.

\subsection{Exact setting and statement}

Fix \(p,k\geq1\), assume
\begin{equation}
u_n=n\delta_n^{p/2}\longrightarrow u\in(0,\infty),
\label{eq:app-critical-assumption}
\end{equation}
and consider query observation \(i=1\).  We do not assume any monotonicity of \(\delta_n\);
\cref{eq:app-critical-assumption} itself implies \(\delta_n\to0\).
Let
\[
A_n=X_1,\text{ } B_n=Z_1.
\]
Conditional on \((A_n,B_n)=(a,b)\), define, for \(j=2,\ldots,n\),
\begin{equation}
\xi_{j,n}=\frac{X_j-a}{\sqrt{\delta_n}},
\text{ }
\eta_{j,n}=\frac{Y_{j,n}-Y_{1,n}}{\sqrt{\delta_n}},
\text{ }
\Xi_n^{a,b}=\sum_{j=2}^n\delta_{(\xi_{j,n},\eta_{j,n})}.
\label{eq:app-local-process}
\end{equation}
For a locally finite point measure \(\xi\) on
\(\R^p\times\R^p\), define
\begin{align}
C_r&=\{(s,t):\|s\|_\infty\leq r,\ \|t\|_\infty\leq r\},\label{eq:app-Cr}\\
\rho_k(\xi)&=\inf\{r\geq0:\xi(C_r)\geq k\},\label{eq:app-rho}\\
\mathcal N_x(\xi)&=\xi\{(s,t):\|s\|_\infty<\rho_k(\xi)\},\label{eq:app-Nx}\\
\mathcal N_y(\xi)&=\xi\{(s,t):\|t\|_\infty<\rho_k(\xi)\}.\label{eq:app-Ny}
\end{align}
The sets in \cref{eq:app-Nx,eq:app-Ny} are noncompact.

\begin{theorem}[Critical local process and strict-count limit]
\label{thm:app-critical-process}
Let \(A,B\stackrel{\mathrm{iid}}{\sim}N(0,I_p)\).  Conditional on
\((A,B)=(a,b)\), let \(\Xi^{a,b}\) be a Poisson point process on
\(\R^p\times\R^p\) with intensity density
\begin{equation}
\lambda_{a,b}(s,t)
=u\phip(a)\phip(b+t-s).
\label{eq:app-limit-intensity}
\end{equation}
Then \(\rho_k(\Xi^{A,B})\), \(\mathcal N_x(\Xi^{A,B})\), and
\(\mathcal N_y(\Xi^{A,B})\) are finite almost surely, and
\begin{equation}
\left(
A_n,B_n,\frac{\varepsilon_{1,n}}{\sqrt{\delta_n}},
N_{x,1},N_{y,1}
\right)
\Rightarrow
\left(
A,B,\rho_k(\Xi^{A,B}),
\mathcal N_x(\Xi^{A,B}),\mathcal N_y(\Xi^{A,B})
\right).
\label{eq:app-critical-joint-convergence}
\end{equation}
For every fixed \((a,b)\), the corresponding regular conditional law given
\((A_n,B_n)=(a,b)\) also converges.
\end{theorem}

\subsection{Exact coordinate map and conditional Poisson convergence}

\begin{lemma}[Exact conditional density]
\label{lem:app-local-density}
Conditional on \((A_n,B_n)=(a,b)\), the points in
\cref{eq:app-local-process} are iid and satisfy
\begin{equation}
\eta_{j,n}
=(1-\delta_n)\xi_{j,n}
+\sqrt{1-\delta_n}\,(Z_j-b).
\label{eq:app-exact-local-channel}
\end{equation}
Their Lebesgue density is
\begin{equation}
q_n^{a,b}(s,t)
=\delta_n^{p/2}(1-\delta_n)^{-p/2}
\phip(a+\sqrt{\delta_n}s)
\phip\!\left(
b+\frac{t-(1-\delta_n)s}{\sqrt{1-\delta_n}}
\right).
\label{eq:app-qn}
\end{equation}
On every compact subset of \(\R^{2p}\),
\begin{equation}
(n-1)q_n^{a,b}(s,t)
\longrightarrow u\phip(a)\phip(b+t-s)
\label{eq:app-local-density-limit}
\end{equation}
uniformly.
\end{lemma}

\begin{proof}
Subtract the query channel equation from the \(j\)th channel equation:
\[
Y_{j,n}-Y_{1,n}
=(1-\delta_n)(X_j-a)
+\sqrt{\delta_n(1-\delta_n)}(Z_j-b).
\]
Division by \(\sqrt{\delta_n}\) proves
\cref{eq:app-exact-local-channel}.  The inverse map from \((s,t)\) to the
independent Gaussian variables \((x,z)\) is
\[
x=a+\sqrt{\delta_n}s,
\text{ }
z=b+\frac{t-(1-\delta_n)s}{\sqrt{1-\delta_n}}.
\]
Its absolute Jacobian determinant is
\(\delta_n^{p/2}(1-\delta_n)^{-p/2}\), which proves
\cref{eq:app-qn}.  Finally,
\((n-1)\delta_n^{p/2}\to u\) and \(\delta_n\to0\).
Both Gaussian factors and the scalar Jacobian correction converge uniformly
on compact sets, proving \cref{eq:app-local-density-limit}.
\end{proof}
\noindent The map is also exact for the KSG metric:
\begin{equation}
\frac{\max\{\|X_j-X_1\|_\infty,\|Y_{j,n}-Y_{1,n}\|_\infty\}}
     {\sqrt{\delta_n}}
=\max\{\|\xi_{j,n}\|_\infty,\|\eta_{j,n}\|_\infty\}.
\label{eq:app-metric-exact}
\end{equation}
Thus
\[
\frac{\varepsilon_{1,n}}{\sqrt{\delta_n}}
=\rho_k(\Xi_n^{a,b}),\text{ }
N_{x,1}=\mathcal N_x(\Xi_n^{a,b}),\text{ }
N_{y,1}=\mathcal N_y(\Xi_n^{a,b})
\]
almost surely. It remains to prove convergence of the radius and marginal counts.

\begin{lemma}[Conditional binomial-to-Poisson convergence]
\label{lem:app-conditional-ppp}
For every fixed \((a,b)\),
\begin{equation}
\Xi_n^{a,b}\Rightarrow\Xi^{a,b}
\label{eq:app-ppp-convergence}
\end{equation}
in the vague topology on Radon point measures on \(\R^{2p}\).
Moreover,
\begin{equation}
(A_n,B_n,\Xi_n^{A_n,B_n})
\Rightarrow(A,B,\Xi^{A,B}).
\label{eq:app-anchor-process-convergence}
\end{equation}
\end{lemma}

\begin{proof}
Fix \(f\in C_c^+(\R^{2p})\) and set
\[
I_n(a,b)
=\int_{\R^{2p}}(1-e^{-f(s,t)})q_n^{a,b}(s,t)\,\dd s\,\dd t.
\]
Conditional on the query values \((X_1,Z_1)=(a,b)\), the \(n-1\) rescaled competitor points are independent and identically distributed. Hence, we have
\[
\E_{a,b}\exp\{-\Xi_n^{a,b}f\}
=\{1-I_n(a,b)\}^{n-1}.
\]

\noindent Since \(f\) has compact support, \cref{eq:app-local-density-limit} gives
\[
(n-1)I_n(a,b)\to
\int_{\R^{2p}}(1-e^{-f(s,t)})\lambda_{a,b}(s,t)\,\dd s\,\dd t.
\]
In particular, \(I_n(a,b)=O(n^{-1})\). Using
\(\log(1-x)=-x+O(x^2)\), we obtain 
\[
\E_{a,b}e^{-\Xi_n^{a,b}f}
\to
\exp\!\left\{
-\int_{\R^{2p}}(1-e^{-f})\lambda_{a,b}
\right\}.
\]
This is the Laplace functional of \(\Xi^{a,b}\).
The Laplace-functional convergence criterion therefore gives convergence in distribution under the vague topology~\citep{resnick,kallenberg}, proving
\cref{eq:app-ppp-convergence}.


To obtain convergence jointly with the query values, let \(H\) be any bounded continuous function of the query values and the point measure, with the latter equipped with the vague topology. For each fixed \((a,b)\), the conditional convergence just proved gives
\[
\E_{a,b}H(a,b,\Xi_n^{a,b})
\longrightarrow
\E H(a,b,\Xi^{a,b}).
\]
The integrands are bounded by \(\|H\|_\infty\), and the query distribution is the same
\(Q=N(0,I_p)\otimes N(0,I_p)\) for every \(n\).
Dominated convergence therefore yields
\[
\begin{aligned}
\E H(A_n,B_n,\Xi_n^{A_n,B_n})
&=\int \E_{a,b}H(a,b,\Xi_n^{a,b})\,Q(\dd a,\dd b)\\
&\longrightarrow
\int \E H(a,b,\Xi^{a,b})\,Q(\dd a,\dd b)\\
&=\E H(A,B,\Xi^{A,B}).
\end{aligned}
\]
Since this holds for every bounded continuous \(H\), it proves
\cref{eq:app-anchor-process-convergence}.
\end{proof}

\subsection{Gaussian Tail Bounds for Marginal Counts}

Vague convergence in \cref{eq:app-ppp-convergence} controls bounded regions, but the marginal-count regions are unbounded in the other coordinate.  For \(0<R<L\), define the two remote portions of the marginal slabs:
\begin{align}
E_x(R,L)
&=\{(s,t):\|s\|_\infty\leq R,\ \|t\|_\infty>L\},\label{eq:app-Ex}\\
E_y(R,L)
&=\{(s,t):\|t\|_\infty\leq R,\ \|s\|_\infty>L\}.\label{eq:app-Ey}
\end{align}
Let \(W\sim N(0,I_p)\), and write
\[
\overline G_p(q)=\Pp(\|W\|_\infty>q),
\]
with \(\overline G_p(q)=1\) for \(q<0\).

\begin{lemma}[Two-sided remote-strip bounds]
\label{lem:app-remote}
For every fixed pair of query values \((a,b)\) and \(R<\infty\),
\begin{align}
\lim_{L\to\infty}\limsup_{n\to\infty}
\E_{a,b}\Xi_n^{a,b}\{E_x(R,L)\}&=0,\label{eq:app-remote-x}\\
\lim_{L\to\infty}\limsup_{n\to\infty}
\E_{a,b}\Xi_n^{a,b}\{E_y(R,L)\}&=0.\label{eq:app-remote-y}
\end{align}
The same two assertions hold with \(\Xi_n^{a,b}\) replaced by
\(\Xi^{a,b}\).
\end{lemma}

\begin{proof}
Choose \(n_0\) so that \(\delta_n\leq1/2\) for \(n\geq n_0\), and set
\[
K=\sup_{n\geq n_0}(n-1)\delta_n^{p/2}<\infty,
\text{ } C_p=(2\pi)^{-p/2}.
\]
We work in source and noise coordinates
\(s=(X_j-a)/\sqrt{\delta_n}\) and \(w=Z_j\).
The corresponding output coordinate is
\[
t=(1-\delta_n)s+\sqrt{1-\delta_n}(w-b).
\]
In these coordinates, the expected number of competitor points
in \(\dd s\,\dd w\) is \\
$(n-1)\delta_n^{p/2} \phip(a+\sqrt{\delta_n}s)\phip(w)\,\dd s\,\dd w$.
If \(\|s\|_\infty\leq R\) and \(\|t\|_\infty>L\), then
\[
\|w\|_\infty>L-R-\|b\|_\infty.
\]
Using \(\phip\leq C_p\) therefore gives
\begin{equation}
\E_{a,b}\Xi_n^{a,b}\{E_x(R,L)\}
\leq
KC_p(2R)^p
\overline G_p(L-R-\|b\|_\infty).
\label{eq:app-remote-x-bound}
\end{equation}
This tends to zero as \(L\to\infty\), uniformly over \(n\geq n_0\). For the reverse strip, fix \(w\).  The set of \(s\) for which
\[
\|(1-\delta_n)s+\sqrt{1-\delta_n}(w-b)\|_\infty\leq R
\]
is a cube of volume
\[
(2R)^p(1-\delta_n)^{-p}\leq(4R)^p.
\]
Writing
\(w=b+t/\sqrt{1-\delta_n}-\sqrt{1-\delta_n}s\),
we see that \(\|t\|_\infty\le R\) and \(\|s\|_\infty>L\) imply
\[
\begin{aligned}
\|w\|_\infty
&>\sqrt{1-\delta_n}\,L
  -\frac{R}{\sqrt{1-\delta_n}}-\|b\|_\infty\\
&\ge \frac{L}{\sqrt2}-\sqrt2R-\|b\|_\infty.
\end{aligned}
\]
Integrating first in \(s\) yields
\begin{equation}
\E_{a,b}\Xi_n^{a,b}\{E_y(R,L)\}
\leq
KC_p(4R)^p
\overline G_p\!\left(
\frac{L}{\sqrt2}-\sqrt2R-\|b\|_\infty
\right),
\label{eq:app-remote-y-bound}
\end{equation}
which also tends to zero. \\
For the limiting process \(\Xi^{a,b}\), the source points form
a homogeneous Poisson process with intensity \(u\phip(a)\).
Each source point \(s\) receives an independent Gaussian noise
vector \(W\sim N(0,I_p)\), and its output coordinate is
\begin{equation}
t=s+W-b.
\label{eq:app-raw-marked-limit}
\end{equation}
With the same calculations, for either \(\ell=x\) or \(\ell=y\), we have:
\[
\E_{a,b}\Xi^{a,b}\{E_\ell(R,L)\}
\le
u\phip(a)(2R)^p
\overline G_p(L-R-\|b\|_\infty).
\]
The right-hand side tends to zero as \(L\to\infty\), proving
both assertions for the limiting process.
\end{proof}

By Markov's inequality, these bounds also control the probability
of any point in \(E_x(R,L)\cup E_y(R,L)\). Whenever the selected radius is at most \(R\), removing points outside \([-L,L]^{2p}\) can change a marginal count only if one of these remote points is present.


\subsection{The Selected Radius and Strict Marginal Counts}

\begin{lemma}[Critical radius tightness]
\label{lem:app-radius-tight}
For every fixed pair of query values \((a,b)\), the conditional sequence
\(\rho_k(\Xi_n^{a,b})\) is tight, and
\(\rho_k(\Xi^{a,b})<\infty\) almost surely.
\end{lemma}

\begin{proof}
The limiting mean in the joint cube \(C_R\) is
\begin{equation}
L_{a,b}(R)
=u\phip(a)
\int_{[-R,R]^p}
\Pp\{s+W-b\in[-R,R]^p\}\,\dd s.
\label{eq:app-Lab}
\end{equation}
The intensity in \cref{eq:app-limit-intensity} is continuous and strictly
positive, so \(L_{a,b}\) is continuous and strictly increasing.
To see that it diverges, interchange expectation and integration in
\cref{eq:app-Lab}. The integral is the expected volume of the intersection
of two radius-\(R\) cubes separated by \(W-b\). For each fixed \(W\),
this overlap volume divided by \((2R)^p\) tends to one and lies between
zero and one. Dominated convergence gives
\begin{equation}
L_{a,b}(R)\sim u\phip(a)(2R)^p.
\label{eq:app-Lab-asymptotic}
\end{equation}

For each fixed \(R\), \cref{lem:app-conditional-ppp} and the zero intensity
of \(\partial C_R\) imply
\[
\Xi_n^{a,b}(C_R)\Rightarrow\Pois\{L_{a,b}(R)\}.
\]
Consequently
\[
\lim_{n\to\infty}
\Pp_{a,b}\{\rho_k(\Xi_n^{a,b})>R\}
=
\Pp\{\Pois(L_{a,b}(R))<k\}.
\]
The right side tends to zero as \(R\to\infty\), proving tightness and
almost-sure finiteness of the limiting radius.
\end{proof}

\begin{lemma}[Genericity of the limiting configuration]
\label{lem:app-generic}
Conditional on every fixed \((a,b)\), the following properties hold almost
surely on every bounded cube:
\begin{enumerate}[label=\textup{(\roman*)}]
\item \(\Xi^{a,b}\) is simple;
\item distinct points have distinct joint radii
\(\max\{\|s\|_\infty,\|t\|_\infty\}\);
\item every point has unequal marginal distances
\(\|s\|_\infty\neq\|t\|_\infty\); and
\item no point other than the \(k\)th joint neighbor has either marginal distance
equal to the \(k\)th joint radius.
\end{enumerate}
\end{lemma}

\begin{proof}
The intensity in \cref{eq:app-limit-intensity} is absolutely continuous and
locally finite, so the Poisson process is simple.  Each equality excluded in
items (ii)--(iv) defines a Lebesgue-null subset of a one- or two-point product
space.  The first or second factorial moment formula shows that the expected
number of violating points or pairs in any bounded cube is zero.  Take a
countable union over integer cubes.
\end{proof}

\begin{lemma}[Convergence of the noncompact strict counts]
\label{lem:app-noncompact}
For every fixed pair of query values \((a,b)\),
\begin{equation}
\left(
\rho_k(\Xi_n^{a,b}),
\mathcal N_x(\Xi_n^{a,b}),
\mathcal N_y(\Xi_n^{a,b})
\right)
\Rightarrow
\left(
\rho_k(\Xi^{a,b}),
\mathcal N_x(\Xi^{a,b}),
\mathcal N_y(\Xi^{a,b})
\right).
\label{eq:app-noncompact-convergence}
\end{equation}
\end{lemma}

\begin{proof}
Fix \(\eta>0\).  By \cref{lem:app-radius-tight}, choose \(R\) so that the
finite and limiting probabilities of a \(k\)th radius exceeding \(R\) have
limsup at most \(\eta\).  By \cref{lem:app-remote}, choose \(L>R\) so that
the finite and limiting probabilities of a point in
\[
E_x(R,L)\cup E_y(R,L)
\]
have limsup at most \(\eta\).

On the complementary good event, the radius and both full marginal counts
are determined by the point measure restricted to the compact cube
\[
K_L=[-L,L]^p\times[-L,L]^p.
\]
Define the corresponding truncated functional, assigning a fixed dummy
value when fewer than \(k\) joint points occur by radius \(R\).
At every configuration satisfying \cref{lem:app-generic} and having no atom
on the deterministic boundaries of \(K_L\) or \(C_R\), this truncated
functional is continuous in the vague topology. To verify this directly,
place disjoint small balls around the finitely many atoms in \(K_L\),
with no atoms on their boundaries. Under vague convergence, each ball
eventually contains exactly one atom, and the total number of atoms in
\(K_L\) is eventually unchanged. Thus there are no additional atoms in
\(K_L\), and the atoms can be labeled so that their locations converge.
If \(C_R\) contains fewer than \(k\) points, its count is eventually
unchanged, so the dummy value is preserved. Otherwise, the positive gaps
in \cref{lem:app-generic} determine the radius order, which
marginal attains the boundary point's joint radius, and all other
marginal-count comparisons. The boundary point's distance in that marginal
equals the selected radius. The same point and marginal remain selected under a sufficiently small
perturbation, so its exclusion from that strict marginal count persists.
All other comparisons persist by their positive gaps.

The two deterministic boundaries have zero limiting intensity.  The
continuous-mapping theorem applied to
\cref{eq:app-ppp-convergence} gives convergence of the truncated vector.
The probability that the truncated and full vectors differ is at most
\(2\eta+o(1)\) for the finite process and \(2\eta\) for the limiting
process.  Letting \(\eta\downarrow0\) proves
\cref{eq:app-noncompact-convergence}.
\end{proof}

\begin{proof}[Proof of \cref{thm:app-critical-process}]
The exact metric relation \cref{eq:app-metric-exact} identifies the
finite-sample scaled radius and counts with the three functionals of
\(\Xi_n^{a,b}\).  The conditional assertion follows from
\cref{lem:app-noncompact}.  For a bounded continuous test function of the
query values and the radius/count vector, condition on the query, apply the
conditional convergence at each \((a,b)\), and then use bounded convergence
against the fixed Gaussian query distribution.  This proves
\cref{eq:app-critical-joint-convergence}. The limiting radius is finite
almost surely by \cref{lem:app-radius-tight}. Conditional on fixed query
values \((a,b)\), each marginal slab of integer radius \(m\ge1\) has
finite intensity \(u\phip(a)(2m)^p\), and therefore contains finitely
many points almost surely. Choosing an integer \(m\) larger than the
selected radius proves that both limiting marginal counts are finite
almost surely.
\end{proof}

\subsection{The exact stopped-Poisson count law}

Once the first \(k\) joint neighbors are determined, the points
outside the selected cube remain conditionally Poisson, with the original
intensity restricted to that region. The following lemma establishes this
property at the randomly selected radius.

\begin{lemma}[Post-arrival Poisson property]
\label{lem:app-post-arrival}
Let \(\Pi\) be a Poisson process with deterministic intensity \(\mu\) on a
Polish space \(E\), and let \(g:E\to(0,\infty)\) be measurable.  Suppose
\(\mu\{g\leq r\}<\infty\) for every \(r\), the pushforward
\(\mu\circ g^{-1}\) is diffuse, and
\(\mu\{g\leq r\}\to\infty\).  If \(\tau_k\) is the \(k\)th smallest
\(g\)-value, then, conditional on the stopped sigma-field generated by points
with \(g\leq\tau_k\), the restriction of \(\Pi\) to
\(\{g>\tau_k\}\) is Poisson with intensity
\(\mu|_{\{g>\tau_k\}}\).
\end{lemma}

\begin{proof}
Let
\[
\mathcal F_r
=\sigma\{\Pi|_{\{g\leq s\}}:s\leq r\}.
\]
Then \(\tau_k\) is an \((\mathcal F_r)\)-stopping time.  We use the
stopped sigma-field
\[
\mathcal F_{\tau_k}
=\{H\in\mathcal F_\infty:
H\cap\{\tau_k\le r\}\in\mathcal F_r\text{ for every }r\ge0\},
\text{ }
\mathcal F_\infty=\sigma\!\left(\bigcup_{r\ge0}\mathcal F_r\right).
\]
The conditional assertion is a statement about a random intensity
restriction. More explicitly, for every nonnegative measurable \(f\)
whose support has finite intensity,
\[
\E\!\left[
e^{-\int_{\{g>\tau_k\}}f\,\dd\Pi}
\,\middle|\,\mathcal F_{\tau_k}\right]
=\exp\!\left[-\int_{\{g>\tau_k\}}(1-e^{-f})\,\dd\mu\right].
\]
First let \(\sigma\) be a stopping time taking values in a
deterministic countable grid. On each event
\(\{\sigma=r_j\}\in\mathcal F_{r_j}\), deterministic-set independent
increments imply that \(\Pi|_{\{g>r_j\}}\) is a Poisson process with the
restricted intensity and is independent of \(\mathcal F_{r_j}\).  Summing
over \(j\) proves the claimed conditional Laplace-functional identity at
\(\sigma\).

Now round \(\tau_k\) upward:
\[
\tau_k^{(m)}=2^{-m}\lceil2^m\tau_k\rceil.
\]
These are grid-valued stopping times and decrease to \(\tau_k\).  If
\[
M(r)=\mu\{g\leq r\},
\]
then \(M\) is continuous.  On every bounded interval it is uniformly
continuous, so for every \(R<\infty\),
\[
\sup_{0\leq r\leq R}
\mu\{r<g\leq r+2^{-m}\}
\longrightarrow0.
\]
The shell is selected from the process, so this deterministic intensity
bound needs one additional step.  On \(\{\tau_k\leq R\}\), a point in
\(\{\tau_k<g\leq\tau_k^{(m)}\}\) produces an ordered pair of radial
points in \([0,R+1]\) whose separation is at most \(2^{-m}\).  With
\(\nu=\mu\circ g^{-1}\), the second factorial-moment formula bounds the
probability of such a pair by
\[
\nu([0,R+1])
\sup_{0\leq r\leq R+1}\nu((r,r+2^{-m}])
\longrightarrow0.
\]
Thus the adaptively selected intervening shell is empty with probability
tending to one.  Its intensity also tends to zero uniformly on
\(\{\tau_k\leq R\}\).  For every bounded
\(\mathcal F_{\tau_k}\)-measurable variable \(H\), \(H\) is also
\(\mathcal F_{\tau_k^{(m)}}\)-measurable.  Multiply the grid-time
conditional Laplace identity by \(H\), let \(m\to\infty\), and use both
the vanishing shell probability and the vanishing change in the restricted
intensity's Laplace exponent to pass to \(\tau_k\).  Then let
\(R\to\infty\), using \(\tau_k<\infty\) almost surely.  This proves the
post-\(\tau_k\) conditional Laplace functional for every nonnegative test
function with finite intensity support.  A monotone-class argument completes
the proof.  This is also the usual strong Markov property of a radially
marked Poisson process; see \citet[Chs.~5 and 7]{lastpenrose}.
\end{proof}

\begin{proposition}[Exact Gamma--Poisson representation]
\label{prop:app-countlaw}
Condition on \((A,B)=(a,b)\), and set
\begin{equation}
c=u\phip(a),\text{ } h=c^{1/p}.
\label{eq:app-c-h}
\end{equation}
Let \(W\sim N(0,I_p)\), and define
\begin{equation}
\Lambda_{h,b}(r)
=\prod_{\ell=1}^p
\E\bigl[(2r-h|W_\ell-b_\ell|)_+\bigr].
\label{eq:app-Lambda}
\end{equation}
Let \(G_k\sim\GP(k,1)\), and let \(R_k\) be the unique solution of
\begin{equation}
\Lambda_{h,b}(R_k)=G_k.
\label{eq:app-Gamma-root}
\end{equation}
Conditional on \((a,b,G_k,R_k)\), let
\[
P_x,P_y\stackrel{\mathrm{ind}}{\sim}
\Pois\{(2R_k)^p-G_k\},
\text{ }
J\sim\operatorname{Bernoulli}(1/2),
\]
with \(J\) independent of \(P_x,P_y,G_k,R_k\).  Then
\begin{equation}
h\rho_k(\Xi^{a,b})\stackrel d=R_k,
\label{eq:app-normalized-radius-law}
\end{equation}
and
\begin{equation}
\bigl(\mathcal N_x(\Xi^{a,b}),\mathcal N_y(\Xi^{a,b})\bigr)
\stackrel d=
\begin{cases}
\bigl(k-1+P_x,\ k+P_y\bigr),&J=0,\\
\bigl(k+P_x,\ k-1+P_y\bigr),&J=1.
\end{cases}
\label{eq:app-exact-countlaw}
\end{equation}
In particular,
\begin{equation}
\Pp\!\left(
\bigl\{\mathcal N_x,\mathcal N_y\bigr\}=\{k-1,k\}
\mid a,b,G_k,R_k
\right)
=\exp[-2\{(2R_k)^p-G_k\}].
\label{eq:app-saturation-law}
\end{equation}
\end{proposition}

\begin{proof}
The intensity in \cref{eq:app-limit-intensity} can be constructed by taking
a homogeneous Poisson process of rate \(c\) in the \(s\)-coordinate,
attaching iid \(N(0,I_p)\) marks \(W\), and mapping
\[
t=s+W-b.
\]
Scale both coordinate blocks by \(h=c^{1/p}\):
\begin{equation}
S'=hs,\text{ } T'=ht=S'+h(W-b).
\label{eq:app-normalized-marked}
\end{equation}
Then \(S'\) has unit Lebesgue intensity.  The \(T'\)-marginal also has unit
Lebesgue intensity because
\[
\int_{\R^p}h^{-p}
\phip\!\left(b+\frac{t'-s'}h\right)\,\dd s'=1.
\]

For the normalized joint cube
\[
B_r=\{(s',t'):\|s'\|_\infty\leq r,\ \|t'\|_\infty\leq r\},
\]
condition on \(W\).  The admissible \(s'\)-set is the intersection of two
radius-\(r\) cubes separated coordinatewise by \(h(W-b)\); its volume is
\[
\prod_{\ell=1}^p(2r-h|W_\ell-b_\ell|)_+.
\]
Taking expectations and using independence of the Gaussian coordinates gives
\cref{eq:app-Lambda}.  The function is continuous by dominated convergence
on bounded \(r\)-intervals.  It is strictly increasing because the intensity
is positive on every nonempty joint annulus.  Finally,
\[
\frac{\Lambda_{h,b}(r)}{(2r)^p}\to1
\]
by dominated convergence, so it tends to infinity.

Map every normalized point to its joint radius
\[
g(s',t')=\max\{\|s'\|_\infty,\|t'\|_\infty\}.
\]
The Poisson mapping theorem \citep{lastpenrose} gives an inhomogeneous
Poisson process on \((0,\infty)\) with cumulative mean
\(\Lambda_{h,b}(r)\).  Its intensity is diffuse.  Time change by
\(\Lambda_{h,b}\) produces a unit-rate Poisson process, whose \(k\)th arrival
is \(\GP(k,1)\).  This proves
\cref{eq:app-Gamma-root,eq:app-normalized-radius-law}.

Apply \cref{lem:app-post-arrival} at \(R_k\).  The two marginal-only regions
\begin{align*}
D_x(r)&=\{\|S'\|_\infty<r,\ \|T'\|_\infty>r\},\\
D_y(r)&=\{\|T'\|_\infty<r,\ \|S'\|_\infty>r\}
\end{align*}
are disjoint and lie strictly outside \(B_r\).  Since both normalized
marginals have unit intensity,
\[
\mu\{D_x(r)\}=\mu\{D_y(r)\}
=(2r)^p-\Lambda_{h,b}(r).
\]
The conditional exterior-process law also applies to these
radius-dependent sets: first restrict to \(R_k\le L\) and intersect the
sets with a deterministic compact box. Approximation by measurable simple
functions of the radius extends the conditional Laplace identity to these
sets; the radial boundaries have zero intensity. Monotone convergence as
the box expands, followed by \(L\uparrow\infty\), gives the same identity
for the full strips, whose displayed intensities are finite.
Conditional on the stopped history at \(R_k\), their counts are therefore
independent Poisson variables with mean
\[
(2R_k)^p-\Lambda_{h,b}(R_k)=(2R_k)^p-G_k.
\]
Their conditional law depends on the stopped history only through
\(R_k\). Conditional on that radius and the fixed query values, they are
therefore independent of the boundary point.

To determine which marginal attains the boundary radius, consider
the symmetry
\[
(s',t')\longmapsto(-t',-s'),
\]
which leaves \(t'-s'\) unchanged. It therefore preserves the conditional
intensity and every joint cube while exchanging the source and output
boundary faces. Let \(\nu_x\)
and \(\nu_y\) be the pushforward intensity measures on the radius coordinate
for boundary points whose active block is, respectively, source and output.
The sign-swap gives \(\nu_x=\nu_y\).  The set
\(\{\|s'\|_\infty=\|t'\|_\infty\}\) has zero intensity, so
\(\nu_x+\nu_y=\nu\), the full radial intensity measure.  Therefore the
pushforward intensity on radius and orientation factorizes as
\(\nu\otimes\operatorname{Bernoulli}(1/2)\).  The Poisson marking theorem
then makes the orientations iid fair marks, independent of the radial
process. Together with the stopped exterior law, this makes the \(k\)th
orientation independent of \((R_k,P_x,P_y)\), conditional on the query values.

There are \(k-1\) earlier joint points strictly inside both marginal cubes.
The boundary point is excluded from the strict count in its active block and
included in the other.  Adding the two marginal-only Poisson counts proves
\cref{eq:app-exact-countlaw}.  Saturation occurs exactly when
\(P_x=P_y=0\), whose conditional probability is the right side of
\cref{eq:app-saturation-law}.
\end{proof}

Together, \cref{thm:app-critical-process,prop:app-countlaw} prove
\cref{thm:phase} part~\textup{(b)} and the exact functional stated in
\cref{prop:countlaw}.

\begin{remark}[Anchor sign convention]
\label{rem:app-sign}
With the actual query noise \(B=Z_1\), the exact conditional intensity is
\[
u\phip(A)\phip(B+T-S),
\]
and the normalized marked relation is \(T'=S'+h(W-B)\).
If instead \(V=-Z_1\), Gaussian evenness rewrites the density as
\[
u\phip(A)\phip(V+S-T).
\]
Since $V \stackrel{d}{=} B$, the unconditional mixture is unchanged, whereas the conditional law given $Z_1 = b$ requires the sign convention in \cref{eq:app-limit-intensity}.

\end{remark}

\subsection{Finite-Sample Score Moments}

To pass from count convergence to convergence of expected scores, we need uniform tail bounds for the finite-sample marginal counts.

\begin{lemma}[Uniform fractional count moments]
\label{lem:app-fractional-moments}
Under \cref{eq:app-critical-assumption}, for every \(0<\alpha<1\),
\begin{equation}
\sup_{n>k}
\E\{(N_{x,1}+1)^\alpha+(N_{y,1}+1)^\alpha\}<\infty.
\label{eq:app-fractional-moments}
\end{equation}
More precisely, for every \(0<\beta<1\), a finite \(C_\beta\) exists such
that, for all integers \(m\geq1\),
\begin{equation}
\sup_{n>k}
\max\{\Pp(N_{x,1}\geq m),\Pp(N_{y,1}\geq m)\}
\leq C_\beta m^{-\beta}.
\label{eq:app-count-tail}
\end{equation}
\end{lemma}

\begin{proof}
We bound a large marginal count by splitting according to whether the
selected joint radius exceeds a deterministic threshold. Below that
threshold, a binomial tail bound controls the marginal count; above it,
we bound the probability that the joint ball contains fewer than \(k\)
points. It suffices to work for sufficiently large \(n\), because finitely
many earlier sample sizes have bounded counts. Choose \(n_0\) so that, for
\(n\geq n_0\),
\begin{equation}
\delta_n\leq\frac12,\text{ }
0<\underline u\leq(n-1)\delta_n^{p/2}\leq\overline u<\infty.
\label{eq:app-u-bounds}
\end{equation}
Increase \(\overline u\), if necessary, so that \(u_n\leq\overline u\) as
well.  Let \(C_p=(2\pi)^{-p/2}\). Conditional on \((A_n,B_n)=(a,b)\), define
\[
Q_{x,n}(r)=\sum_{j=2}^n\ind\{\|\xi_{j,n}\|_\infty\leq r\},
\text{ }
Q_{y,n}(r)=\sum_{j=2}^n\ind\{\|\eta_{j,n}\|_\infty\leq r\}.
\]
The \(\xi\)-density is
\(\delta_n^{p/2}\phip(a+\sqrt{\delta_n}s)\).
The other output samples are independent of the query and have law
\(N(0,(1-\delta_n)I_p)\).  After centering at \(Y_{1,n}\) and scaling by
\(\sqrt{\delta_n}\), the conditional \(\eta\)-density is bounded by
\[
\delta_n^{p/2}(1-\delta_n)^{-p/2}C_p.
\]
Thus a deterministic \(K<\infty\) exists such that, for every query,
\(r>0\), and \(n\geq n_0\),
\begin{equation}
\max\{\E_{a,b}Q_{x,n}(r),\E_{a,b}Q_{y,n}(r)\}
\leq Kr^p.
\label{eq:app-marginal-cube-mean}
\end{equation}
Choose \(\gamma>0\) so small that \(K\gamma^p\leq1/4\), and for
\(1\leq m\leq n-1\) put
\[
r_m=\gamma m^{1/p}.
\]
Both \(Q\)-variables are binomial conditional on the query, and
\cref{eq:app-marginal-cube-mean} bounds their means at \(r_m\) by \(m/4\).
The elementary binomial Chernoff bound therefore gives
\begin{equation}
\max\{
\Pp_{a,b}(Q_{x,n}(r_m)\geq m),
\Pp_{a,b}(Q_{y,n}(r_m)\geq m)
\}
\leq(e/4)^m.
\label{eq:app-marginal-chernoff}
\end{equation}

Write
\(\rho_{n,k}=\varepsilon_{1,n}/\sqrt{\delta_n}\).
If \(\rho_{n,k}\leq r_m\), every point counted by \(N_{x,1}\) is counted by
\(Q_{x,n}(r_m)\), and similarly for the output block.  Hence
\begin{equation}
\max\{\Pp_{a,b}(N_{x,1}\geq m),\Pp_{a,b}(N_{y,1}\geq m)\}
\leq(e/4)^m+\Pp_{a,b}(\rho_{n,k}>r_m).
\label{eq:app-count-radius-split}
\end{equation}
To bound the joint radius, let \(W\sim N(0,I_p)\) and set
\[
q_p=\Pp(\|W\|_\infty\leq1)>0.
\]
Consider points satisfying
\begin{equation}
\|\xi_{j,n}\|_\infty\leq r_m/4,
\text{ }
\|Z_j\|_\infty\leq1.
\label{eq:app-favorable}
\end{equation}
If \(r_m\geq2(1+\|b\|_\infty)\), then
\cref{eq:app-exact-local-channel} gives
\[
\|\eta_{j,n}\|_\infty
\leq r_m/4+(1+\|b\|_\infty)
\leq3r_m/4<r_m.
\]
Every favorable point is therefore inside the joint radius-\(r_m\) cube. Because \(m\leq n-1\),
\[
\sqrt{\delta_n}r_m
\leq\gamma(n\delta_n^{p/2})^{1/p}
\leq\gamma\overline u^{1/p}.
\]
Set \(H=\sqrt p\,\gamma\overline u^{1/p}/4\).
On the source cube in \cref{eq:app-favorable},
\[
\phip(a+\sqrt{\delta_n}s)
\geq
C_p\exp[-(\|a\|_2+H)^2/2].
\]
The number of favorable points is consequently binomial with mean at least
\begin{equation}
\lambda_m(a)
\geq
d_0m\exp[-(\|a\|_2+H)^2/2],
\text{ }
d_0=\frac{\underline u\,q_pC_p\gamma^p}{2^p}>0.
\label{eq:app-favorable-mean}
\end{equation}
\noindent If \(B_0\) is any binomial variable of mean \(\lambda\), then for fixed
\(k\) there is a finite \(D_k\) such that
\begin{equation}
\Pp(B_0<k)\leq D_ke^{-\lambda/2}.
\label{eq:app-binomial-lower-tail}
\end{equation}
Indeed, if \(B_0\sim\operatorname{Bin}(N,q)\), then for \(0\leq j<k\),
\[
\binom Njq^j(1-q)^{N-j}
\leq e^ke^{-Nq}(Nq)^j/j!,
\]
and \(\sum_{j<k}\lambda^j/j!\) is bounded by a constant times
\(e^{\lambda/2}\). The event \(\rho_{n,k}>r_m\) means that the full joint cube contains fewer than \(k\) points, and therefore implies that the favorable count is below \(k\).  From \cref{eq:app-favorable-mean,eq:app-binomial-lower-tail},
\begin{equation}
\begin{split}
\Pp_{a,b}(\rho_{n,k}>r_m)
\leq{}&
\ind\{r_m<2(1+\|b\|_\infty)\}
+D_k\exp\!\left[
-d_1m e^{-(\|a\|_2+H)^2/2}
\right],
\end{split}
\label{eq:app-radius-tail-conditional}
\end{equation}
where \(d_1=d_0/2\).
Average over the independent standard Gaussian query values \(A,B\).  Gaussian
tails imply, after enlarging constants to cover small \(m\),
\begin{equation}
\Pp\{\gamma m^{1/p}<2(1+\|B\|_\infty)\}
\leq Ce^{-cm^{2/p}}.
\label{eq:app-B-tail}
\end{equation}
For the remaining mixture, fix \(0<\beta<1\), select
\(\beta<\beta_1<1\), and let
\[
R_m=(2\beta_1\log m)^{1/2},\text{ } m\geq2.
\]
On \(\{\|A\|_2\leq R_m\}\),
\[
m e^{-(\|A\|_2+H)^2/2}
\geq
m^{1-\beta_1}
e^{-H(2\beta_1\log m)^{1/2}-H^2/2}
\geq m^{(1-\beta_1)/2}
\]
for all sufficiently large \(m\).  On the complement, choose
\(\theta\in(\beta/(2\beta_1),1/2)\).  Since
\(\E e^{\theta\|A\|_2^2}=(1-2\theta)^{-p/2}\),
\[
\Pp(\|A\|_2>R_m)
\leq
(1-2\theta)^{-p/2}e^{-\theta R_m^2}
\leq C_\beta m^{-\beta}.
\]
Splitting at \(R_m\) therefore gives
\begin{equation}
\E\exp\!\left[
-d_1m e^{-(\|A\|_2+H)^2/2}
\right]
\leq C_\beta m^{-\beta}.
\label{eq:app-A-mixture-tail}
\end{equation}
Combining
\cref{eq:app-count-radius-split,eq:app-radius-tail-conditional,eq:app-B-tail,eq:app-A-mixture-tail}
proves \cref{eq:app-count-tail} for all sufficiently large \(n\).
The finitely many earlier \(n\)'s and the case \(m>n-1\) are absorbed into
\(C_\beta\). Finally, choose \(0<\alpha<\beta<1\).  For either marginal count \(N\),
\[
\begin{split}
\E(N+1)^\alpha
&=1+\sum_{m\geq1}
\{(m+1)^\alpha-m^\alpha\}\Pp(N\geq m)\\
&\leq
1+C\sum_{m\geq1}m^{\alpha-1-\beta}<\infty
\end{split}
\]
uniformly in \(n\).  This proves
\cref{eq:app-fractional-moments}.
\end{proof}

\begin{corollary}[Uniform integrability of the neighborhood score]
\label{cor:app-score-ui}
Let
\[
H_n=\psi(N_{x,1}+1)+\psi(N_{y,1}+1)
\]
and
\[
H=\psi\{\mathcal N_x(\Xi^{A,B})+1\}
 +\psi\{\mathcal N_y(\Xi^{A,B})+1\}.
\]
By \cref{prop:app-countlaw}, \(H\) has the same distribution as
\(\psi(M_x+1)+\psi(M_y+1)\).
Then \(\{H_n^2\}\) is uniformly integrable and
\begin{equation}
\E H_n\to\E H,\text{ }
\E H_n^2\to\E H^2.
\label{eq:app-score-moment-convergence}
\end{equation}
The same assertions hold when each digamma term is replaced by the logarithm
of its integer argument.
\end{corollary}

\begin{proof}
Fix \(\alpha\in(0,1)\) and \(q>1\).  Every fixed power of
\(\log(m+1)\) is bounded by a constant times \((m+1)^\alpha\).  Hence
\cref{lem:app-fractional-moments} and
\((x+y)^{2q}\leq2^{2q-1}(x^{2q}+y^{2q})\) imply
\[
\sup_n
\E\left|
\log(N_{x,1}+1)+\log(N_{y,1}+1)
\right|^{2q}<\infty.
\]
For integers \(m\geq0\),
\[
|\psi(m+1)-\log(m+1)|\leq C
\]
with a universal finite constant \(C\).  Thus
\(\sup_n\E|H_n|^{2q}<\infty\), which implies uniform integrability of
\(\{H_n^2\}\).
\Cref{thm:app-critical-process} gives \(H_n\Rightarrow H\).
Uniform integrability yields convergence of the first and second moments in
\cref{eq:app-score-moment-convergence}. Replacing the digamma terms
by logarithms gives the same convergence, using the count limit and
the same moment bound.
\end{proof}

\section{Thin and regular finite-sample endpoints}
\label{app:thin-regular}
\label{app:thinregular}

We prove the thin-count collapse and regular radius/count scaling used in
\cref{thm:phase} directly for the finite Gaussian array.

\subsection{Conditional local coordinates}

It suffices initially to study query observation \(1\).  Put \(m_n=n-1\), condition
on
\[
U=X_1=a,\text{ } V=Z_1=b,
\]
and let \(\phip\) be the standard \(p\)-variate Gaussian density.  Define
\begin{equation}
\alpha_n=\{m_n\phip(a)\}^{1/p},
\text{ }
h_n=\alpha_n\sqrt{\delta_n(1-\delta_n)}.
\label{eq:tr-anchor-scales}
\end{equation}
For \(j=2,\ldots,n\), introduce
\begin{equation}
\mathsf S_{j,n}=\alpha_n(X_j-a),
\text{ }
\mathsf T_{j,n}=\alpha_n(Y_{j,n}-Y_{1,n}).
\label{eq:tr-local-coordinates}
\end{equation}
The channel identity \cref{eq:channel} gives, exactly,
\begin{equation}
\mathsf T_{j,n}
=(1-\delta_n)\mathsf S_{j,n}+h_n(Z_j-b).
\label{eq:tr-local-channel}
\end{equation}
Write
\[
\mathsf A_{j,n}=\|\mathsf S_{j,n}\|_\infty,\text{ }
\mathsf B_{j,n}=\|\mathsf T_{j,n}\|_\infty,\text{ }
\mathsf J_{j,n}=\mathsf A_{j,n}\vee\mathsf B_{j,n},
\]
and let \(\rho_n=\alpha_n\varepsilon_{1,n}\), the \(k\)th order statistic
of \(\{\mathsf J_{j,n}:2\le j\le n\}\).  The common positive rescaling does
not change the strict marginal counts:
\begin{equation}
N_{x,1}=\sum_{j=2}^n\ind\{\mathsf A_{j,n}<\rho_n\},
\text{ }
N_{y,1}=\sum_{j=2}^n\ind\{\mathsf B_{j,n}<\rho_n\}.
\label{eq:tr-rescaled-counts}
\end{equation}
For \(r\ge0\), define the conditional mean measures
\begin{align}
M_{x,n}(r;a,b)
&=m_n\Pp(\mathsf A_{2,n}\le r\mid U=a,V=b),\nonumber\\
M_{y,n}(r;a,b)
&=m_n\Pp(\mathsf B_{2,n}\le r\mid U=a,V=b),\nonumber\\
\Lambda_n(r;a,b)
&=m_n\Pp(\mathsf J_{2,n}\le r\mid U=a,V=b).
\label{eq:tr-mean-measures}
\end{align}
The conditioning on \(b\) is redundant in the first line but is retained to
keep the notation uniform.

\begin{lemma}[Gaussian general position]
\label{lem:tr-general-position}
For every \(n>k\) and \(0<\delta_n<1\), conditionally on every finite query
pair \((a,b)\), the variables
\(\{\mathsf J_{j,n}:2\le j\le n\}\) are almost surely distinct, and
\[
\Pp(\mathsf A_{j,n}=\mathsf B_{j,n}\mid U=a,V=b)=0.
\]
Consequently, the \(k\)th joint shell has one boundary observation and that
observation is strictly inside exactly one of the two projected marginal
balls.
\end{lemma}

\begin{proof}
For \(0<\delta_n<1\), the linear map
\[
(X_j,Z_j)\longmapsto(\mathsf S_{j,n},\mathsf T_{j,n})
\]
has a nonsingular block-triangular matrix: its first diagonal block is
\(\alpha_nI_p\), and its second is \(h_nI_p\).  Hence
\((\mathsf S_{j,n},\mathsf T_{j,n})\) has a Lebesgue density on
\(\R^{2p}\), conditionally on \((a,b)\).

After choosing the coordinates and signs attaining the two max norms, the
event
\(\|\mathsf S_{j,n}\|_\infty=\|\mathsf T_{j,n}\|_\infty\) lies in a finite
union of nontrivial affine hyperplanes and therefore has Lebesgue measure
zero.  The same coordinate-and-sign decomposition shows that equality
between \(\mathsf J_{j,n}\) and \(\mathsf J_{\ell,n}\), for \(j\ne\ell\),
lies in a finite union of null hypersurfaces in the product sample.  A finite
union over pairs \(j,\ell\) remains null.  The last assertion follows because
the unique boundary observation satisfies
\(\mathsf A_{j,n}\vee\mathsf B_{j,n}=\rho_n\), while the two block radii are
unequal.
\end{proof}

\begin{lemma}[Monotone local-uniform convergence]
\label{lem:tr-monotone-uniform}
Let \(f_n:[0,\infty)\to\R\) be nondecreasing.  If \(f_n(r)\to f(r)\) at
every \(r\ge0\), where \(f\) is continuous, then, for every \(M<\infty\),
\[
\sup_{0\le r\le M}|f_n(r)-f(r)|\longrightarrow0.
\]
\end{lemma}

\begin{proof}
Fix \(\epsilon>0\).  Uniform continuity of \(f\) on \([0,M]\) gives a
partition
\[
0=r_0<r_1<\cdots<r_L=M
\]
such that \(f(r_\ell)-f(r_{\ell-1})<\epsilon\) for every \(\ell\).  Pointwise
convergence at the finitely many grid points gives an \(n_0\) such that
\[
\max_{0\le\ell\le L}|f_n(r_\ell)-f(r_\ell)|<\epsilon
\]
for \(n\ge n_0\).  If \(r\in[r_{\ell-1},r_\ell]\), monotonicity gives
\[
f_n(r_{\ell-1})\le f_n(r)\le f_n(r_\ell).
\]
The upper and lower bounds differ from \(f(r)\) by at most \(2\epsilon\).
Thus the required supremum is at most \(2\epsilon\) for \(n\ge n_0\).
Letting \(\epsilon\downarrow0\) proves the result.
\end{proof}

\begin{lemma}[Conditional decomposition at the \(k\)th radius]
\label{lem:tr-order-decomposition}
Let \((A_j,B_j)\), \(1\le j\le m\), be iid with a joint density.  Put
\(D_j=A_j\vee B_j\), suppose \(D_j\) has a continuous density, and let
\(R=D_{(k)}\).  Define
\[
q_x(r)=\Pp(A<r,B>r),\text{ }
q_y(r)=\Pp(B<r,A>r).
\]
Conditionally on \(R=r\), the boundary observation, and the labels of
the observations above and below the boundary, the \(m-k\) upper
observations are independent with the original law conditioned on
\(D>r\). After averaging over those labels, their expected total number
entering either strict marginal ball is
\begin{equation}
(m-k)\frac{q_x(r)+q_y(r)}{1-F_D(r)}.
\label{eq:tr-intruder-mean}
\end{equation}
If additionally \(\Pp(A=B)=0\), the unordered strict-count pair at radius
\(R\) is \(\{k-1,k\}\) if and only if no above-boundary observation enters
either marginal ball.
\end{lemma}

\begin{proof}
Let \(f_{A,B}\) be the joint density and \(f_D\) the density of \(D\).  The
density of \(R=D_{(k)}\) is
\begin{equation}
f_R(r)
=\frac{m!}{(k-1)!(m-k)!}
F_D(r)^{k-1}\{1-F_D(r)\}^{m-k}f_D(r).
\label{eq:tr-order-density}
\end{equation}
To see the conditional factorization explicitly, first choose which
\(k-1\) observations lie below the boundary and which observation lies on
the boundary.  Before integrating the individual observations, the
corresponding joint density contains \(k-1\) factors proportional to $f_{A,B}(s,t)\ind\{s\vee t<r\}
$
and \(m-k\) factors proportional to $f_{A,B}(s,t)\ind\{s\vee t>r\}.$
For any fixed choice of labels, conditioning on these labels and the
boundary radius normalizes each lower factor by \(F_D(r)\) and each
upper factor by \(1-F_D(r)\); the combinatorial prefactor in
\cref{eq:tr-order-density} accounts for the possible label choices.
This constructs a regular conditional distribution under which the
labelled upper observations are conditionally iid with the law of
\((A,B)\) given \(D>r\).

\noindent An upper observation enters the \(x\)-marginal ball precisely on
\(\{A<r,B>r\}\), and enters the \(y\)-marginal ball precisely on
\(\{B<r,A>r\}\), up to null boundary events.  Summing these two conditional
probabilities over the \(m-k\) upper observations proves
\cref{eq:tr-intruder-mean}. Exactly \(k-1\) observations have \(D<r\), and each lies strictly inside both marginal balls.  The unique boundary observation has \(A\vee B=r\).  When \(A\ne B\), it lies on one marginal boundary and strictly inside the other, so it contributes to exactly one strict count.  Hence the base unordered pair is \(\{k-1,k\}\), and it remains that pair exactly when no upper observation contributes an additional marginal count.
\end{proof}

\subsection{The Thin Regime}

\begin{lemma}[Diagonal conditional small-ball limit]
\label{lem:tr-thin-small-ball}
If \(u_n=n\delta_n^{p/2}\to0\), then, for every finite pair of query values
\((a,b)\) and every \(M<\infty\),
\begin{align}
\sup_{0\le r\le M}|M_{x,n}(r;a,b)-(2r)^p|&\to0,
\label{eq:tr-thin-mx}\\
\sup_{0\le r\le M}|M_{y,n}(r;a,b)-(2r)^p|&\to0,
\label{eq:tr-thin-my}\\
\sup_{0\le r\le M}|\Lambda_n(r;a,b)-(2r)^p|&\to0.
\label{eq:tr-thin-lambda}
\end{align}
In particular,
\begin{equation}
\sup_{0\le r\le M}\{M_{x,n}(r)-\Lambda_n(r)\}
+
\sup_{0\le r\le M}\{M_{y,n}(r)-\Lambda_n(r)\}
\longrightarrow0.
\label{eq:tr-thin-excess}
\end{equation}
\end{lemma}

\begin{proof}
The assumption \(u_n\to0\) implies \(\delta_n\to0\).  For fixed \(a\),
\[
\alpha_n\longrightarrow\infty
\]
and, by \cref{eq:tr-anchor-scales},
\begin{equation}
h_n^p
=m_n\phip(a)\delta_n^{p/2}(1-\delta_n)^{p/2}
=\frac{m_n}{n}\phip(a)u_n(1-\delta_n)^{p/2}
\longrightarrow0.
\label{eq:tr-thin-h}
\end{equation}
Let \(C_r=[-r,r]^p\).  The change of variables
\(s=\alpha_n(x-a)\) gives
\begin{equation}
M_{x,n}(r;a,b)
=\frac1{\phip(a)}
\int_{C_r}\phip(a+s/\alpha_n)\dd s
\longrightarrow(2r)^p.
\label{eq:tr-thin-source-integral}
\end{equation}

The output marginal is \(N(0,(1-\delta_n)I_p)\).  Denote its density by
\[
g_{\delta_n}(y)
=(1-\delta_n)^{-p/2}
\phip\!\left(\frac{y}{\sqrt{1-\delta_n}}\right)
\]
and the fixed query output by
\[
y_{1,n}(a,b)
=(1-\delta_n)a+\sqrt{\delta_n(1-\delta_n)}\,b.
\]
Then \(y_{1,n}(a,b)\to a\), and the change of variables
\(t=\alpha_n(y-y_{1,n})\) gives
\begin{equation}
M_{y,n}(r;a,b)
=\frac1{\phip(a)}
\int_{C_r}g_{\delta_n}(y_{1,n}+t/\alpha_n)\dd t
\longrightarrow(2r)^p.
\label{eq:tr-thin-output-integral}
\end{equation}

For the joint ball, let \(W\sim N(0,I_p)\).  From
\cref{eq:tr-local-channel},
\begin{align}
\Lambda_n(r;a,b)
=\frac1{\phip(a)}
\int_{C_r}\phip(a+s/\alpha_n)
&\Pp\bigl((1-\delta_n)s+h_n(W-b)\in C_r\bigr)
\dd s.
\label{eq:tr-thin-joint-integral}
\end{align}
For every \(s\) in the interior of \(C_r\), the random vector inside the
probability converges in probability to \(s\), by
\cref{eq:tr-thin-h}; hence the probability tends to one.  The boundary of
\(C_r\) has Lebesgue measure zero.  On a fixed box,
\(\phip(a+s/\alpha_n)/\phip(a)\) is bounded for all sufficiently large
\(n\), while the probability is bounded by one.  Dominated convergence
therefore proves
\[
\Lambda_n(r;a,b)\longrightarrow(2r)^p
\]
for every fixed \(r\). Both calculations integrate the Gaussian noise
over its full support.

Each mean measure is nondecreasing in \(r\), and their common pointwise limit
is continuous.  Lemma \ref{lem:tr-monotone-uniform} proves
\cref{eq:tr-thin-mx,eq:tr-thin-my,eq:tr-thin-lambda}.  A joint ball is
contained in each marginal ball, so both differences in
\cref{eq:tr-thin-excess} are nonnegative.  The three uniform convergences
then imply \cref{eq:tr-thin-excess}.
\end{proof}

\begin{theorem}[Saturation in the thin regime]
\label{thm:tr-thin-collapse}
If \(u_n\to0\), then
\begin{equation}
\Pp\{\{N_{x,1},N_{y,1}\}=\{k-1,k\}\}\longrightarrow1.
\label{eq:tr-thin-row}
\end{equation}
Consequently,
\begin{equation}
\overline S_{n,k}\longrightarrow1\text{ }\text{in }L^1
\label{eq:tr-thin-fraction}
\end{equation}
and
\begin{equation}
\frac{\Ihat_{n,k}}{\log n}\longrightarrow1
\text{ }\text{in }L^1.
\label{eq:tr-thin-ksg}
\end{equation}
\end{theorem}

\begin{proof}
Fix finite query values \((a,b)\).  Let \(F_n(\,\cdot\,;a,b)\) be the conditional
CDF of one \(\mathsf J_{j,n}\).  By definition and
\cref{lem:tr-thin-small-ball},
\[
m_nF_n(r;a,b)=\Lambda_n(r;a,b)\longrightarrow(2r)^p.
\]
The binomial order-statistic identity gives, for \(r>0\),
\begin{align}
\Pp(\rho_n\le r\mid a,b)
=\Pp\{\operatorname{Bin}(m_n,F_n(r;a,b))\ge k\}\nonumber
\longrightarrow
\Pp\{\Pois((2r)^p)\ge k\}
=\Pp(G_k\le(2r)^p),
\label{eq:tr-thin-radius-cdf}
\end{align}
where \(G_k\sim\GP(k,1)\).  Therefore
\begin{equation}
\rho_n\Rightarrow R_0:=\frac12G_k^{1/p}
\text{ }\text{conditionally on }(a,b).
\label{eq:tr-thin-radius-limit}
\end{equation}
In particular, \(\rho_n\) is conditionally tight.
For deterministic \(r\), define
\begin{align*}
q_{x,n}(r)
&=\Pp(\mathsf A_{2,n}<r,\mathsf B_{2,n}>r\mid a,b),\\
q_{y,n}(r)
&=\Pp(\mathsf B_{2,n}<r,\mathsf A_{2,n}>r\mid a,b).
\end{align*}
All relevant distributions are continuous, so strict and non-strict
versions agree, and
\begin{equation}
m_nq_{x,n}(r)=M_{x,n}(r)-\Lambda_n(r),\text{ }
m_nq_{y,n}(r)=M_{y,n}(r)-\Lambda_n(r).
\label{eq:tr-thin-q}
\end{equation}
Fix \(M<\infty\), and let \(\mathcal E_n\) be the event that an observation
above the \(k\)th joint boundary enters one of the two strict marginal balls.
The local-uniform convergence in \cref{lem:tr-thin-small-ball} gives
\[
\sup_{r\le M}m_n\{q_{x,n}(r)+q_{y,n}(r)\}\longrightarrow0.
\]
It also gives
\[
\sup_{r\le M}F_n(r)=\frac{\Lambda_n(M)}{m_n}=O(m_n^{-1}),
\]
so \(1-F_n(r)\ge1/2\) uniformly on \([0,M]\) for all sufficiently large
\(n\).  Applying \cref{lem:tr-order-decomposition} conditionally at
\(\rho_n\), followed by Markov's inequality, yields
\begin{align}
\Pp(\mathcal E_n,\rho_n\le M\mid a,b)
&\le
2\sup_{0\le r\le M}
m_n\{q_{x,n}(r)+q_{y,n}(r)\}\longrightarrow0.
\label{eq:tr-thin-intruders}
\end{align}
On \(\mathcal E_n^c\), the \(k-1\) observations below the joint boundary
enter both marginal balls, and the unique boundary observation enters
exactly one by \cref{lem:tr-general-position}.  The count pair is then
\(\{k-1,k\}\). \\
Given \(\epsilon>0\), conditional tightness in
\cref{eq:tr-thin-radius-limit} permits a finite \(M\) for which
\[
\limsup_n\Pp(\rho_n>M\mid a,b)<\epsilon.
\]
With this \(M\), \cref{eq:tr-thin-intruders} shows that the conditional
probability of a nonsaturated neighborhood is at most \(\epsilon+o(1)\).  Let
\(\epsilon\downarrow0\).  This proves
\cref{eq:tr-thin-row} conditionally for every finite \((a,b)\).  Since
\((U,V)\) is finite almost surely and conditional failure probabilities lie
in \([0,1]\), dominated convergence removes the conditioning and proves
\cref{eq:tr-thin-row}. Exchangeability and \(0\le\overline S_{n,k}\le1\) give
\[
\E|1-\overline S_{n,k}|
=\E(1-\overline S_{n,k})
=\Pp(S_{1,n}=0)\longrightarrow0,
\]
which is \cref{eq:tr-thin-fraction}. \\
To prove the estimator limit, we bound how far each neighborhood's score
can fall below its saturated value. General position implies, for every query,
\[
N_{x,i}\ge k-1,\text{ } N_{y,i}\ge k-1,\text{ }
\max(N_{x,i},N_{y,i})\ge k.
\]
Because \(\psi\) is increasing, the neighborhood's deficit
\begin{equation}
\mathcal L_{i,n}
=\psi(N_{x,i}+1)+\psi(N_{y,i}+1)-\psi(k)-\psi(k+1)
\label{eq:tr-row-deficit}
\end{equation}
is nonnegative and vanishes exactly on a saturated neighborhood.  Since both counts
are at most \(n-1\),
\begin{equation}
0\le\mathcal L_{i,n}
\le H_{n,k}:=2\psi(n)-\psi(k)-\psi(k+1).
\label{eq:tr-deficit-bound}
\end{equation}
Substitution into the exact statistic \cref{eq:ksg} gives
\begin{equation}
C_{n,k}-\Ihat_{n,k}
=\frac1n\sum_{i=1}^n\mathcal L_{i,n},
\text{ } C_{n,k}=\psi(n)-\psi(k+1).
\label{eq:tr-ceiling-deficit}
\end{equation}
By exchangeability, \cref{eq:tr-thin-row}, and
\cref{eq:tr-deficit-bound},
\begin{align}
0
&\le
\E\frac{C_{n,k}-\Ihat_{n,k}}{\log n}
=\frac{\E\mathcal L_{1,n}}{\log n}\nonumber\\
&\le
\frac{H_{n,k}}{\log n}\Pp(S_{1,n}=0)
\longrightarrow0,
\label{eq:tr-normalized-deficit}
\end{align}
because \(H_{n,k}/\log n\to2\) for fixed \(k\).  Also
\(C_{n,k}/\log n\to1\).  Since the difference in
\cref{eq:tr-ceiling-deficit} is nonnegative,
\[
\left|\frac{\Ihat_{n,k}}{\log n}-1\right|
\le
\left|\frac{C_{n,k}}{\log n}-1\right|
+\frac{C_{n,k}-\Ihat_{n,k}}{\log n}.
\]
Taking expectations and using \cref{eq:tr-normalized-deficit} proves
\cref{eq:tr-thin-ksg}.
\end{proof}

\subsection{The Regular Regime}

The regular result concerns a shrinking channel.  We therefore assume both
\begin{equation}
\delta_n\longrightarrow0,
\text{ }
u_n=n\delta_n^{p/2}\longrightarrow\infty.
\label{eq:tr-regular-assumptions}
\end{equation}
The first condition is not implied by the second. Write \(b_n=u_n^{1/(2p)}\).

\begin{lemma}[Regular conditional mean measures]
\label{lem:tr-regular-means}
Under \cref{eq:tr-regular-assumptions}, for every finite pair of query values
\((a,b)\) and every \(M<\infty\), locally uniformly for \(q\in[0,M]\),
\begin{equation}
\Lambda_n(b_nq;a,b)
\longrightarrow
\lambda_{a,b}(q)
:=4^pq^{2p}\frac{\phip(b)}{\phip(a)},
\label{eq:tr-regular-joint-mean}
\end{equation}
and
\begin{equation}
\frac{M_{x,n}(b_nq;a,b)}{\sqrt{u_n}}\longrightarrow(2q)^p,
\text{ }
\frac{M_{y,n}(b_nq;a,b)}{\sqrt{u_n}}\longrightarrow(2q)^p.
\label{eq:tr-regular-marginal-means}
\end{equation}
\end{lemma}

\begin{proof}
For the joint mean measure, start from the conditional competitor variables
\((X_2,Z_2)=(x,z)\).  Use the exact change of variables
\[
s=\alpha_n(x-a),\text{ }
t=(1-\delta_n)s+h_n(z-b).
\]
Its Jacobians are \(\dd x=\alpha_n^{-p}\dd s\) and
\(\dd z=h_n^{-p}\dd t\).  Since
\(m_n/\alpha_n^p=1/\phip(a)\),
\begin{equation}
\Lambda_n(r;a,b)
=\frac1{\phip(a)h_n^p}
\int_{C_r\times C_r}
\phip(a+s/\alpha_n)
\phip\!\left(b+\frac{t-(1-\delta_n)s}{h_n}\right)
\dd s\,\dd t.
\label{eq:tr-regular-joint-integral}
\end{equation}
Set \(s=b_n\sigma\) and \(t=b_n\tau\).  The ratios appearing in the
integrand satisfy
\begin{align}
\left(\frac{b_n}{\alpha_n}\right)^p
&=\frac{\sqrt{u_n}}{m_n\phip(a)}
\longrightarrow0,
\label{eq:tr-b-over-alpha}\\
\left(\frac{b_n}{h_n}\right)^p
&=\frac{\sqrt{u_n}}
{m_n\phip(a)\delta_n^{p/2}(1-\delta_n)^{p/2}}
\longrightarrow0.
\label{eq:tr-b-over-h}
\end{align}
The first ratio tends to zero because \(u_n<n\) and \(m_n=n-1\).
For the second, \(m_n\delta_n^{p/2}\sim u_n\) and
\(\delta_n\to0\), so its denominator is asymptotic to
\(\phip(a)u_n\). Moreover,
\begin{equation}
\frac{u_n}{h_n^p}
=\frac{n}{m_n\phip(a)(1-\delta_n)^{p/2}}
\longrightarrow\frac1{\phip(a)}.
\label{eq:tr-u-over-h}
\end{equation}
Since \(b_n^{2p}=u_n\), \cref{eq:tr-regular-joint-integral} becomes
\begin{align}
\Lambda_n(b_nq;a,b)
=\frac{u_n}{\phip(a)h_n^p}
\int_{C_q\times C_q}
&\phip(a+(b_n/\alpha_n)\sigma)\nonumber\\
{}\times&
\phip\!\left(
b+\frac{b_n}{h_n}\{\tau-(1-\delta_n)\sigma\}
\right)
\dd\sigma\,\dd\tau.
\label{eq:tr-regular-rescaled-integral}
\end{align}
On every fixed box, the two densities in the integrand converge uniformly to
\(\phip(a)\) and \(\phip(b)\), respectively.  They are bounded by the
finite global supremum of the Gaussian density, and the prefactor converges
by \cref{eq:tr-u-over-h}.  Dominated convergence gives
\[
\Lambda_n(b_nq;a,b)
\longrightarrow
|C_q|^2\frac{\phip(b)}{\phip(a)}
=4^pq^{2p}\frac{\phip(b)}{\phip(a)}.
\]

For the source marginal, the same change of variables used in
\cref{eq:tr-thin-source-integral}, followed by \(s=b_n\sigma\), gives
\begin{equation}
\frac{M_{x,n}(b_nq;a,b)}{\sqrt{u_n}}
=\frac1{\phip(a)}
\int_{C_q}\phip(a+(b_n/\alpha_n)\sigma)\dd\sigma
\longrightarrow(2q)^p.
\label{eq:tr-regular-source-mean}
\end{equation}
For the output marginal, use the density \(g_{\delta_n}\) and query output
\(y_{1,n}(a,b)\) defined in the proof of
\cref{lem:tr-thin-small-ball}:
\begin{equation}
\frac{M_{y,n}(b_nq;a,b)}{\sqrt{u_n}}
=\frac1{\phip(a)}
\int_{C_q}
g_{\delta_n}(y_{1,n}+(b_n/\alpha_n)\tau)\dd\tau
\longrightarrow(2q)^p.
\label{eq:tr-regular-output-mean}
\end{equation}
Here \(b_n/\alpha_n\to0\), \(\delta_n\to0\), and
\(y_{1,n}(a,b)\to a\).

The three limits have been proved pointwise in \(q\).  The left sides are
nondecreasing functions of \(q\), and the limits are continuous.  Applying
\cref{lem:tr-monotone-uniform} to each proves local-uniform convergence.
\end{proof}

\begin{lemma}[Uniform law for the marginal empirical counts]
\label{lem:tr-regular-empirical}
Under \cref{eq:tr-regular-assumptions}, conditionally on every finite
\((a,b)\), for each \(M<\infty\),
\begin{align}
\sup_{0\le q\le M}
\left|
\frac1{\sqrt{u_n}}\sum_{j=2}^n
\ind\{\mathsf A_{j,n}<b_nq\}-(2q)^p
\right|
&\xrightarrow{\Pp}0,
\label{eq:tr-regular-empirical-x}\\
\sup_{0\le q\le M}
\left|
\frac1{\sqrt{u_n}}\sum_{j=2}^n
\ind\{\mathsf B_{j,n}<b_nq\}-(2q)^p
\right|
&\xrightarrow{\Pp}0.
\label{eq:tr-regular-empirical-y}
\end{align}
\end{lemma}

\begin{proof}
We prove the source statement; the output proof is identical.  Define
\[
\mathcal N_{x,n}(q)
=\sum_{j=2}^n\ind\{\mathsf A_{j,n}<b_nq\},
\text{ }
\nu_{x,n}(q)=\E[\mathcal N_{x,n}(q)\mid a,b].
\]
Continuity of the conditional distribution implies
\(\nu_{x,n}(q)=M_{x,n}(b_nq;a,b)\).  For fixed \(M\),
\cref{lem:tr-regular-means} supplies a constant \(K_M(a,b)<\infty\) such
that, for all sufficiently large \(n\),
\[
\sup_{0\le q\le M}\nu_{x,n}(q)
\le K_M(a,b)\sqrt{u_n}.
\]
At a fixed \(q\), \(\mathcal N_{x,n}(q)\) is binomial conditionally on
\((a,b)\), so
\[
\Var(\mathcal N_{x,n}(q)\mid a,b)
\le\nu_{x,n}(q)
\le K_M(a,b)\sqrt{u_n}.
\]

Fix \(\epsilon>0\).  Since \(f(q)=(2q)^p\) is uniformly continuous on
\([0,M]\), choose a finite grid
\[
0=q_0<q_1<\cdots<q_L=M
\]
such that
\[
\max_{1\le\ell\le L}\{f(q_\ell)-f(q_{\ell-1})\}<\epsilon.
\]
Chebyshev's inequality and a union bound over this fixed grid give
\begin{align}
&\Pp\left(
\max_{0\le\ell\le L}
\frac{|\mathcal N_{x,n}(q_\ell)-\nu_{x,n}(q_\ell)|}
{\sqrt{u_n}}>\epsilon
\;\middle|\;a,b\right)\nonumber\\
&\text{ }\le
\frac{(L+1)K_M(a,b)}{\epsilon^2\sqrt{u_n}}
\longrightarrow0.
\label{eq:tr-grid-chebyshev}
\end{align}
By the local-uniform mean convergence in
\cref{eq:tr-regular-marginal-means},
\[
\sup_{0\le q\le M}
\left|\frac{\nu_{x,n}(q)}{\sqrt{u_n}}-f(q)\right|
\longrightarrow0.
\]

For \(q\in[q_{\ell-1},q_\ell]\), monotonicity yields
\[
\mathcal N_{x,n}(q_{\ell-1})
\le\mathcal N_{x,n}(q)
\le\mathcal N_{x,n}(q_\ell).
\]
On the complement of the event in \cref{eq:tr-grid-chebyshev}, the upper
bound, divided by \(\sqrt{u_n}\), is at most
\[
f(q_\ell)+\epsilon+o(1)
\le f(q)+2\epsilon+o(1),
\]
and the corresponding lower bound is at least
\[
f(q_{\ell-1})-\epsilon-o(1)
\ge f(q)-2\epsilon-o(1).
\]
Thus the supremum in \cref{eq:tr-regular-empirical-x} is at most
\(2\epsilon+o(1)\) with conditional probability tending to one.  Let
\(\epsilon\downarrow0\).  This proves the source statement.  Replacing
\(\mathsf A\) by \(\mathsf B\) and using the second marginal mean limit proves
\cref{eq:tr-regular-empirical-y}.  A union bound makes the two statements
joint.
\end{proof}

\begin{theorem}[Regular \(\sqrt{u_n}\)-count law]
\label{thm:tr-regular-counts}
Under \cref{eq:tr-regular-assumptions}, let \(U=X_1\) and \(V=Z_1\)
be the query's source and noise values, and let \(G_k\sim\GP(k,1)\)
be independent of \((U,V)\). Define
\begin{equation}
Q_{p,k}
=\left\{
\frac{G_k\phip(U)}{4^p\phip(V)}
\right\}^{1/(2p)},
\text{ }
L_{p,k}=(2Q_{p,k})^p
=\left\{G_k\frac{\phip(U)}{\phip(V)}\right\}^{1/2}.
\label{eq:tr-Q-L}
\end{equation}
Then
\begin{equation}
\left(
U,V,\frac{\rho_n}{b_n},
\frac{N_{x,1}}{\sqrt{u_n}},
\frac{N_{y,1}}{\sqrt{u_n}}
\right)
\Rightarrow
(U,V,Q_{p,k},L_{p,k},L_{p,k}).
\label{eq:tr-regular-stable}
\end{equation}
In particular, \cref{eq:regular-count} holds,
\[
\Pp(S_{1,n}=1)\longrightarrow0,
\]
and
\[
\overline S_{n,k}\longrightarrow0\text{ }\text{in }L^1.
\]
\end{theorem}

\begin{proof}
Fix finite query values \((a,b)\) and put \(Q_n=\rho_n/b_n\).  For each \(q>0\),
\cref{lem:tr-regular-means} gives
\[
m_n\Pp(\mathsf J_{2,n}\le b_nq\mid a,b)
\longrightarrow
\lambda_{a,b}(q)
=4^pq^{2p}\frac{\phip(b)}{\phip(a)}.
\]
The binomial order-statistic identity and the Poisson limit therefore give
\begin{align}
\Pp(Q_n\le q\mid a,b)
&\longrightarrow
\Pp\{\Pois(\lambda_{a,b}(q))\ge k\}\nonumber\\
&=\Pp\{G_k\le\lambda_{a,b}(q)\}.
\label{eq:tr-regular-radius-cdf}
\end{align}
Equivalently,
\begin{equation}
Q_n\Rightarrow
Q(a,b)
:=\left\{\frac{G_k\phip(a)}{4^p\phip(b)}\right\}^{1/(2p)}
\text{ }\text{conditionally on }(a,b).
\label{eq:tr-regular-radius-limit}
\end{equation}
In particular, \(Q_n\) is conditionally tight.

At the random radius \(Q_n\), the identities
\cref{eq:tr-rescaled-counts} read
\[
N_{x,1}=\mathcal N_{x,n}(Q_n),\text{ }
N_{y,1}=\mathcal N_{y,n}(Q_n),
\]
where the empirical functions are those of
\cref{lem:tr-regular-empirical}.  Fix \(M<\infty\).  On
\(\{Q_n\le M\}\), the uniform empirical law gives
\begin{equation}
\frac{N_{x,1}}{\sqrt{u_n}}-(2Q_n)^p
\xrightarrow{\Pp}0,\text{ }
\frac{N_{y,1}}{\sqrt{u_n}}-(2Q_n)^p
\xrightarrow{\Pp}0
\label{eq:tr-count-slutsky-bounded}
\end{equation}
conditionally on \((a,b)\).  Conditional tightness of \(Q_n\) lets
\(M\uparrow\infty\), so \cref{eq:tr-count-slutsky-bounded} holds without
the restriction \(Q_n\le M\).  Combining it with
\cref{eq:tr-regular-radius-limit} and the continuous mapping theorem proves
\[
\left(
Q_n,\frac{N_{x,1}}{\sqrt{u_n}},\frac{N_{y,1}}{\sqrt{u_n}}
\right)
\Rightarrow
\left(Q(a,b),(2Q(a,b))^p,(2Q(a,b))^p\right)
\]
conditionally on every finite \((a,b)\). The conditional law of \(G_k\) in this display does not depend on the query values, so \(G_k\) may be realized independently of \((U,V)\). For any
bounded continuous test function of the query values, radius, and two counts, the conditional expectation converges pointwise for every finite \((a,b)\) and is uniformly bounded.  Dominated convergence over \((U,V)\sim N(0,I_p)^{\otimes2}\) proves the joint convergence \cref{eq:tr-regular-stable}.

\noindent The variable \(L_{p,k}\) is strictly positive and finite almost surely.
For \(\epsilon>0\), once \(k/\sqrt{u_n}<\epsilon\), the portmanteau theorem
and \cref{eq:regular-count} imply
\[
\limsup_n\Pp(N_{x,1}\le k)
\le
\limsup_n\Pp\left(
\frac{N_{x,1}}{\sqrt{u_n}}\le\epsilon
\right)
\le\Pp(L_{p,k}\le\epsilon).
\]
Letting \(\epsilon\downarrow0\) gives
\(\Pp(N_{x,1}\le k)\to0\).  Saturation forces
\(N_{x,1}\in\{k-1,k\}\), hence
\[
\Pp(S_{1,n}=1)\le\Pp(N_{x,1}\le k)\longrightarrow0.
\]
Finally, exchangeability and \(0\le\overline S_{n,k}\le1\) give
\[
\E|\overline S_{n,k}|
=\E\overline S_{n,k}
=\Pp(S_{1,n}=1)\longrightarrow0.
\]
\end{proof}

\begin{remark}[Why \(\delta_n\to0\) cannot be omitted]
\label{rem:tr-delta-necessary}
Suppose instead that \(\delta_n\to\delta_0\in(0,1)\).  Repeating the exact
calculation in \cref{eq:tr-regular-joint-integral} gives
\[
\Lambda_n(b_nq;a,b)\longrightarrow
4^pq^{2p}
\frac{\phip(b)}
{\phip(a)(1-\delta_0)^{p/2}}.
\]
The source marginal still has normalized mean \((2q)^p\), whereas the output
marginal has normalized mean
\[
(2q)^p
\frac{
g_{\delta_0}((1-\delta_0)a+
\sqrt{\delta_0(1-\delta_0)}\,b)}
{\phip(a)}.
\]
These two constants need not agree.  Bare \(u_n\to\infty\) therefore does not
imply the common-limit law in \cref{eq:regular-count}; the shrinking-channel
assumption \(\delta_n\to0\) changes the theorem rather than merely simplifying
its proof.
\end{remark}

\section{Score moments and replacement variance}
\label{app:score-variance}
\label{app:scorevariance}

We prove the exact-convention replacement variance bound and score-moment
convergence used in \cref{thm:bias}. The variance bound is uniform in the
positive noise level and does not assume independent query scores.

\subsection{A geometric reverse-neighbor bound}

We first isolate the deterministic geometry used by the variance proof.

\begin{lemma}[Reverse nearest-neighbor bound]
\label{lem:app-reverse-knn}
For every fixed \(d\), there is a finite constant \(\kappa_d\) with the
following property.  Let \(\mathcal Z\subset\R^d\) be finite and suppose the
relevant distances are distinct.  For a fixed \(z\in\mathcal Z\), the number
of centers \(x\in\mathcal Z\setminus\{z\}\) for which \(z\) is among the
first \(m\) max-norm nearest neighbors of \(x\) is at most
\(\kappa_d m\).
\end{lemma}

\begin{proof}
Partition the max-norm unit sphere into finitely many sets of max-norm
diameter strictly less than one, and extend each set radially to a cone.
Such a finite partition exists by compactness.  Suppose
\(a=r\theta\) and \(b=s\eta\), where \(0<s\le r\), are in one cone and
\(\|\theta\|_\infty=\|\eta\|_\infty=1\).  Then
\begin{equation}
\|a-b\|_\infty<r.
\label{eq:app-cone-contraction}
\end{equation}
Indeed, put \(t=s/r\).  If \(|\theta_j|<1\), convexity of
\(t\mapsto|\theta_j-t\eta_j|\) bounds it by
\[
\max\{|\theta_j|,|\theta_j-\eta_j|\}<1.
\]
If \(|\theta_j|=1\), diameter less than one forces \(\eta_j\) to have the
same sign and to be nonzero, and hence
\(|\theta_j-t\eta_j|<1\) for every \(0<t\le1\).  This proves
\cref{eq:app-cone-contraction}.

Fix a cone and order its candidate centers by increasing distance from
\(z\).  Every earlier center in that cone is closer to a later center than
\(z\) is, by \cref{eq:app-cone-contraction}.  Thus \(z\) cannot be among
the first \(m\) neighbors of the \((m+1)\)st or any later center in that
ordering.  There are at most \(m\) qualifying centers per cone.  Taking
\(\kappa_d\) to be the finite number of cones proves the claim.
\end{proof}

\begin{lemma}[Harmonic reverse-rank bound]
\label{lem:app-harmonic-reverse}
Suppose a fixed point \(z\) is among the first \(m_i\) nearest neighbors of
each center indexed by \(i\in A\), where
\(1\le m_i\le n-1\).  Then
\begin{equation}
\sum_{i\in A}\frac1{m_i}
\le 2\kappa_d\{1+\lceil\log_2 n\rceil\}.
\label{eq:app-harmonic-reverse}
\end{equation}
\end{lemma}

\begin{proof}
Group the indices according to
\(2^\ell\le m_i<2^{\ell+1}\).  By
\cref{lem:app-reverse-knn}, a group has at most
\(\kappa_d2^{\ell+1}\) elements, while every summand in it is at most
\(2^{-\ell}\).  There are at most
\(1+\lceil\log_2n\rceil\) nonempty groups, proving
\cref{eq:app-harmonic-reverse}.
\end{proof}

\subsection{A Variance Bound Independent of the Sampling Law}

\begin{theorem}[Exact strict-radius KSG variance]
\label{thm:app-ksg-variance}
Fix \(p,k\), and let \(n>k\).  For an iid sample from any law having a
Lebesgue density on
\(\R^p\times\R^p\), evaluated with the query-excluded, strict-radius
KSG-1 convention in \cref{eq:counts,eq:ksg},
\begin{equation}
\Var(\Ihat_{n,k})
\le C_{p,k}\frac{(1+\log n)^2}{n},
\label{eq:app-ksg-variance}
\end{equation}
where \(C_{p,k}<\infty\) is independent of the sampling law.  The same
constant therefore applies uniformly to all sample sizes and laws in the
shrinking-noise Gaussian triangular array.
\end{theorem}

\begin{proof}
For a data set \(\mathcal W=(W_1,\ldots,W_n)\), write
\[
Q(\mathcal W)
=\sum_{i=1}^n
\{\psi(N_{x,i}+1)+\psi(N_{y,i}+1)\}.
\]
Replace \(W_j\) by an independent copy \(W_j'\), producing
\(\mathcal W^{(j)}\).  We first bound
\(|Q(\mathcal W)-Q(\mathcal W^{(j)})|\) on the probability-one event on
which, in both samples, the distances from each query are distinct in the
joint space and in each marginal space, and every joint boundary point is
active in exactly one block. These exclusions follow from the joint
Lebesgue density: after selecting coordinates and signs attaining the max
norms, each excluded equality lies in a nontrivial affine hyperplane.
A finite union over queries, competitors, and the two samples is still
null. Marginal ties are excluded explicitly because marginal ranks are
used below.

For a query \(i\ne j\) common to both samples, its joint \(k\)th radius can change only if
the old point \(W_j\) is among its first \(k\) joint neighbors in
\(\mathcal W\), or the new point \(W_j'\) is among its first \(k\) joint
neighbors in \(\mathcal W^{(j)}\).  Applying
\cref{lem:app-reverse-knn} in dimension \(2p\), there are at most
\(2\kappa_{2p}k\) such queries. \\
General position and the strict, query-excluded convention give
\[
k-1\le N_{x,i},N_{y,i}\le n-1.
\]
Thus each neighborhood's two-digamma summand lies in
\([2\psi(k),2\psi(n)]\). The replaced observation itself and all queries whose
joint radius changes contribute at most $2(1+2\kappa_{2p}k)\{\psi(n)-\psi(k)\}$
to the absolute change in \(Q\). \\
Consider now the other queries, whose joint radii are unchanged. In a given
marginal block, if the old and new points are both inside or both outside
the strict ball, the count does not change. If only the old point is
inside, write \(m\) for the original count. Its removal changes
\(\psi(m+1)\) to \(\psi(m)\), an exact change of \(1/m\).
The old point has marginal rank at most \(m\) in the original sample, so
\cref{lem:app-harmonic-reverse} bounds the sum of these changes by $2\kappa_p\{1+\lceil\log_2 n\rceil\}.$ \\
If only the new point is inside, write \(m\) for the final count. The
increment is again \(1/m\), and the new point has rank at most \(m\)
in the replacement sample. The identical reverse-rank bound therefore
applies to insertion, and separately to the \(x\)- and \(y\)-blocks.
The total contribution of all
queries with unchanged radii is therefore at most four times this display.  Combining
it with the bound and using
\(\psi(n)-\psi(k)\le C_k+\log n\), we obtain the deterministic bound
\begin{equation}
|Q(\mathcal W)-Q(\mathcal W^{(j)})|
\le C'_{p,k}(1+\log n).
\label{eq:app-Q-sensitivity}
\end{equation}

The two deterministic digamma terms in \cref{eq:ksg} do not change.
The independent-replacement Efron--Stein inequality
\citep{efronstein}, followed by \cref{eq:app-Q-sensitivity}, gives
\[
\begin{split}
\Var(\Ihat_{n,k})
&\le \frac12\sum_{j=1}^n
\E\left[
\left\{\frac{Q(\mathcal W)-Q(\mathcal W^{(j)})}{n}\right\}^2
\right]\\
&\le \frac{(C'_{p,k})^2}{2}\frac{(1+\log n)^2}{n}.
\end{split}
\]
This proves \cref{eq:app-ksg-variance}.
\end{proof}

\begin{remark}[Why a plain bounded-degree argument is insufficient]
Replacing one observation changes only \(O_{p,k}(1)\) joint \(k\)th
radii, but that observation can belong to many projected marginal balls.
The reciprocal digamma increment and the harmonic reverse-rank bound
control these additional score changes.  This is also why the proof retains
the strict-radius and \(+1\) conventions explicitly.
\end{remark}

\subsection{Critical score moments}

Let \((N_{x,n},N_{y,n})\) denote the strict counts for a uniformly selected
query and set
\[
\ell_n
=\psi(k)-\psi(N_{x,n}+1)-\psi(N_{y,n}+1).
\]
With \((M_x,M_y)\) as in \cref{prop:countlaw}, define
\[
\ell^{(u)}
=\psi(k)-\psi(M_x+1)-\psi(M_y+1).
\]

\begin{lemma}[Critical score moment convergence]
\label{lem:app-score-moment}
If \(u_n\to u\in(0,\infty)\), then
\begin{equation}
\E\ell_n\longrightarrow\E\ell^{(u)},
\text{ }
\E[\ell_n^2]\longrightarrow\E[(\ell^{(u)})^2].
\label{eq:app-score-moment}
\end{equation}
\end{lemma}

\begin{proof}
\Cref{thm:app-critical-process} gives count convergence in the discrete
count space. Applying its score map gives convergence in distribution of
\(\ell_n\). By \cref{cor:app-score-ui}, the scores and their squares
are uniformly integrable and have moments of every fixed finite order
bounded uniformly in \(n\). Truncation at a fixed score magnitude,
followed by uniform integrability as the truncation increases, gives both
limits in \cref{eq:app-score-moment}.
\end{proof}

\subsection{Relation to prior KSG variance results}

\Citet{zhao2020} state an \(O((\log n)^2/n)\) variance rate for continuous
distributions; their proof refers to the fixed-\(k\) analysis of
\citet{gao2018}, and the untruncated extension is not written out there in
the exact strict-radius convention used here.  We therefore do not claim the rate in
\cref{thm:app-ksg-variance} as novel.  Its role is to make the convention
and the required uniformity explicit: \(C_{p,k}\) contains no density
bound, derivative bound, support parameter, or factor depending on
\(\delta_n\).  In contrast, the fixed-law bias and consistency results of
\citet{gao2018,zhao2020} cannot simply be applied to the present triangular
array, because their regularity constants need not remain uniform as
\(\delta_n\downarrow0\).

\section{Overlap and integrability estimates for the bias endpoints}
\label{app:endpoints}

We prove the overlap, logarithmic integrability, and Poisson-score
estimates used in \cref{thm:endpoints}, all for the limiting Poisson
functional.

\subsection{A common overlap representation}

Let \(A,B\stackrel{\mathrm{iid}}{\sim}N(0,I_p)\), let
\(G\sim\GP(k,1)\) be independent, and put
\[
h=\{u\phip(A)\}^{1/p}.
\]
Following \cref{prop:countlaw}, let \(R\) be the unique solution of
\[
\Lambda_{h,B}(R)
=\prod_{j=1}^p\E(2R-h|W_j-B_j|)_+=G,
\]
and define
\begin{equation}
M=(2R)^p,\text{ } \mu=M-G.
\label{eq:app-endpoint-M-mu}
\end{equation}
Conditional on \((A,B,G)\), the two marginal-only Poisson counts are
independent with common mean \(\mu\). It is strictly positive almost surely for every \(u>0\).
Indeed, with \(a=2R/h\),
\[
\frac{G}{M}
=\prod_{j=1}^p\frac{\ell_{B_j}(a)}a,
\]
and every factor is strictly less than one because
\(\ell_v(a)/a=\E(1-|W-v|/a)_+\) and a nondegenerate Gaussian displacement
is nonzero almost surely.  Hence \(G/M<1\) and \(\mu=M-G>0\), so all
logarithms below are well defined.

For later use, define, for \(a>0\) and \(v\in\R\),
\begin{equation}
\ell_v(a)
=\E(a-|W-v|)_+
=\int_{-a}^a(a-|z|)\varphi(v+z)\,\dd z.
\label{eq:app-scalar-overlap}
\end{equation}
With \(a=2R/h\), the radius equation and marginal volume are
\begin{equation}
G=h^p\prod_{j=1}^p\ell_{B_j}(a),
\text{ }
M=(ha)^p.
\label{eq:app-overlap-equation}
\end{equation}

Recall from \cref{eq:gm,eq:B-definition} that
\[
g_m(\mu)=\E\psi\{m+\Pois(\mu)\},
\text{ }
B_{p,k}(u)
=\log u+\psi(k)-\E\{g_k(\mu)+g_{k+1}(\mu)\}.
\]

\subsection[The thin endpoint with a quantitative remainder]
{The thin endpoint, including the \(O(u^{1/p})\) remainder}

\begin{lemma}[Expected marginal-only mass at small \(u\)]
\label{lem:app-small-u-mu}
For \(0<u\le1\),
\begin{equation}
\E\mu\le C_{p,k}u^{1/p}.
\label{eq:app-small-u-mu}
\end{equation}
\end{lemma}

\begin{proof}
For a deterministic displacement \(z\in\R^p\), a telescoping product
bound gives
\begin{equation}
(2R)^p-\prod_{j=1}^p(2R-h|z_j|)_+
\le h(2R)^{p-1}\|z\|_1.
\label{eq:app-cube-overlap-loss}
\end{equation}
To verify it, replace the \(p\) factors one at a time and bound every
unchanged factor by \(2R\).  Each coordinate loss is at most
\(\min\{h|z_j|,2R\}\). If \(h|z_j|\ge2R\), its contribution
to the right-hand side of \cref{eq:app-cube-overlap-loss} is already at
least \((2R)^p\), so the same inequality still holds.

Average \cref{eq:app-cube-overlap-loss} over \(W\), evaluate it at the
random solution \(R\), and set
\[
L(B)=\E\|W-B\|_1\le C_p+\|B\|_1,
\text{ } q=(p-1)/p.
\]
Using \(M=G+\mu\), we obtain
\begin{equation}
\mu
=M-\Lambda_{h,B}(R)
\le hL(B)M^q
=hL(B)(G+\mu)^q.
\label{eq:app-mu-implicit}
\end{equation}
For \(p=1\), this directly gives \(\mu\le hL(B)\).  For \(p>1\),
concavity and Young's inequality yield
\[
\begin{split}
\mu
&\le hL(B)G^q+hL(B)\mu^q\\
&\le hL(B)G^q+\frac12\mu+C_p\{hL(B)\}^p.
\end{split}
\]
Thus, in every dimension,
\begin{equation}
\mu\le
C_p\left[
hL(B)G^{(p-1)/p}+\{hL(B)\}^p
\right].
\label{eq:app-mu-upper}
\end{equation}
Since \(h=u^{1/p}\phip(A)^{1/p}\), the Gaussian and Gamma moments on the
right are finite.  For \(0<u\le1\), the first term has expectation
\(O(u^{1/p})\), and the second has expectation \(O(u)\), which is no
larger.  This proves \cref{eq:app-small-u-mu}.
\end{proof}

\subsection{The regular endpoint: almost-sure overlap scaling}

\begin{lemma}[Large-\(u\) scaling of the marginal-only mean]
\label{lem:app-large-u-scaling}
Under the coupling above,
\begin{equation}
\frac{M}{\sqrt u}
\longrightarrow
Q:=\left\{\frac{G\phip(A)}{\phip(B)}\right\}^{1/2},
\text{ }
\frac{\mu}{\sqrt u}\longrightarrow Q,
\label{eq:app-large-u-scaling}
\end{equation}
almost surely as \(u\to\infty\).
\end{lemma}

\begin{proof}
For each fixed \(v\), the triangular kernel in
\cref{eq:app-scalar-overlap} has total mass \(a^2\), and continuity of the
Gaussian density gives
\begin{equation}
\frac{\ell_v(a)}{a^2}\longrightarrow\varphi(v),
\text{ } a\downarrow0.
\label{eq:app-overlap-local}
\end{equation}
The product on the right of
\[
\frac{G}{h^p}=\prod_{j=1}^p\ell_{B_j}(a)
\]
is continuous and strictly increasing from zero to infinity.  Since
\(G/h^p=G/\{u\phip(A)\}\to0\) almost surely, its solution \(a=a_u\)
tends to zero. \Cref{eq:app-overlap-equation,eq:app-overlap-local} then give
\[
G
=\frac{M^2}{h^p}\phip(B)\{1+o(1)\},
\]
which proves the first convergence in
\cref{eq:app-large-u-scaling}.  The second follows from
\(\mu=M-G\) and \(G/\sqrt u\to0\).
\end{proof}

\subsection{Uniform integrability of the logarithmic scaling}

Pointwise convergence in \cref{lem:app-large-u-scaling} is not enough for
the bias, because \(A,B,G\) are unbounded and the score is logarithmic.
The next lemma supplies the needed envelope.

\begin{lemma}[Logarithmic moment convergence]
\label{lem:app-log-mu}
As \(u\to\infty\),
\begin{equation}
\E\log\frac{\mu}{\sqrt u}
\longrightarrow
\frac12\E\{\log G+\log\phip(A)-\log\phip(B)\}.
\label{eq:app-log-mu}
\end{equation}
Equivalently,
\begin{equation}
2\E\log\mu-\log u\longrightarrow\psi(k).
\label{eq:app-log-mu-cancel}
\end{equation}
\end{lemma}

\begin{proof}
We first control \(\log(M/\sqrt u)\), then show that replacing
\(M\) by \(\mu=M-G\) does not change the limiting expectation.
Let \(\varphi_0=\varphi(0)\).  From
\cref{eq:app-scalar-overlap},
\begin{equation}
\ell_v(a)\le a^2\varphi_0.
\label{eq:app-overlap-upper}
\end{equation}
For a lower bound, set \(b_0=\min(a/2,1/2)\).  On
\(|z|\le b_0\), we have \(a-|z|\ge a/2\) and
\[
\varphi(v+z)
\ge\varphi(v)e^{-|v|/2-1/8}.
\]
Therefore, with
\[
c(v)=\frac12\varphi(v)e^{-|v|/2-1/8},
\text{ } C(B)=\prod_{j=1}^pc(B_j),
\]
\begin{equation}
\ell_v(a)\ge
\begin{cases}
c(v)a^2,&0<a\le1,\\
c(v)a,&a>1.
\end{cases}
\label{eq:app-overlap-lower}
\end{equation}

Combine \cref{eq:app-overlap-equation,eq:app-overlap-upper} to obtain
\[
\frac M{\sqrt u}
\ge
\left\{\frac{G\phip(A)}{\varphi_0^p}\right\}^{1/2}.
\]
If \(a\le1\), \cref{eq:app-overlap-lower} gives
\[
\frac M{\sqrt u}
\le
\left\{\frac{G\phip(A)}{C(B)}\right\}^{1/2};
\]
if \(a>1\), it gives
\[
\frac M{\sqrt u}\le\frac{G}{C(B)\sqrt u}\le\frac G{C(B)}
\text{ }(u\ge1).
\]
Hence, for all \(u\ge1\),
\begin{equation}
\left\{\frac{G\phip(A)}{\varphi_0^p}\right\}^{1/2}
\le\frac M{\sqrt u}
\le
\left\{\frac{G\phip(A)}{C(B)}\right\}^{1/2}
+\frac G{C(B)}.
\label{eq:app-M-envelope}
\end{equation}
The logarithms of the two envelopes are bounded in absolute value by a
constant times
\begin{equation}
1+\|A\|_2^2+\|B\|_2^2+\|B\|_1
+|\log G|+\log(1+G).
\label{eq:app-integrable-envelope}
\end{equation}
This variable has moments of every finite order.  Thus
\(\{\log(M/\sqrt u):u\ge1\}\) is uniformly integrable, and
\cref{lem:app-large-u-scaling} implies
\begin{equation}
\E\log\frac M{\sqrt u}\longrightarrow\E\log Q.
\label{eq:app-log-M}
\end{equation}

\noindent It remains to replace \(M\) by \(\mu=M-G\).  Define
\[
\vartheta=\frac{\mu}{M}
=1-\prod_{j=1}^p\frac{\ell_{B_j}(a)}a.
\]
For one coordinate,
\begin{equation}
1-\frac{\ell_v(a)}a
=\E\min\{|W-v|/a,1\}.
\label{eq:app-overlap-fraction-loss}
\end{equation}
Since \(1-\prod_jq_j\ge1-q_1\) for \(0\le q_j\le1\),
\cref{eq:app-overlap-fraction-loss} and Gaussian unimodality give,
for \(a\le1\),
\[
\vartheta
\ge \Pp(|W-B_1|\ge a)
\ge \Pp(|W|\ge a)
\ge c_0:=\Pp(|W|\ge1)>0.
\]
The Gaussian probability of an interval of fixed length is largest
when the interval is centered at zero.
For \(a>1\),
\begin{equation}
\vartheta
\ge\frac{d(B_1)}a,
\text{ }
d(v)
=\E[|W-v|\ind\{|W-v|\le1\}]
\ge\frac14\varphi(|v|+1).
\label{eq:app-vartheta-lower}
\end{equation}
On the same event \(a>1\), the second line of
\cref{eq:app-overlap-lower} and the radius equation give
\begin{equation}
a^p
\le\frac{G}{u\phip(A)C(B)}
\le\frac{G}{\phip(A)C(B)}
\text{ }(u\ge1).
\label{eq:app-a-envelope}
\end{equation}
It follows from
\cref{eq:app-vartheta-lower,eq:app-a-envelope} that
\((-\log\vartheta)_+\) is also bounded by an integrable random variable of
the form \cref{eq:app-integrable-envelope}, uniformly over \(u\ge1\).
Because \(a_u\to0\), \(\vartheta_u\to1\) almost surely.  Dominated
convergence and \cref{eq:app-log-M} now yield
\[
\E\log\frac{\mu}{\sqrt u}
=\E\log\frac M{\sqrt u}+\E\log\vartheta
\longrightarrow\E\log Q,
\]
which is \cref{eq:app-log-mu}.

Finally, \(A\) and \(B\) are iid, so their expected log-density terms
cancel.  Differentiating the Gamma integral shows
\(\E\log G=\psi(k)\).  Multiplying
\cref{eq:app-log-mu} by two proves
\cref{eq:app-log-mu-cancel}.
\end{proof}

\subsection{The Poisson--digamma correction}

\begin{lemma}[Poisson--digamma identity]
\label{lem:app-poisson-digamma}
If \(N\sim\Pois(\mu)\), with \(\mu>0\), then
\begin{equation}
\E\psi(N+1)=\log\mu+E_1(\mu),
\text{ }
E_1(\mu)=\int_\mu^\infty\frac{e^{-t}}t\,\dd t.
\label{eq:app-poisson-digamma-one}
\end{equation}
Consequently, for every integer \(m\ge1\),
\begin{equation}
g_m(\mu)-\log\mu
=E_1(\mu)+\sum_{j=1}^{m-1}\E\frac1{N+j}.
\label{eq:app-poisson-digamma-m}
\end{equation}
The right side tends to zero as \(\mu\to\infty\) and is bounded by
\begin{equation}
C_m+(-\log\mu)_+.
\label{eq:app-poisson-digamma-envelope}
\end{equation}
\end{lemma}

\begin{proof}
Using \(\psi(N+1)=H_N-\gamma\) and
\[
H_N=\int_0^1\frac{1-t^N}{1-t}\,\dd t,
\]
Tonelli's theorem and \(\E t^N=e^{-\mu(1-t)}\) give
\[
\E H_N
=\int_0^\mu\frac{1-e^{-s}}s\,\dd s
=\gamma+\log\mu+E_1(\mu).
\]
This proves \cref{eq:app-poisson-digamma-one}.  The recurrence
\[
\psi(N+m)=\psi(N+1)+\sum_{j=1}^{m-1}\frac1{N+j}
\]
proves \cref{eq:app-poisson-digamma-m}.

\noindent For \(\mu\le1\),
\[
E_1(\mu)
\le-\log\mu+\int_1^\infty e^{-t}/t\,\dd t,
\]
and the reciprocal terms are bounded by \(\sum_{j<m}1/j\).  For
\(\mu\ge1\), all terms are bounded, and each tends to zero as
\(\mu\to\infty\).  For the reciprocal terms, couple
\(N_\mu\) as the count of a unit-rate Poisson process on \([0,\mu]\).
Then \(N_\mu\uparrow\infty\) almost surely, and dominated convergence
applies to \(1/(N_\mu+j)\).  These observations
give both the limit and \cref{eq:app-poisson-digamma-envelope}.
\end{proof}

\section{Saturation, Bias Correction, and the Gaussian Comparator}
\label{app:saturation}

We establish the saturation variance and bias continuity used in
\cref{cor:saturation,cor:calibration}, then give sufficient numerical-lookup
conditions and an exact Gaussian parametric comparator.

\subsection{Variance of the empirical fraction}

\begin{lemma}[Distribution-uniform saturation variance]
\label{lem:saturation-variance}
For an iid sample of size \(n>k\) from any law having a Lebesgue density
on \(\R^p\times\R^p\), evaluated with the strict, query-excluded
convention \cref{eq:counts},
\begin{equation}
\Var(\overline S_{n,k})\le\frac{D_{p,k}^2}{2n},
\text{ }
D_{p,k}=1+2\kappa_{2p}k+4\kappa_p(k+1),
\label{eq:saturation-variance}
\end{equation}
where \(\kappa_d\) is the geometric constant in
\cref{lem:app-reverse-knn}. In particular, the bound is independent of
the sampling density.
\end{lemma}

\begin{proof}
Write \(T(\mathcal W)=\sum_{i=1}^nS_{i,n}\), and replace observation
\(W_j\) by an independent copy \(W_j'\), giving
\(\mathcal W^{(j)}\). Work on the probability-one general-position
event used in \cref{thm:app-ksg-variance}, including absence of marginal
distance ties in both samples.

As in the variance proof for KSG, a common query's joint \(k\)th radius
can change only if the old point is among its first \(k\) joint
neighbors in \(\mathcal W\), or the new point is among its first
\(k\) joint neighbors in \(\mathcal W^{(j)}\).
\Cref{lem:app-reverse-knn} bounds the number of such queries by
\(2\kappa_{2p}k\). Together with the replaced query itself, they
account for at most \(1+2\kappa_{2p}k\) changed indicators.

Consider any remaining query. Its radius is unchanged and each marginal
count changes by at most one. If its saturation indicator changes, one of
the two configurations is saturated. Both counts in that configuration
are at most \(k\), and both counts in the other are therefore at most
\(k+1\). At least one marginal count changes. The old or new point
responsible for that change belongs to the corresponding strict marginal
ball in one of the two samples, whose count is at most \(k+1\).
Its marginal nearest-neighbor rank in that sample is consequently at most
\(k+1\). Applying \cref{lem:app-reverse-knn} to the old and new
points in each of the two marginal spaces bounds the number of these
radius-stable queries by \(4\kappa_p(k+1)\). A union bound suffices
even if a query is included more than once. \\
Each indicator changes by at most one, so
\[
|T(\mathcal W)-T(\mathcal W^{(j)})|\le D_{p,k}.
\]
The independent-replacement Efron--Stein inequality gives
\[
\Var(\overline S_{n,k})
\le\frac12\sum_{j=1}^n
\E\left[\left\{
\frac{T(\mathcal W)-T(\mathcal W^{(j)})}{n}
\right\}^2\right]
\le\frac{D_{p,k}^2}{2n}.
\]
This proves \cref{eq:saturation-variance}.
\end{proof}

\subsection{Continuity and integrability of the bias curve}

\begin{lemma}[Continuity of the critical bias]
\label{lem:bias-continuity}
For fixed \(p,k\), \(B_{p,k}\) is finite and continuous on
\((0,\infty)\), and hence bounded and uniformly continuous on every
compact subinterval.
\end{lemma}

\begin{proof}
First, for integers \(m\ge1\) and \(0\le s\le t\), couple
\(P_s\sim\Pois(s)\) and \(P_t=P_s+Q\), where
\(Q\sim\Pois(t-s)\) is independent. The digamma recurrence gives
\[
0\le\psi(m+P_t)-\psi(m+P_s)
=\sum_{j=0}^{Q-1}\frac1{m+P_s+j}\le\frac Qm.
\]
Every term is integrable, since
\(|\psi(m+P_s)|\le|\psi(m)|+P_s/m\). Recall from \cref{eq:gm} that
\(g_m(\mu)=\E[\psi(m+P)]\), where \(P\sim\Pois(\mu)\).
Taking expectations gives
\begin{equation}
|g_m(t)-g_m(s)|\le\frac{|t-s|}{m},
\text{ }
|g_m(t)|\le|\psi(m)|+\frac tm.
\label{eq:g-lipschitz}
\end{equation}
(Note that \(g_m\) is continuous)\\
Next, fix \(0<a<b<\infty\) and use the common coupling
\((A,B,G_k)\) from the proof of \cref{cor:saturation}, which gives
continuous paths \(u\mapsto\mu_u\), increasing in \(u\).
The pointwise estimate \cref{eq:app-mu-upper} was obtained before
restricting \(u\le1\); thus it holds at \(u=b\) as well. Since
\(\phip(A)\le(2\pi)^{-p/2}\), it implies
\[
0\le\mu_u\le\mu_b
\le C_p\left[b^{1/p}(C_p+\|B\|_1)G_k^{(p-1)/p}
             +b(C_p+\|B\|_1)^p\right],
\text{ } u\in[a,b].
\]
The right-hand side is integrable: all Gaussian moments are finite,
the indicated nonnegative Gamma moment is finite, and the query values and
arrival are independent. Consequently \(\E\mu_b<\infty\). \\
If \(u_j\to u\) within this interval, pathwise continuity and
\cref{eq:g-lipschitz} give \(g_m(\mu_{u_j})\to g_m(\mu_u)\)
almost surely, dominated by \(|\psi(m)|+\mu_b/m\).
Dominated convergence proves continuity of both expectations in
\cref{eq:B-definition}. Together with continuity of \(\log u\),
this proves continuity and finiteness of \(B_{p,k}\). Every point of
\((0,\infty)\) lies in such a compact interval. Boundedness and
uniform continuity on compact intervals then follow from continuity.
\end{proof}

\subsection{Consistency with Numerical Lookup Tables}
\begin{proposition}[A sufficient deterministic approximation condition]
\label{prop:numerical-calibration}
Under the assumptions of \cref{cor:calibration}, let
\(q_n:K\to\R\) be continuous and strictly decreasing, and let
\(B_n:K\to\R\) be measurable deterministic functions. Suppose
\begin{equation}
\epsilon_n=\sup_{v\in K}|q_n(v)-q(v)|\longrightarrow0,
\text{ }
\eta_n=\sup_{v\in K}|B_n(v)-B(v)|\longrightarrow0.
\label{eq:numerical-uniform}
\end{equation}
Set
\[
\widetilde u_n=q_n^{-1}\!\left(
\Pi_{[q_n(b),q_n(a)]}(\overline S_{n,k})\right).
\]
Then \(\widetilde u_n\to u\) and
\(\Ihat_{n,k}-B_n(\widetilde u_n)-I_n\to0\), both in \(L^2\).
\end{proposition}

\begin{proof}
If two closed intervals have corresponding endpoints within
\(\epsilon\), their clipping maps differ by at most \(\epsilon\)
at every real argument. This follows directly from
\(\Pi_{[c,d]}(s)=\min\{d,\max\{c,s\}\}\): clipping is
nondecreasing in each endpoint, and shifting both endpoints by
\(\pm\epsilon\) changes its value by at most \(\epsilon\).
Apply this fact to \(J_n=[q_n(b),q_n(a)]\) and
\(J=[q(b),q(a)]\). Since projection onto an interval is
1-Lipschitz and \(q(u)\in J\),
\begin{align*}
|q(\widetilde u_n)-q(u)|
&\le |q(\widetilde u_n)-q_n(\widetilde u_n)|
    +|\Pi_{J_n}(\overline S_{n,k})-\Pi_J(\overline S_{n,k})|\\
&\hspace{1em}+|\Pi_J(\overline S_{n,k})-\Pi_J(q(u))|\\
&\le2\epsilon_n+|\overline S_{n,k}-q(u)|.
\end{align*}
The right-hand side tends to zero in \(L^2\). Uniform continuity
of \(q^{-1}:J\to K\), followed by the bounded splitting argument
in the proof of \cref{cor:calibration}, gives \(\widetilde u_n\to u\) in \(L^2\).
The same argument gives \(B(\widetilde u_n)\to B(u)\) in \(L^2\).
Moreover,
\[
|B_n(\widetilde u_n)-B(u)|
\le\eta_n+|B(\widetilde u_n)-B(u)|.
\]
This tends to zero in \(L^2\), and the final decomposition in the
proof of \cref{cor:calibration} completes the proof.
\end{proof}
The proposition requires uniform convergence of both approximating curves
on a fixed interval \(K=[a,b]\subset(0,\infty)\). Exact values interpolated
linearly on increasingly fine grids satisfy \cref{eq:numerical-uniform}
by uniform continuity. The same conclusion holds when the errors at the
grid points vanish uniformly, provided \(q_n\) remains strictly decreasing.
Appendix~\ref{app:repro} reports interpolation and Monte Carlo sensitivity
checks for the fixed tables used in the experiments.

\subsection{An exact Gaussian parametric comparator}

When the channel family is known, its residual variance gives a direct
information estimate. The following Gaussian calculation gives an exact
benchmark.

\begin{proposition}[Residual-variance comparator]
\label{prop:gaussian-comparator}
For the Gaussian channel \cref{eq:channel}, with iid observations, define
\[
\widehat\delta_n=\frac1{np}\sum_{i=1}^n\|X_i-Y_{i,n}\|_2^2,
\text{ } \nu=np,
\text{ } c_\nu=\psi(\nu/2)-\log(\nu/2).
\]
Then \(\nu\widehat\delta_n/\delta_n\sim\chi^2_\nu\).
The untruncated plug-in estimator
\(\widehat I^{\mathrm{plug}}_n=-(p/2)\log\widehat\delta_n\)
has mean error \(-pc_\nu/2\). Its corrected version
\begin{equation}
\widehat I^{\mathrm G}_n
=-\frac p2\log\widehat\delta_n+\frac p2 c_\nu
\label{eq:gaussian-comparator}
\end{equation}
is exactly unbiased for \(I_n\), with
\begin{equation}
\E(\widehat I^{\mathrm G}_n-I_n)^2
=\frac{p^2}{4}\psi_1(\nu/2)
\le\frac{p}{2n}+\frac1{n^2},
\label{eq:gaussian-comparator-mse}
\end{equation}
where \(\psi_1\) is the trigamma function. For fixed \(p\), this
implies \(L^2\) consistency along every sequence
\(0<\delta_n<1\), with no restriction on \(u_n\).
\end{proposition}

\begin{proof}
The difference for observation \(i\) is
\[
X_i-Y_{i,n}
=\delta_nX_i-\sqrt{\delta_n(1-\delta_n)}Z_i.
\]
It is a centered Gaussian vector with covariance
\(\{\delta_n^2+\delta_n(1-\delta_n)\}I_p=\delta_nI_p\).
Its coordinates are independent, as are the different observations. Thus the
sum of all \(np\) squared standardized residuals is \(\chi^2_\nu\),
proving the claimed distribution. \\
Write \(z=\nu/2\). If \(T\sim\GP(z,1)\), the chi-square
identity yields \(\widehat\delta_n/\delta_n\stackrel d=T/z\).
For real \(s>-z\), integration of the Gamma density gives
\[
\E T^s=\frac{\Gamma(z+s)}{\Gamma(z)}.
\]
Differentiate twice at zero. The differentiations are justified by
domination on any sufficiently small closed interval around zero:
near zero, the integrands are bounded by a constant times
\(t^{z-1-\varepsilon}(1+|\log t|^2)\), with
\(0<\varepsilon<z\); near infinity, the exponential Gamma tail
dominates the powers and logarithms. Consequently
\[
\E\log T=\psi(z),\text{ }
\Var(\log T)=\psi_1(z).
\]
It follows that \(\E\log(T/z)=c_\nu\). Subtracting
\(I_n=-(p/2)\log\delta_n\) from the plug-in estimator proves
its stated mean error, and adding \(pc_\nu/2\) removes that error
exactly. The variance is \(p^2\psi_1(z)/4\).
Finally, the trigamma series and an integral comparison give
\[
\psi_1(z)=\sum_{j=0}^\infty\frac1{(z+j)^2}
\le\frac1{z^2}+\int_0^\infty\frac{\dd t}{(z+t)^2}
=\frac1{z^2}+\frac1z.
\]
Substituting \(z=np/2\) proves \cref{eq:gaussian-comparator-mse}.
The bound is independent of \(\delta_n\), so it applies to any
allowed sequence of channel parameters.
\end{proof}
This comparator uses the known Gaussian channel model and standardized
source covariance. The stated mean and variance formulas apply to the
untruncated estimator. Projecting the estimate onto \([0,\infty)\)
can only reduce its squared error because \(I_n>0\).

\section{Numerical evaluation and experimental methods}
\label{app:repro}

\subsection{Computing the Poisson Predictions}

To compute the predictions, we draw source and noise values \(A,B\) and a
Gamma arrival \(G_k\), then solve for the joint radius. In the notation
of Appendix \ref{app:endpoints},
\begin{align}
\ell_v(a)&=\int_0^a(a-z)
\{\varphi_1(v+z)+\varphi_1(v-z)\}\,\dd z,
\label{eq:compute-overlap}\\
\Lambda_{h,B}(r)&=h^p\prod_{j=1}^p\ell_{B_j}(2r/h).
\nonumber
\end{align}
The integrand is nonnegative. The reported calculations use a fixed upper
radius \(w\). If \(G_k<\Lambda_{h,B}(w)\), bisection on \([0,w]\)
determines \(R\); otherwise the draw is recorded as unresolved.
Appendix \ref{sec:simulation-settings} gives the search limits and their
checks. The resolved radius gives
\(\mu_u=(2R)^p-G_k>0\). \\
For saturation, one can average \(e^{-2\mu_u}\) over the independent
Monte Carlo draws. For bias, one may draw the Poisson marginal-only counts or integrate
them out. \Cref{lem:app-poisson-digamma} gives
\begin{equation}
g_m(\mu)=\log\mu+E_1(\mu)+
\sum_{j=1}^{m-1}\int_0^1t^{j-1}e^{\mu(t-1)}\,\dd t,
\text{ } \mu>0.
\label{eq:compute-gm}
\end{equation}
Indeed, for \(N\sim\Pois(\mu)\), Tonelli's theorem and the Poisson
generating function give
\[
\E\frac1{N+j}
=\int_0^1t^{j-1}\E t^N\,\dd t
=\int_0^1t^{j-1}e^{\mu(t-1)}\,\dd t.
\]
At zero, use \(g_m(0)=\psi(m)\). Near zero, direct summation of
Poisson-weighted digamma values avoids cancellation between
\(\log\mu\) and \(E_1(\mu)\).

\subsection{Reducing Monte Carlo Noise}

For the large-\(u\) calculation, we reduce Monte Carlo noise by subtracting
a random quantity whose expectation is exactly zero. Let the bias score
from one draw, in nats, be
\[
T_u=\log u+\psi(k)-\psi(M_x+1)-\psi(M_y+1),
\]
where the counts follow \cref{prop:countlaw}. Its expectation is
\(B_{p,k}(u)\). The control \(Z\) in \cref{eq:mc-control} has
exact mean zero because
\[
\E\log G_k=\frac{1}{\Gamma(k)}
\int_0^\infty g^{k-1}e^{-g}\log g\,\dd g
=\frac{\Gamma'(k)}{\Gamma(k)}=\psi(k)
\]
and \(\E\|A\|_2^2=\E\|B\|_2^2=p\). Therefore
\(\E(T_u-Z)=B_{p,k}(u)\). Both the raw and adjusted scores are
divided by \(\log2\) for bit units. The coefficient of \(Z\) is
fixed at one rather than estimated from the draws. The raw and adjusted
averages therefore estimate the same bias. The choice of \(Z\) follows
the large-\(u\) limit in Appendix \ref{app:endpoints}.

\subsection{Simulation Settings and Numerical Checks}
\label{sec:simulation-settings}

\textbf{Finite-sample simulations:}
We use dimensions \(p=1,2,4\), neighbor indices \(k=3,5,10\),
and sample sizes \(n=250,500,1000,2000,4000,8000\), with
30 independent datasets per dimension and sample size.
Table~\ref{tab:finite-u-grid} lists the values of \(u\).
For fixed \(u<n\), we set \(\delta_n=(u/n)^{2/p}\).
Within each repetition, the same Gaussian source and noise draws
\((X_i,Z_i)\) are reused across channel settings and values of \(k\).
\begin{table}[htbp]
\centering
\caption{Fixed-\(u\) Gaussian simulation grid. Each listed exponent
\(e\) specifies \(u=10^e\), at every sample size and neighbor index.}
\label{tab:finite-u-grid}
\small
\begin{tabular}{@{}cl@{}}
\toprule
\(p\) & Exponents \(e=\log_{10}u\)\\
\midrule
1 & \(-6,-5,-4,-3,-2.5,-2,-1.5,-1.25,-1,\)\\
  & \(-0.75,-0.5,-0.25,0,0.5,1,2\)\\[2pt]
2 & \(-9,-8,-7.5,-7,-6,-5,-4.5,-4,-3.5,-3,\)\\
  & \(-2.75,-2.5,-2.25,-2,-1.75,-1.5,-1,0,1,2\)\\[2pt]
4 & \(-12,-11,-10,-9,-8,-7.5,-7,-6.5,-6,\)\\
  & \(-5.75,-5.5,-5.25,-5,-4.75,-4.5,-4,-3,-2,-1,0,1,2\)\\
\bottomrule
\end{tabular}
\end{table}
We generate channel samples using \cref{eq:channel} and compute
the strict-radius counts in \cref{eq:counts} using max-norm
nearest-neighbor searches. The query is excluded, and we add
no jitter or marginal rescaling. To enforce strict inequalities,
marginal searches use the largest floating-point radius below
the selected joint radius. 


\textbf{Poisson calculations:}
Table~\ref{tab:poisson-settings} gives the numerical settings.
The original reference uses \(p=1,2,4\) and \(k=3,5,10\);
the extension and endpoint calculations use \(p=4,k=5\).
The upper search radius \(w\) is measured in the normalized
coordinates where each marginal has unit intensity.

\begin{table}[htbp]
\centering
\small
\setlength{\tabcolsep}{4pt}
\caption{Poisson calculation settings. Draw counts are per grid point.
The endpoint calculation uses two independent sets of 30,000 draws.
Order is the quadrature order; steps is the number of bisection steps.}
\label{tab:poisson-settings}
\begin{tabular}{@{}>{\raggedright\arraybackslash}p{1.05in}
                  >{\raggedright\arraybackslash}p{2.25in}rrrr@{}}
\toprule
Calculation & \(\log_{10}u\) grid & Draws & \(w\) & Order & Steps\\
\midrule
Original reference & \([-4,1]\), \([-7,1]\), \([-10,0]\) for
\(p=1,2,4\), respectively; spacing \(0.5\)
& 30,000 & 24 & 40 & 52\\
\addlinespace
Dimension-four extension & \(0,0.5,\ldots,8\)
& 30,000 & 128 & 40 & 52\\
\addlinespace
Endpoint & \(4,6,8,10,12\)
& \(2\times30{,}000\) & 512 & 40 & 56\\
\bottomrule
\end{tabular}
\end{table}
Within each calculation, the same draws of \((A,B,G_k)\)
are reused across \(u\) at fixed \((p,k)\).
The dimension-four extension is independent of the original reference.
The endpoint calculation reuses the extension's draws and adds
a second, independent set. \\
Overlaps are evaluated using Gaussian distribution functions when
\(2r/h>1\), and Gauss--Legendre quadrature of
\cref{eq:compute-overlap} otherwise. We solve the radius equation
by bisection. Bias estimates average scores from sampled marginal
counts, while saturation estimates average \(e^{-2\mu_u}\). \\
\textbf{Numerical checks:}
We solve the radius equation for every draw used in the Poisson
curves, endpoint estimates, and correction tables.
For the original reference, dimension-four extension, and endpoint
calculations, the largest normalized residual $\frac{|\Lambda_{h,B}(R)-G_k|}{\max\{1,G_k\}}$
was below \(1.2\times10^{-13}\).
Increasing the quadrature order from 20 to 40 produced negligible
changes in the joint intensities at the checked settings
for the original reference and extension. \\
\textbf{Reported intervals:}
For the main transition study, finite-sample error bars are
pointwise 95\% Student-\(t\) intervals across 30 independent datasets.
Poisson bands show Monte Carlo means plus or minus
1.96 standard errors, with probability intervals clipped to \([0,1]\).
These intervals describe sampling uncertainty only.
The convergence comparisons combine uncertainty in the finite-sample
means and Poisson references as described in
\cref{sec:fixed-transition-residuals}.

\subsection{Uncertainty in the Convergence Comparisons}
\label{sec:fixed-transition-residuals}

\textbf{Experimental design and intervals:}
The convergence comparisons reuse the simulations at all six sample sizes for \(p=1,2,4\), \(u=0.01,0.1,1\), and \(k=5\). These three \(u\) values are present in every reference grid and were selected before inspecting their residuals. Both the mean-residual and mean-square analyses use all 30 repetitions at these settings. Within each dimension and sample size, the datasets are paired across \(u\), while the finite and Poisson simulations are independent. The bias reference averages sampled-count scores, and the saturation reference averages \(e^{-2\mu}\). Residuals retain the finite-\(n\) term \(\psi(n)-\log n\) without subtraction.

Let \(s_f\) be the standard error across 30 independent finite datasets, and \(s_P\) the standard error across 30,000 independent Monte Carlo draws from the Poisson law. The displayed residual interval has half-width
\[
t_{0.975,\nu_W}\sqrt{s_f^2+s_P^2},\text{ with }
\nu_W=\frac{(s_f^2+s_P^2)^2}
{s_f^4/29+s_P^4/29999}.
\]
This approximate Welch--Satterthwaite interval combines the two independent sources of sampling error, with whole datasets serving as the finite-sample replicates. Because the same Poisson reference is reused across sample sizes and settings computed from the same source and noise samples are paired, we report intervals for individual settings rather than pooling across settings as independent. Section~\ref{sec:evidence} reports the results and the mean-square decomposition. \\
\textbf{Reference uncertainty:}
If \(\widetilde b\) is an independent unbiased Monte Carlo mean with variance \(v_b\), then
\[
\mathbb E\widetilde M_n
=\mathbb E(T_1-b)^2+v_b.
\]
Thus \(\widetilde M_n-s_b^2\) estimates the mean-squared discrepancy from the limiting bias, and \(\bar d^{\,2}-s_T^2/R-s_b^2\) estimates its squared-bias component, where \(s_b\) is the reference standard error. Every dataset uses the same numerical reference, so its estimated variance is subtracted once as \(s_b^2\). Subtracting the estimated variance terms can produce negative values through sampling variation; we retain these values. This adjustment accounts for Monte Carlo variance; numerical approximation errors are checked separately.

\subsection{Implementing and Testing Bias Correction}

\textbf{Lookup tables:}
We apply \cref{cor:calibration} using two independently computed tables of the Poisson bias and saturation curves: the first supplies the correction, and the second checks sensitivity to Monte Carlo noise. The two lookup tables are independent of one another and of the correction datasets. Both tables and the correction datasets are independent of the main transition study. Each table uses 30,000 independent draws of \((A,B,G_k)\) per dimension, reused across its grid, and evaluates the Poisson-averaged scores \(g_m\) directly. The \(\log_{10}u\) ranges are \([-1,1]\), \([-2.5,1]\), and \([-4.5,0.5]\) for \(p=1,2,4\), with spacing \(0.25\), fixed before the lookup simulation. \\
The finite-sample tests evaluate all midpoints between adjacent \(\log_{10}u\) grid values at \(n=1000,4000,8000\) with \(k=5\) and 16 independent dataset repetitions per dimension and sample size. Within a repetition, the source and noise samples are reused across \(u\). We interpolate both curves linearly in \(\log u\). To correct an estimate, we clip any saturation fraction outside the table's range to the nearest endpoint, interpolate \(\log u\) from that fraction, and subtract the interpolated bias. \\
\textbf{Numerical checks:}
The tables use \(w=64\), quadrature order 20, and 48 bisection steps. Checks at the endpoints, center, and off-grid interpolation points use \(w=128\), order 40, and 56 steps with the same Monte Carlo draws, so the comparison adds no variation from a new random sample. We also halve the \(\log_{10}u\) spacing to \(0.125\), using the same Monte Carlo draws and finite-sample datasets. The original test points then lie on the refined grid. We check interpolation at five additional midpoints per dimension. The main results use the original tables, and Section~\ref{sec:evidence} reports the refinement comparison. \\
Mean-error intervals are 95\% Student-\(t\) intervals across the 16 independent datasets per setting. RMSE intervals use the percentile bootstrap with 3,000 resamples of those datasets, sharing resampled datasets across methods. Both interval calculations hold the primary lookup table fixed; the second table checks sensitivity to its Monte Carlo error. All datasets are retained, including those with zero saturation or saturation fractions outside the table's range.

\subsection{Additional Bias-Correction Results}
\label{sec:additional-correction-results}

\begin{table}[!htbp]
\centering
\caption{RMSE in bits, pooled over off-grid \(u\) settings at \(k=5\).
Test windows differ across dimensions. All trials, including
clipped inversions, are retained.}
\label{tab:calibration}
\begin{tabular}{@{}cccccc@{}}
\toprule
\(p\) & \(n\) & Raw KSG & Saturation & Oracle & Gaussian\\
\midrule
1 & 1000 & 3.209 & 0.1353 & 0.0157 & 0.0216\\
1 & 4000 & 3.207 & 0.0688 & 0.0131 & 0.0136\\
1 & 8000 & 3.207 & 0.0453 & 0.0137 & 0.0124\\
\addlinespace
2 & 1000 & 6.176 & 0.1922 & 0.0185 & 0.0425\\
2 & 4000 & 6.177 & 0.1028 & 0.0095 & 0.0211\\
2 & 8000 & 6.175 & 0.0774 & 0.0090 & 0.0133\\
\addlinespace
4 & 1000 & 10.792 & 0.6215 & 0.0288 & 0.0635\\
4 & 4000 & 10.796 & 0.4922 & 0.0121 & 0.0307\\
4 & 8000 & 10.797 & 0.4118 & 0.0082 & 0.0258\\
\bottomrule
\end{tabular}
\end{table}

Table~\ref{tab:calibration} gives pooled results, while
Table~\ref{tab:calibration-matched} compares methods at matched \(u\).

At \(p=4,n=8000,k=5,u\approx0.75\), the lookup predicts about
84 saturated neighborhoods per dataset, whereas the observed mean
is about 44. None of the 16 estimates is clipped. Halving the
lookup-grid spacing changes the corrected mean error only from
\(-0.832\) to \(-0.822\) bit, and the refined table still predicts
nearly twice the observed saturation count. Together with the
accurate true-\(u\) correction, these checks point to estimating
\(u\) from finite-sample saturation as the main difficulty
in this example.

At the two largest dimension-four test settings with \(n=8000\),
13 and 16 of the 16 inversions are clipped to the upper boundary.
The decrease in corrected RMSE at the right edge of
Figure~\ref{fig:calibration} therefore does not by itself indicate
more accurate recovery of \(u\).

At the checked off-grid points, the largest interpolation
discrepancies were \(0.019\) bit for bias and \(0.55\) percentage
points for saturation probability. Switching to the independently
computed lookup table gives a root-mean-square difference of
\(0.039\) bit between corrected estimates, pooled over the tested
\(u\) settings at \(p=4,n=8000\).

\FloatBarrier

\end{document}